\documentclass[11pt,a4paper]{article}
\usepackage[a4paper, total={7in, 10in}]{geometry}
\usepackage{mathtools, mathrsfs, xcolor, hyperref, amssymb, fancyhdr}
\usepackage{blkarray}
\usepackage{float}
\usepackage{pdfpages}
\usepackage[T1]{fontenc}
\usepackage{datetime}
\usepackage{tikz}
\usetikzlibrary{positioning}
\usepackage{cite}

\newcommand*{\Comb}[2]{{}^{#1}C_{#2}}%

\hypersetup{
                    colorlinks=true,
                    linkcolor=blue,
                    urlcolor=blue,
                    citecolor=red,
                    }

\allowdisplaybreaks
\numberwithin{equation}{section}

\begin{document}
\begin{titlepage}
\title{\Huge{One-loop integrand from generalised scattering equations}}

\author{\Large{Md. Abhishek, Subramanya Hegde, Arnab Priya Saha} \\ 
\vspace{0.5cm}\\
Harish-Chandra Research Institute, \\
Homi Bhaba National Institute,\\
Chhatnag Road, Jhunsi, Allahabad 211019, India \\
\vspace{0.5cm}\\
E-mail: \href{mdabhishek@hri.res.in}{mdabhishek@hri.res.in}, \href{subramanyahegde@hri.res.in}{subramanyahegde@hri.res.in}, \href{arnabpriyasaha@hri.res.in}{arnabpriyasaha@hri.res.in}}

\date{}
 \maketitle
 
 \vspace{ 2cm}
\begin{abstract}
	
Generalised bi-adjoint scalar amplitudes, obtained from integrations over moduli space of punctured $\mathbb{CP}^{k-1}$, are novel extensions of the CHY formalism. These amplitudes have realisations in terms of  Grassmannian cluster algebras. Recently connections between one-loop integrands for bi-adjoint cubic scalar theory and $\mathcal{D}_n$ cluster polytope have been established.  In this paper using the $\text{Gr}\left(3,6\right)$ cluster algebra, we relate the singularities of $\left(3,6\right)$ amplitude to four-point one-loop integrand in the bi-adjoint cubic scalar theory through the $\mathcal{D}_{4}$ cluster polytope. We also study factorisation properties of the $(3,6)$ amplitude at various boundaries in the worldsheet. 

\end{abstract}
\thispagestyle{empty}
\end{titlepage}

\tableofcontents

\section{Introduction}

One of the important aspects of the S-matrix program is to develop efficient techniques for calculating scattering amplitudes. In the eighties, Parke and Taylor observed that MHV gluon amplitudes, which in the conventional Lagrangian formulation require computation of numerous Feynman diagrams, can be expressed in a nice compact form \cite{Parke:1986gb} using spinor helicity variables. Inspired by string theory, worldsheet formalisms for S-matrix computations are of interest as they offer novel insights  into quantum field theories. Witten's seminal work on twistor string theory \cite{Witten:2003nn} led to the formulation of integral representations of the S-matrix for $\mathcal{N}=4$ super Yang-Mills theory \cite{Roiban:2004yf}. BCFW recursion relations \cite{Cachazo:2004kj, Britto:2005fq, ArkaniHamed:2008yf} and ambitwistor string models \cite{Mason:2013sva, Geyer:2014fka, Casali:2015vta, Geyer:2015bja, Geyer:2015jch, Geyer:2016wjx, Geyer:2017ela, Geyer:2018xwu, Berkovits:2019bbx} are notable developments in the contemporary  S-matrix program.

Cachazo, He and Yuan (CHY) pioneered a remarkable formalism \cite{Cachazo:2013iaa, Cachazo:2013gna, Cachazo:2013hca, Cachazo:2013iea, Cachazo:2014fwa, Cachazo:2014nsa, Cachazo:2014xea} of computing the tree-level S-matrix for a plethora of non-supersymmetric theories of scalars, gluons and gravitons, by performing localised integrations over moduli space of punctured Riemann spheres,  $\mathfrak{M}_{0,n}$. The punctures correspond to external states of the S-matrix and are labelled by $\bigl\{z_{1}, z_{2}, \ldots z_{n}\bigr\}$ for an $n$-point amplitude. Singularities of the S-matrix in the kinematic space are mapped to marked points on $\mathbb{CP}^{1}$ via scattering equations, which form an essential ingredient in the CHY formalism. There are $n$ number of these equations,
\begin{equation}\label{scattering-eqn-chy}
E_{a} := \sum\limits_{\substack{b=1\\b\ne a}}^{n} \frac{s_{ab}}{z_{a} - z_{b}} = 0, \qquad \forall a \in \{1,2,\ldots n\}.
\end{equation}
The scattering equations above are invariant under $\text{SL}\left(2,\mathbb{C}\right)$ transformations. Therefore one can fix three of the punctures, and only $n-3$ equations are linearly independent. These equations are manifestations of momentum conservation. In the above, $s_{ab}$ are Mandelstam invariants and for massless fields, $s_{ab} = 2\; k_{a}.k_{b}$, where $k_{a}$ is the momentum of the $a$-th external state. In CHY formalism, the tree-level S-matrix with $n$-external states is expressed as,
\begin{equation}
M_{n} = \int \frac{\mathrm{d}z_{1}\; \mathrm{d}z_{2} \ldots \mathrm{d}z_{n}}{\mathrm{vol}\left[\text{SL}(2,\mathbb{C})\right]} \sideset{}{'}\prod\limits_{a} \delta \left( \sum\limits_{b \ne a}\frac{s_{ab}}
		      {z_{a}-z_{b}}\right) I_{n}(\{k,\epsilon,\sigma\}),
\end{equation}
where $I_{n}$ is a function of the kinematic data like momenta, polarizations of external states as well as positions of the punctures while its functional form depends on the particular theory under consideration. An example for the integrand is a product of two Parke-Taylor factors with orderings $\alpha$ and $\beta$,
\begin{equation}
I_{n} = \frac{1}{\left[\left(z_{\alpha_{1}} - z_{\alpha_{2}}\right)\left(z_{\alpha_{2}} - z_{\alpha_{3}}\right)\ldots \left(z_{\alpha_{n}} - z_{\alpha_{1}}\right)\right]\left[\left(z_{\beta_{1}} - z_{\beta_{2}}\right)\left(z_{\beta_{2}} - z_{\beta_{3}}\right)\ldots \left(z_{\beta_{n}} - z_{\beta_{1}}\right)\right]}.
\end{equation} 
This integrand gives rise to partial amplitudes, $m_{n}^{(2)}\left(\alpha|\beta\right)$, which have their origins in bi-adjoint cubic scalar theory. These amplitudes find an important application in KLT orthogonality relations between gauge theory and gravity \cite{Cachazo:2013gna, Cachazo:2016sdc}.

CHY constructions for massive fields \cite{Naculich_2014, Naculich_2015} and fermionic fields \cite{Weinzierl:2014ava} have also been studied. Loop amplitudes from forward limits have been studied in ambitwistor string theory in \cite{Adamo:2013tsa, Adamo:2015hoa}.  One-loop integrands for $n$-point amplitudes in the CHY formalism have been obtained from the forward limit of $\left(n+2\right)$-point tree-level amplitudes in \cite{He:2015yua, Cachazo:2015aol} and further explored in \cite{Feng:2016nrf, Feng:2019xiq, Farrow_2020}. Various factorisation properties of the CHY amplitudes under soft and collinear limits have been studied in \cite{Schwab, Afkhami, Zlotnikov, Kalousios, DoubleSoftPRD, VolovichZlotnikov, Saha:2017yqi, Saha:2016kjr, Chakrabarti:2017zmh, Nandan_2017}.

Recently, in \cite{Cachazo:2019ngv} Cachazo, Early, Guevara and Mizera (CEGM) introduced a generalisation of the scattering equations (\ref{scattering-eqn-chy}), where the punctures are defined on higher dimensional projective spaces, $\mathbb{CP}^{k-1}$, with $k\ge 3$. The generalized Mandelstam variables, $s_{a_{1}\; a_{2}\cdots a_{k}}$, are introduced that are symmetric in $k$ indices and satisfy the properties of,
\begin{eqnarray}
\sum\limits_{\substack{a_{2}, a_{3}\cdots a_{k}\\ a_{i} \ne a_{j}}} s_{a_{1} a_{2}\cdots a_{k}} &= & 0, \quad \forall a_{1} \in \{1,2, \cdots n\}, \qquad \text{momentum conservation and} \nonumber\\
s_{a_{1} a_{2}\cdots a_i\cdots a_j\cdots a_{k}} &= & 0, \quad \text{when}\; a_{i} = a_{j}, \qquad \text{massless condition}.
\end{eqnarray}
Let $\sigma_{a}$ denote the $a$-th puncture on $\mathbb{CP}^{k-1}$ and has $k-1$ inhomogeneous coordinates which we denote as $\sigma_{a} = \left(1\; x_{a}^{1}\; x_{a}^{2} \cdots x_{a}^{k-1}\right)^{\text{T}}$.  The generalised scattering potential function, introduced in \cite{Cachazo:2016ror}, is given as follows,
\begin{equation}\label{generalised-potential}
\mathcal{S}^{\left(k\right)} = \sum\limits_{1\le a_{1} < a_{2} < \ldots < a_{k} \le n} s_{a_{1}a_{2}\ldots a_{k}} \log |a_{1}\; a_{2} \ldots a_{k}|,
\end{equation}
where $|a_{1}\; a_{2} \ldots a_{k}|$ are determinants of the $k\times k$ minors formed by taking any $k$ punctures. Extremisation of  this potential function leads to a set of $n\left(k-1\right)$ equations, known as generalised scattering equations,
\begin{equation}\label{scattering-eqn-cegm}
E_{a}^{(i)} : = \frac{\partial \mathcal{S}^{(k)}}{\partial x_{a}^{i}} = 0, \qquad \forall a \in \{1,2, \ldots n\}, \quad \forall i = \{1,2,\ldots k-1\}.
\end{equation}
The above equations are invariant under $\text{SL}\left(k, \mathbb{C}\right)$ transformations, which implies that we can gauge fix the positions of any  $k+1$ punctures. Therefore the dimension of the moduli space of $n$ punctured $\mathbb{CP}^{k-1}$ is $\left(k-1\right)\left(n-k-1\right)$. So far, the only available example for amplitudes computed in the CEGM formalism is that of generalised bi-adjoint scalars. The $\text{SL}\left(k, \mathbb{C}\right)$-covariant Parke-Taylor factor with a particular ordering $\alpha = \left(a_{1}, a_{2}, \ldots a_{n}\right)$ can be defined as,
\begin{equation}
\text{PT}^{(k)}\left(\alpha\right) = \frac{1}{|a_{1}a_{2}\ldots a_{k}||a_{2}a_{3}\ldots a_{k+1}|\ldots |a_{n}a_{1}\ldots a_{k-1}|}.
\end{equation}
The generalised bi-adjoint scalar amplitude with $\alpha$ and $\beta$ orderings is then given by, 
\begin{equation}
m_{n}^{(k)}\left(\alpha|\beta\right) =  \int \left(\frac{1}{\text{Vol}\left[\text{SL}\left(k, \mathbb{C}\right)\right]}\prod\limits_{a=1}^{n}\prod\limits_{i=1}^{k-1}dx_{a}^{i}\right)\prod\limits_{a=1}^{n}\sideset{}{'}\prod\limits_{i=1}^{k-1} \delta\left(\frac{\partial \mathcal{S}^{(k)}}{\partial x_{a}^{i}}\right)\text{PT}^{(k)}\left(\alpha\right) \text{PT}^{(k)}\left(\beta\right). 
\end{equation}
The primed product denotes that $k+1$ delta functions have been removed. Various properties of these amplitudes have been investigated in \cite{Cachazo:2019apa, Sepulveda:2019vrz,  Borges:2019csl, Cachazo:2019ble, Cachazo:2019xjx,  Guevara:2020lek, Abhishek:2020xfy, Cachazo:2020wgu}.

CEGM amplitudes have realisations in terms of the Grassmannian spaces $\text{Gr}\left(k,n\right)$, and the $k=2$  case reduces to the CHY construction. In \cite{Drummond:2019qjk, Drummond:2020kqg} Drummond, Foster, G\"urdo\u{g}an and Kalousios used cluster algebra \cite{fomin2002cluster, Fomin_2003, 2006math......2259F, fomin2003systems} to study the CEGM amplitudes. The Grassmannian cluster algebra $\text{Gr}\left(2,n\right)$ is related to the $\mathcal{A}_{n-3}$ cluster algebra, whereas $\text{Gr}\left(3,6\right) \simeq \mathcal{D}_{4}$ and $\text{Gr}\left(3,7\right) \simeq \mathcal{E}_{6}$, are some examples of finite cluster algebras.

The realisation of scattering amplitudes as differential forms on positive geometries have paved the way to unravel beautiful geometric structures associated with the S-matrix. Amplituhedron program \cite{Arkani-Hamed:2013jha}  explored the connection between S-matrix in $\mathcal{N}=4$ SYM and positive Grassmannian spaces \cite{Arkani-Hamed:2016byb}. In \cite{Arkani-Hamed:2017tmz, Arkani-Hamed:2017mur} Arkani-Hamed, Bai, He, Lam and Yan discovered that the tree-level bi-adjoint amplitudes in the CHY formalism can be expressed in terms of canonical forms of certain polytopes, called the associahedron. This polytope resides in the kinematic space spanned by Mandelstam invariants. Moduli space of open string worldsheet is an associahedron, and the scattering equations of CHY act as diffeomorphism between the associahedron in the worldsheet and that described in the kinematic space. This led to a fascinating series of investigations into the connection between scattering amplitudes and positive geometries for various scalar theories \cite{He:2018pue, Banerjee:2018tun, Raman:2019utu, Aneesh:2019ddi, Kalyanapuram:2019nnf, Aneesh:2019cvt, Salvatori:2019phs, Arkani-Hamed:2019plo, Arkani-Hamed:2019vag,  He:2020ray, Arkani-Hamed:2020cig, He:2020onr, Kalyanapuram:2020vil, Kalyanapuram:2020tsr, Kalyanapuram:2020axt}.  Stringy deformations of the scattering forms have been considered in \cite{Arkani-Hamed:2019mrd}. Loop integrands from generalised associahedra \cite{Arkani-Hamed:2019vag, Jagadale:2020qfa} and other positive geometries \cite{Salvatori:2018aha} have also been studied. 
 
\subsection{Summary} 

CEGM amplitudes for arbitrary values of $k$ and $n$ are beautiful mathematical constructions. However, a satisfactory field theoretic formulation of these amplitudes is yet to be discovered. There are speculations in the literature that $\text{Gr}\left(4,n\right)$ amplitudes are related to singularities of loop-level amplitudes in $\mathcal{N}=4$ SYM theory. This motivates us to explore the first non-trivial example of CEGM amplitudes, the $\text{Gr}\left(3,6\right)$ amplitude. 

In \cite{Arkani-Hamed:2019mrd}, cluster string integral corresponding to the $\mathcal{D}_{4}$ cluster algebra has been considered, which in the $\alpha ' \rightarrow 0$ limit produces the four-point one-loop integrand for planar cubic scalar theory \cite{Arkani-Hamed:2019vag}. The positive geometry associated to the $\mathcal{D}_{4}$ cluster algebra is a four-dimensional polytope with sixteen co-dimension one facets. In \cite{Drummond:2020kqg}, by introducing two additional Mandelstam invariants, a basis of sixteen kinematic variables was obtained that captured the singularities of $\text{Gr}\left(3,6\right)$ amplitude.  The basis variables are in one-to-one correspondence with the facets of the $\mathcal{D}_{4}$ cluster polytope. However, the implication of this correspondence for the one-loop integrand in planar $\phi^{3}$ theory was left unexplored. In this work, we exploit this map to identify the singularities of $\text{Gr}\left(3,6\right)$ amplitude with the variables describing the $\mathcal{D}_{4}$ polytope in the kinematic space and subsequently provide Feynman diagrammatic representations of the amplitude with the prescription given in \cite{Arkani-Hamed:2019vag}. Consequently, we interpret the $\text{Gr}\left(3,6\right)$ amplitude as providing the four-point one-loop integrand in the planar $\phi^{3}$ theory.

The paper is organised as follows: In Sec.(\ref{sec:review}), we begin with a review of the relation between the moduli space of $n$ punctured $\mathbb{CP}^{k-1}$ and the Grassmannian $\text{Gr}(k,n)$ space, their tropicalisation, and their relation to the kinematic polytope and the amplitude. In Sec.(\ref{sec:Gr-36}), we present the realisation of the kinematic $\mathcal{D}_{4}$ polytope through the ray vectors of $\text{Gr}\left(3,6\right)$ cluster algebra. We present a relation between the kinematic variables for the one-loop polytope and the generalised Mandelstam variables. We map the singularities of the $\text{Gr}\left(3,6\right)$ amplitude with the facets of the $\mathcal{D}_{4}$ cluster polytope in Sec.(\ref{sec:1-loop}). Detailed results for clusters and related Feynman diagrams for this amplitude are provided in Appendix (\ref{sec:Gr-36-appendix}). In Sec.(\ref{sec:kinematic-polytope}) we find the precise constraints for the polytope in the  kinematic space arising from the correspondence between the generalised Mandelstam variables and the kinematic variables for the one-loop polytope.   We discuss factorisation properties of the amplitude in Sec.(\ref{Sec:factorisations}). In Sec.(\ref{sec:polytope-boundary}) we show the relations between different boundaries of $\mathcal{D}_{4}$ polytope in the kinematic space and boundaries in the worldsheet $u$-space. We consider an example of a forward limit in Appendix(\ref{sec:u2-boundary}) and show that the CHY representation of a six-point tree-level amplitude emerges at this boundary. In Sec.(\ref{sec:soft-limits}) we obtain sub-algebras related to soft limits of the $\text{Gr}\left(3,6\right)$ amplitude. Finally, we conclude with Sec.(\ref{sec:conclusion}).

\section{Cluster algebras and scattering potentials}
\label{sec:review}

In this section, we will review the relationship between the moduli space of $n$ punctured $\mathbb{CP}^{k-1}$ and the $\text{Gr}(k,n)$ space. We will review the tropicalisation of these spaces and the correspondence of positive tropical hypersurfaces to the amplitude. We will review how these tropical hypersurfaces can be obtained from cluster algebras and their relation with the polytope in the kinematic space. We will take the example of $\text{Gr}(2,6)$ throughout to illustrate various features. 

\subsection{Tropical Grassmannian and tropical fans}
We have discussed how the generalised kinematic space is mapped to an $n$ punctured $\mathbb{CP}^{k-1}$ in the CEGM formalism. To constuct the $\text{SL}(k,\mathbb{C})$ invariant integral for the CEGM amplitude a crucial object was the $k \times k$ determinant, which involved the coordinates of $k$ of the $n$ punctures on $\mathbb{CP}^{k-1}$. Let us forget for a moment the projective nature of the space that allows us to scale each puncture as $\sigma_a \sim t_a \sigma_a$ and consider each puncture as a $k$ component vector. The determinants can then be interpreted as $k \times k$ minors of a $k \times n$ matrix, where each column is a vector with $k$ complex components.  However, a $k \times n$ matrix can also describe a $k$ plane in $n$ dimensions up to an over-all rescaling. Thus if we ignore the scaling of each puncture, up to an over-all rescaling, the moduli space of $n$ punctured $\mathbb{CP}^{k-1}$ is the space of $k$ planes in $n$ dimensions i.e., the space $\text{Gr}(k,n)$. We can now take into account, scaling of the individual punctures. As an overall scaling is already accounted for, there are now $n-1$ independent scalings. This amounts to moding out the space $\text{Gr}(k,n)$ by $(\mathbb{C}^*)^{n-1}$ where each $\mathbb{C}^*=\mathbb{C}-\{0\}$ represents a scaling. Thus the moduli space of $n$ punctured $\mathbb{CP}^{k-1}$ is $\text{Conf}(k,n)$ defined as,
\begin{align}
	\text{Conf}(k,n)=\text{Gr}(k,n)/(\mathbb{C}^*)^{n-1}.
\end{align}
To obtain the dimensionality of this space note that we originally have a $k\times n$ matrix and we have the symmetries $\text{SL}(k,\mathbb{C})$ and $n$ rescalings naturally defined on $\mathbb{CP}^{k-1}$. Therefore the dimensionality is $(k-1)(n-k-1)$.

The $k\times k$ determinants of punctures discussed above, coordinatise this space in a highly redundant fashion. This is because, as they are minors of a $k\times n$ matrix, they satisfy the quadratic Pl\"ucker relations given as,
\begin{align}
	|a_1a_2\cdots [a_{r+1}\cdots a_k||b_1b_2\cdots b_{r+1}]b_{r+2}\cdots b_k|=0,
\end{align}
where there is an antisymmetrisation over $k+1$ indices\footnote{Even though the indices run over $n$ column values, one can shift this antisymmetrisation to the antisymmetrisation over $k$ row entries which makes the relation obvious.}. Pl\"ucker relations along with the rescalings bring down the number of coordinates from $\Comb{n}{k}$ to $(k-1)(n-k-1)$. To coordinatise the space more efficiently, consider the $k\times n$ matrix we discussed above. Using the $\text{SL}(k,\mathbb{C})$ transformation and the rescalings, we can gauge fix $k+1$ punctures and write the remaining $n-k-1$ punctures in terms of inhomogeneous coordinates with $k-1$ components. For simplicity, the first $k\times k$ block is set to identity. The rest of the columns are given by the matrix,
\begin{align}
	M_{ij}=(-1)^{i+k}\sum_{0\leq\lambda_{k-i}<\cdots <\lambda_2<\lambda_1\leq j-1} \prod_{r=1}^{k-i}\prod_{s=1}^{\lambda_r}x_{rs}.
\end{align}
The determinants $|a_1a_2\ldots a_k|$ are then given by determinants of $k\times k$ minors of the full $k\times n$ matrix where we choose the columns corresponding to the indices $a_1,a_2,\cdots,a_k$. These determiants are known as the Pl\"ucker coordinates or $\mathcal{A}$ coordinates for $\text{Gr}(k,n)$\footnote{We will interchangably use the notations $|a_1a_2\cdots a_k|=\langle a_1a_2\cdots a_k\rangle.$}. The $k\times n$ matrix is known as the web matrix, $W=(\mathbb{I}_{k\times k}|M)$.  Although we have gauge fixed the $\text{SL}(k,\mathbb{C})$ transformation, the coordinates $x_{rs}$ are defined in such a way that they are $\text{SL}(k,\mathbb{C})$ invariant ratios of the $\mathcal{A}$ coordinates. These are refered to as the $\chi$ coordinates of $\text{Gr}(k,n)$.

Let us consider the example of $\text{Gr}(2,6)$. The web matrix is given by,
\begin{align}\label{26web}
	W=\begin{pmatrix}
		1 & 0 & -1 & -1-x_{11} &-1-x_{11}-x_{11}x_{12} & -1-x_{11}x_{12}-x_{11}x_{12}x_{13} \\
		0 & 1 & 1 & 1 & 1 & 1
	\end{pmatrix}.
\end{align}
We can now compute the determinants to express the $\mathcal{A}$ coordinates in terms of the $\chi$ coordinates. We will obtain,
\begin{align}
	\langle 1i \rangle &=1,\nonumber\\
	\langle 23 \rangle &=1, \quad \langle 24\rangle =1+x_{11}, \quad \langle 25\rangle=1+x_{11}+x_{11}x_{12}, \quad \langle 26\rangle =1+x_{11}+x_{11}x_{12}++x_{11}x_{12}x_{13},\nonumber\\
	\langle 34\rangle &=x_{11}, \quad \langle 35\rangle=x_{11}(1+x_{12}), \quad \langle 36\rangle =x_{11}(1+x_{12}+x_{12}x_{13}),\nonumber\\
	\langle 45\rangle &=x_{11}x_{12},\quad \langle 46\rangle=x_{11}x_{12}(1+x_{13}),\nonumber\\
	\langle 56 \rangle &=x_{11}x_{12}x_{13}.
\end{align}
We can conversely write the $\chi$ coordinates in terms of $\mathcal{A}$ coordinates as,
\begin{align}
	x_{11}=\frac{\langle 12 \rangle \langle 34 \rangle}{\langle 14 \rangle \langle 23 \rangle }, \quad x_{12}=\frac{\langle 13 \rangle \langle 45 \rangle}{\langle 34 \rangle \langle 15 \rangle }, \quad x_{13}=\frac{\langle 14 \rangle \langle 56 \rangle}{\langle 45 \rangle \langle 16 \rangle }.
\end{align}
The guide for writing this in the above form is that the $\chi$ coordinates are written in terms of  $SL(2,\mathbb{C})$ invariant ratios of the determinants, as discussed earlier for general $k$. We can thus fix the punctures $1,2$ and $3$ using $\text{SL}(2,\mathbb{C})$ to any set of fixed values. However, the fact that we choose these particular punctures for gauge fixing remains, and we will comment on this choice later. Independent Pl\"ucker relations now read,
\begin{align}\label{plucker-relations}
	|a_1a_2||a_3a_4|-|a_1a_3||a_2a_4|+|a_1a_4||a_2a_3|=0, \hspace{2cm} 1 \leq a_1 < a_2 < a_3 < a_4 \leq 6.
\end{align}

So far, the relations we have considered such as Pl\"ucker relations or the relations between $\mathcal{A}$ coordinates and $\chi$ coordinates are nonlinear. To simplify the analysis, it is useful to  tropicalise these relations by replacing multiplication with addition and addition with minimum. Looking for solutions to the Pl\"ucker relations now translates to looking for tropical hypersurfaces that lie between different regions of linearity of the tropicalised Pl\"ucker expression, which reads,
\begin{align}\label{tropical-plucker-relations}
	\text{min}(w_{a_1a_2}+w_{a_3a_4},w_{a_1a_3}+w_{a_2a_4},w_{a_1a_4}+w_{a_2a_3}), \hspace{2cm} 1 \leq a_1 < a_2 < a_3 < a_4 \leq 6,
\end{align}
for the case $\text{Gr}(2,6)$. Here, $w_{a_1a_2}$ are tropicalised Pl\"ucker coordinates. The expression above has different regions of linearity in this tropicalised Pl\"ucker space depending on which of the entries are minimum. Hypersurfaces that separate these regions are known as tropical hypersurfaces, and are supposed to contain the same information as solving the Pl\"ucker relations. However, note that while tropicalising, we have lost the information on the relative sign between the different terms in the Pl\"ucker relations. To remedy this, one defines the notion of positive tropical hypersurfaces. For $\text{Gr}(2,6)$, we will have the tropical hypersurfaces,
\begin{align}
	w_{a_1a_2}+w_{a_3a_4}&=w_{a_1a_3}+w_{a_2a_4}\leq w_{a_1a_4}+w_{a_2a_3}, \nonumber\\
	w_{a_1a_2}+w_{a_3a_4}&=w_{a_1a_4}+w_{a_2a_3} \leq  w_{a_1a_3}+w_{a_2a_4},\nonumber\\
	w_{a_1a_4}+w_{a_2a_3} &=  w_{a_1a_3}+w_{a_2a_4} \leq w_{a_1a_2}+w_{a_3a_4}.
\end{align}
The first and the third inequalities above are said to define positive tropical hypersurfaces as the expressions being equal in these relations come from terms with a relative sign difference in Eq.(\ref{plucker-relations}). Solutions to the positive tropical hypersurface relations are collectively referred to as a fan. 

Although the equations are now linear, these hypersurfaces exist in the tropical Pl\"ucker space with $\Comb{n}{k}$ dimensions. We can instead study the tropical hypersurfaces in the space of tropicalised $\chi$ coordinates with $(k-1)(n-k-1)$ dimensions. For $\text{Gr}(2,6)$, the tropicalised Pl\"ucker coordinates are given in terms of tropicalised $\chi$ coordinates as,
\begin{align}\label{tropical-achi-relation}
	w_{1i}&=0,\nonumber\\
	w_{23}&=0, \quad w_{24}=\text{min}(0,\tilde{x}_{11}), \quad w_{25}=\text{min}(0,\tilde{x}_{11},\tilde{x}_{11}+\tilde{x}_{12}), \quad w_{26}=\text{min}(0,\tilde{x}_{11},\tilde{x}_{11}+\tilde{x}_{12},\tilde{x}_{11}+\tilde{x}_{12}+\tilde{x}_{13}),\nonumber\\
	w_{34}&=\tilde{x}_{11}, \quad w_{35}=\text{min}(\tilde{x}_{11},\tilde{x}_{11}+\tilde{x}_{12}), \quad w_{36}=\text{min}(\tilde{x}_{11},\tilde{x}_{11}+\tilde{x}_{12},\tilde{x}_{11}+\tilde{x}_{12}+\tilde{x}_{13}),\nonumber\\
	w_{45}&= \tilde{x}_{11}+\tilde{x}_{12}, \quad w_{46}=\text{min}(\tilde{x}_{11}+\tilde{x}_{12},\tilde{x}_{11}+\tilde{x}_{12}+\tilde{x}_{13}),\nonumber\\
	w_{56}&=\tilde{x}_{11}+\tilde{x}_{12}+\tilde{x}_{13}.
\end{align}
Using the above, we ask what are the different regions of linearity for tropicalised Pl\"ucker coordinates in the tropicalised $\chi$ space. These regions are separated by rays, which are collectively referred to as a fan in this space. For $\text{Gr}(2,6)$ a simple analysis of the above equations leads to the rays,
\begin{align}\label{26fans}
	\{\mathbf{e}_1, \; \mathbf{e}_2,\;  \mathbf{e}_3, \; -\mathbf{e}_1, \; -\mathbf{e}_2 \; ,-\mathbf{e}_3, \; \mathbf{e}_1-\mathbf{e}_2, \; \mathbf{e}_1-\mathbf{e}_3, \; \mathbf{e}_2-\mathbf{e}_3\},
\end{align}
where,
\begin{align}
	\mathbf{e}_1=(1,0,0),\quad \mathbf{e}_2=(0,1,0), \quad \mathbf{e}_3=(0,0,1).
\end{align} 
The rays are given as vectors in the tropicalised $\chi$ coordinate space. Using the relations in Eq.(\ref{tropical-achi-relation}), one can obtain the corresponding vectors in the tropicalised Pl\"ucker coordinate space, which precisely give us the positive tropical hypersurfaces. 

For eg.,
\begin{align}
	\text{ev}(\mathbf{e}_1)=(0,0,0,0,0,0,0,0,0,1,1,1,1,1,1).
\end{align}
We can construct a corresponding variable in the kinematic space by taking the dot product of the above vector with,
\begin{align}
	\mathbf{y}=(s_{12},s_{13},s_{14},s_{15},s_{16},s_{23},s_{24},s_{25},s_{26},s_{34},s_{35},s_{36},s_{45},s_{46},s_{56}).
\end{align}
By using momentum conservation, we obtain,
\begin{align}
	\mathbf{y}\cdot\text{ev}(\mathbf{e}_1)=s_{12}.
\end{align}
Corresponding to the rays in the fan Eq.(\ref{26fans}), we obtain the kinematic variables to be\footnote{For this mapping we had to use the relation between tropicalised $\chi$ coordinates and tropicalised Pl\"ucker coordinates. This relation depends on the gauge choice in the moduli space while defining Eq.(\ref{26web}).},
\begin{align}\label{26basis}
	\{s_{12},s_{123},s_{56},s_{23},s_{234},s_{16},s_{34},s_{345},s_{45}\},
\end{align}
in the same ordering. These are precisely the poles one would obtain in a six-point tree-level amplitude in canonical ordering for the biadjoint cubic scalar theory. In general, for $\text{Gr}(2,n)$, the above process gives the basis variables for planar amplitudes for this theory with $n$ external particles. We can further ask if one can obtain the full amplitude using the positive tropical Grassmannian. For this, we will need to obtain the compatible set of basis variables appearing in the amplitude. This is facilitated by the Grassmannian cluster algebra. 

\subsection{Cluster algebra and fans}

As discussed earlier, Pl\"ucker coordinates coordinatise the $\text{Gr}(k,n)$ manifold in a highly redundant fashion. Finding a choice of independent Pl\"ucker coordinates involves the tedious task of solving the quadratic Pl\"ucker relations. For the positive Grassmannian  $\text{Gr}^+(k,n)$ defined by demanding all the Pl\"ucker coordinates to be positive, the $\text{Gr}(k,n)$ cluster algebra gives an efficient way to find the independent Pl\"ucker coordinates. 

The fundamental object in a cluster algebra is the cluster quiver, which involves a set of frozen and unfrozen nodes connected by arrows. The algebra is then defined by mutations over unfrozen nodes, which lead to a different cluster quiver. For finite cluster algebras, this operation closes after a finite number of mutations. We will review below these concepts and how they can be used to find the compatible set of rays in the fan or basis variables leading us to the amplitude. We will illustrate it with the example of $\text{Gr}(2,6)$ cluster algebra. 

For $\text{Gr}(2,6)$, we can take the initial cluster to be as follows.
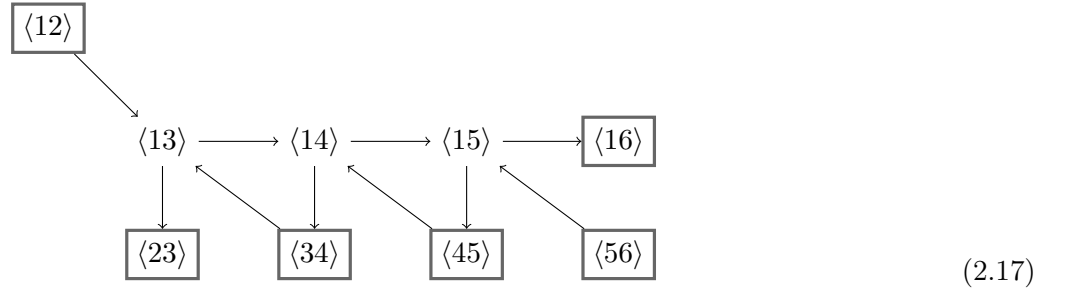
\begin{eqnarray}\label{26initialcluster}                            
	\begin{tikzpicture}[squarednode/.style={rectangle, draw=black!60, very thick, minimum size=5mm},
		]                      
		\node at (-2,0) (1)  {$\langle 13\rangle$};
		\node at (0,0) (2)  {$\langle 14\rangle$};
		\node at (2,0) (3) {$\langle 15\rangle$};
		\node[squarednode] at (4,0) (4) {$\langle 16\rangle$};
		\node[squarednode] at (-2,-1.5) (5)  {$\langle 23\rangle$};
		\node[squarednode] at (0,-1.5) (6)  {$\langle 34\rangle$};
		\node[squarednode] at (2,-1.5) (7) {$\langle 45\rangle$};
		\node[squarednode] at (4,-1.5) (8) {$\langle 56\rangle$};
		\node[squarednode] at (-3.5,1.5) (9) {$\langle 12\rangle$};
		\draw [->] (1) -- (2);
		\draw [->] (2) -- (3);
		\draw [->] (3)-- (4);
		\draw [->] (1) -- (5);
		\draw [->] (6) -- (1);
		\draw [->] (9) -- (1);
		\draw [->] (2) -- (6);
		\draw [->] (7) -- (2);
		\draw [->] (3) -- (7);
		\draw [->] (8) -- (3);
	\end{tikzpicture}
\end{eqnarray}
Several comments are in order. Notice firstly that the frozen nodes are the Pl\"ucker coordinates of the form $\langle ii+1\rangle$. This is so because when the Pl\"ucker coordinates are positive, these coordinates can not be made dependent. The unfrozen nodes give a choice of independent coordinates among the Pl\"cuker coordinates, which are not of this form. This is equivalent to specifying the edges of a cyclic polytope and triangulating it with non-intersecting chords. The Pl\"ucker relations then correspond to Ptolemy's theorem. Notice that the above initial cluster has a correspondence with the $\chi$ coordinates we defined earlier. Let us recall,
\begin{align}
	x_{11}=\frac{\langle 12 \rangle \langle 34 \rangle}{\langle 14 \rangle \langle 23 \rangle }, \quad x_{12}=\frac{\langle 13 \rangle \langle 45 \rangle}{\langle 34 \rangle \langle 15 \rangle }, \quad  x_{13}=\frac{\langle 14 \rangle \langle 56 \rangle}{\langle 45 \rangle \langle 16 \rangle }.
\end{align}
Consider the expression for $x_{11}$, the Pl\"ucker coordinates in the numerator(denominator) are those that are flowing into (out of) the unfrozen node $\langle 13 \rangle$ in the quiver. Simialrly $x_{12}$ and $x_{13}$ are associated with $\langle 14 \rangle $ and $\langle 15 \rangle $ unfrozen nodes respectively. Thus the choice of which punctures to gauge fix in the web matrix Eq.(\ref{26web}) is related to the choice of the initial cluster. Notice that the unfrozen nodes are connected by arrows which resemble an $\mathcal{A}_3$ Dynkin diagram, hence the $\text{Gr}(2,6) $ cluster algebra is known as the $\mathcal{A}_3$ cluster algebra. For general $n$, $\text{Gr}(2,n)$ cluster algebra becomes the $\mathcal{A}_{n-3}$ cluster algebra. 

Given the initial cluster, we need to define the mutation rules to obtain a subsequent cluster. Consider mutating over the $k^{\text{th}}$ unfrozen node with the $\mathcal{A}$ coordinate $a_k$, the mutation rules are as follows, to be followed in sequence.
\begin{itemize}
	\item In the rules to follow, ignore any rule that would connect two unfrozen nodes.
	\item If any two nodes are connected via the node $k$ by following the arrows, connect them by drawing an arrow between them in the same net direction.
	\item Reverse the direction of all the arrows connected to the node $k$.
	\item If there are two arrows in opposite directions between any two nodes, delete both the arrows.
	\item Replace the Pl\"ucker coordinate for the node $k$ by using the Pl\"ucker relations as follows\footnote{Note that this is particular to the $\text{Gr}(k,n)$ cluster algebras which have the Pl\"ucker cordinates at each node and the Pl\"ucker relations to mutate them. We can drop this mutation rule for more general cluster algebras, and mutate the quivers with abstract nodes. Even for $\text{Gr}(k,n)$ cluster algebras, we can choose not to associate the Pl\"ucker coordinates to the nodes this way. We will discuss this briefly in Sec.(\ref{sec:Gr-36}).}. 
	\begin{align}\label{amutation}
		a_k a^\prime_k=\prod_{i,b_{ik}<0}a_i^{-b_{ik}}+\prod_{i,b_{ik}>0}a_i^{b_{ik}},
	\end{align}
	where $a_i$ are $\mathcal{A}$ coordinates and $b_{ij}$ is the adjacency matrix defined as,
	\begin{align}
		b_{ij}=\text{No. of arrows from $i$ to $j$} - \text{No. of arrows from $j$ to $i$}.
	\end{align}
\end{itemize}
After this mutation, we get another quiver, which gives another consistent choice for independent Pl\"ucker coordinates. Note that mutation of the cluster mutates the adjacency matrix to a new adjacency matrix corresponding to the subsequent quiver. This mutation can be obtained by,
\begin{align}\label{bmutation}
	b^\prime_{ij}&=-b_{ij}, \quad \text{if $i=k$ or $ j=k$}\nonumber\\
	b^\prime_{ij}&=b_{ij}+\text{sign}[b_{ik}]\text{max}[0,b_{ik}b_{kj}], \quad  \text{if $i\neq k$ and $j \neq k$},
\end{align}
where each term above can easily be understood from the quiver mutation rules defined above. To generate a fan from the cluster algebra, we need to associate the unfozen nodes in the initial cluster to rays, which are the basis of $\mathbb{R}^m$, where $m$ is the number of unfrozen nodes. i.e., associate
\begin{align}
	\mathbf{g}_a=\mathbf{e}_a, \quad a=1,\cdots,m.
\end{align}
For $\text{Gr}(2,6)$ we have $\mathbf{e}_1=(1,0,0), \; \mathbf{e}_2=(0,1,0)$ and $\mathbf{e}_3=(0,0,1)$ associated with the unfrozen nodes $\langle 13 \rangle, \langle 14 \rangle $ and $\langle 15 \rangle $. Mutation of the rays under cluster mutations is then given by,
\begin{align}\label{fanmutation}
	\mathbf{g}^\prime_a&=\mathbf{g}_a, \quad \text{if $a \neq k$}\nonumber\\
	\mathbf{g}^\prime_k&=-\mathbf{g}_k+\sum_{i=1}^m \text{max}[0,-b_{ik}]\mathbf{g}_i+\sum_{i=1}^m\text{max}[0,c_{jk}]\mathbf{b^0}_j,
\end{align}
where $\mathbf{b^0}_j$ is the $j^{\text{th}}$ column of the adjacency matrix for the initial cluster and $c_{ij}$ is the coefficient matrix defined as the identity matrix for the initial cluster and follows the mutation rule,
\begin{align}\label{cmutation}
	c^\prime_{ij}&=-c_{ij},\quad\text{if $ j=k$}\nonumber\\
	c^\prime_{ij}&=c_{ij}-\text{sign}[c_{ik}]\text{max}[0,c_{ik}b_{kj}],\quad\text{if $j \neq k$}.
\end{align}
For $\text{Gr}(2,6)$, the above mutations also lead to the fan given in Eq.(\ref{26fans}). Further, as the rays are associated with basis variables for the amplitude, the set of rays corresponding to each cluster give a compatible set of basis kinematic variables leading to a particular term in the canonically ordered amplitude. From Eq.(\ref{26basis}), we can see that the initial cluster corresponds to the term,
\begin{align}
	\frac{1}{s_{12}s_{123}s_{56}},
\end{align}
which corresponds to the "caterpillar" Feynman diagram. Notice that we associated rays to the $\mathcal{A}$ coordinates of the cluster quiver, and we have already seen their association to the basis kinematic variables\footnote{Evidently, these associations depend on the choice of the initial cluster or equivalently the gauge choice in the moduli space made in Eq.(\ref{26web}).}. Thus the basis kinematic variables in Eq.(\ref{26basis}) are associated with the $\mathcal{A}$ coordinates,
\begin{align}\label{26acoord}
	\{\langle 13 \rangle, \langle 14 \rangle, \langle 15 \rangle, \langle 24 \rangle, \langle 25 \rangle, \langle 26 \rangle, \langle 35 \rangle, \langle 36 \rangle, \langle 46 \rangle \},
\end{align}
in the same ordering. In the triangulation picture, this tells us that we associate the chord $\langle ij \rangle$ is associated with the basis variable $s_{ii+1\cdots j-1}$. This mapping is precisely that obtained in the context of the kinematic associahedron in \cite{Arkani-Hamed:2017mur}.

Given a cluster algebra, we can ask what are the sub algebras of the cluster algebra. These correspond to freezing a particular unfrozen node and mutating over the rest of the unfrozen nodes. This way, one obtains the subalgebras that form the facets of the cluster polytope. For $\text{Gr}(2,6)$, there are $6 \; \mathcal{A}_2$ subalgebras and $3 \; \mathcal{A}_1\times \mathcal{A}_1$ subalgebras. In the triangulation picture, these correspond to holding a particular chord fixed in the cyclic hexagon. In the kinematic associahedron to be discussed later, these correspond to the pentagonal and quadrilateral facets of the associahedron.  

In the $\text{Gr}(2,6)$ case considered above, we associated the rays to both Pl\"ucker cordinates and the basis kinematic variables. Indeed, there are $9$ rays in the fan and $9$ unfrozen nodes that made the first association possible. This matching is true for all $\text{Gr}(2,n)$. However for $\text{Gr}(3,6)$, it was observed in \cite{Drummond:2020kqg} that this is no longer the case. Although one can associate the rays to basis generalised kinematic variables, the number of Pl\"ucker determinants no longer match the number of rays in the fan or the basis variables. This is reflected in the fact that the basis kinematic variables obtained from the fan are overcomplete. They satisfy conditions between them that render the cluster polytope non-simplicial. To overcome this, in \cite{Drummond:2020kqg}, the authors suggested including two quadratic $\mathcal{A}$ coordinates that appear in the central nodes of $\mathcal{D}_4$ shaped quivers that occur in the mutation of $\text{Gr}(3,6)$. Further, two new generalised Mandelstam variables were introduced, which made the entire mapping possible, and obtain a simplicial cluster polytope. We will review their results relevant for this paper in section \ref{sec:Gr-36}.

\subsection{Dihedral \texorpdfstring{$u$}- coordinates and boundaries of the moduli space}
So far, we have discussed the notion of cluster algebra and how they can be used to compute the amplitude by illustration. However, the fact that the $\text{Gr}(2,n)$ cluster algebra is related to the CHY integral over $n$ punctures can be understood from the perspective of the worldsheet associahedron. We will review the relevant aspects here through the example of $\text{Gr}(2,6)$.

Recall that for $\text{Gr}(2,6)$, we have the Mandelstam variables,
\begin{align}
	\{s_{12},s_{13},s_{14},s_{15},s_{16},s_{23},s_{24},s_{25},s_{26},s_{34},s_{35},s_{36},s_{45},s_{46},s_{56}\},
\end{align}
which obey the momentum conservation relation,
\begin{align}
	\sum_{b\neq a} s_{ab}=0, \quad a=1,2,\cdots,6.
\end{align}
We have the basis Mandelstam variables given by,
\begin{align}
	v_\alpha\equiv \{s_{12},s_{123},s_{56},s_{23},s_{234},s_{16},s_{34},s_{345},s_{45}\},
\end{align} 
where we have,
\begin{align}
	s_{123}&=s_{12}+s_{23}+s_{13},\nonumber\\
	s_{234}&=s_{23}+s_{34}+s_{24},\nonumber\\
	s_{345}&=s_{34}+s_{45}+s_{35}.
\end{align}
We can invert these relations to obtain the dependent Mandelstam variables in terms of basis variables using the above definitions and the momentum conservation relations to get,
\begin{align}
	s_{14}&=s_{23}+s_{56}-s_{123}-s_{234},\nonumber\\
	s_{15}&=-s_{56}-s_{16}+s_{234},\nonumber\\
	s_{25}&=s_{34} + s_{61} - s_{234} - s_{345},\nonumber\\
	s_{26}&= -s_{12} - s_{61} + s_{345},\nonumber\\
	s_{36}&=s_{12} + s_{45} - s_{123}-s_{345},\nonumber\\
	s_{46}&=-s_{45}-s_{56}+s_{123}.
\end{align}
We can now obtain the dihedral $u$ coordinates by writing the scattering potential in terms of the basis variables as,
\begin{equation}
	\sum\limits_{1\le i<j\le 6} s_{ij} \log \langle ij\rangle = \sum\limits_{\alpha} v_{\alpha}\log u_{\alpha},
\end{equation}
to obtain\footnote{Association between $u_\alpha$ and $v_\alpha$ does not depend on any gauge fixing in moduli space as we have seen here by obtaining the mapping in a gauge invariant fashion.},
\begin{align}
	u_{\alpha}=\biggl\{\frac{\langle 36 \rangle  \langle 12 \rangle}{\langle 13 \rangle \langle 26 \rangle}, \frac{\langle 13 \rangle  \langle 46 \rangle}{\langle 14 \rangle \langle 36 \rangle},\frac{\langle 56 \rangle  \langle 14 \rangle}{\langle 15 \rangle \langle 46 \rangle}, \frac{\langle 23 \rangle  \langle 14 \rangle}{\langle 13 \rangle \langle 24 \rangle}, \frac{\langle 15 \rangle  \langle 24 \rangle}{\langle 14 \rangle \langle 25 \rangle},\frac{\langle 16 \rangle  \langle 25 \rangle}{\langle 15 \rangle \langle 26 \rangle},\frac{\langle 34 \rangle  \langle 25 \rangle}{\langle 24 \rangle \langle 35 \rangle}, \frac{\langle 26 \rangle  \langle 35 \rangle}{\langle 25 \rangle \langle 36 \rangle}, \frac{\langle 45 \rangle  \langle 36 \rangle}{\langle 35 \rangle \langle 46 \rangle}\biggr\},
\end{align}
in the same ordering as $v_\alpha$ given in Eq.(\ref{26basis}) or the $\mathcal{A}$ coordinates in Eq.(\ref{26acoord}). The label $\alpha$ will be assigned according to the labels on the Pl\"ucker coordinates in Eq.(\ref{26acoord}). Notice that the $u_\alpha$ variables are cross ratios on the positive moduli space that run between $0$ and $1$. Their $\text{SL}(2,\mathbb{C})$ invariance is manifest.

The $u_\alpha$ variables obey compatibility relations due to the Pl\"ucker relations. For example, consider $u_{13}=\frac{\langle 36 \rangle  \langle 12 \rangle}{\langle 13 \rangle \langle 26 \rangle} $. We can look for products of $u$ variables which will give the same denominator as that of $u_{13}$. It is straight forward to obtain,
\begin{align}
	u_{24}u_{25}u_{26}=\frac{  \langle 16 \rangle \langle 23 \rangle}{\langle 13 \rangle \langle 26 \rangle}.
\end{align}
Using the Pl\"ucker relation $\langle 12 \rangle \langle 36 \rangle - \langle 13 \rangle \langle 26 \rangle + \langle 23 \rangle \langle 16 \rangle =0$, we obtain,
\begin{align}
	u_{13}+ u_{24}u_{25}u_{26}=1.
\end{align}
Another way to see this is to look for all the chords that cross the chord $13$ in the triangulation picture. All the $u$ variables which appear in the second term are said to be incompatible with the $u$ variable in the first term. From the compatibility relation above, it is clear that when $u_{13}\rightarrow 0$, all the incompatible $u$ variables must approach $1$. As the $u$ variables run between $0$ and $1$, the limit $u_{13}\rightarrow 0, u_{24} \rightarrow 1, u_{25} \rightarrow 1, u_{26} \rightarrow 1$ corresponds to a boundary of the positive moduli space. To see this in terms of the moduli space coordinates recall, $u_{13}=\frac{\langle 36 \rangle  \langle 12 \rangle}{\langle 13 \rangle \langle 26 \rangle}$. If punctures $1$ and $2$ collide, then $u_{13}\rightarrow 0$. However, in the same limit,
\begin{align}
	u_{24}=\frac{\langle 23 \rangle  \langle 14 \rangle}{\langle 13 \rangle \langle 24 \rangle} \rightarrow \frac{\langle 13 \rangle  \langle 14 \rangle}{\langle 13 \rangle \langle 14 \rangle }=1.
\end{align}
Thus the $u$ compatibility relations are useful in studying the boundary of positive moduli space. For $\text{Gr}(2,6)$, there are $9$ $u$ compatibility relations. With the compatibility relations, it can be found that any maximal set of compatible variables contains $3$ $ u$-coordinates as expected from a six-point tree-level amplitude. The compatible sets are,
\begin{align}
	&\{u_{13},u_{14},u_{15}\}, \{u_{24},u_{14},u_{15}\},\{u_{13},u_{35},u_{15}\}, \{u_{25},u_{34},u_{15}\}, \{u_{24},u_{25},u_{15}\}\nonumber\\
	&\{u_{13},u_{14},u_{46}\}, \{u_{13},u_{36},u_{35}\}, \{u_{13},u_{36},u_{46}\}, \{u_{24},u_{14},u_{46}\},\{u_{26},u_{25},u_{35}\}\nonumber\\
	&\{u_{24},u_{25},u_{26}\}, \{u_{26},u_{36},u_{35}\}, \{u_{13},u_{36},u_{35}\}, \{u_{26},u_{36},u_{46}\}.
\end{align}
The fourteen terms above correspond to the fourteen Feynman diagrams for the six-point tree-level cubic scalar amplitude when the $u$ variables above are replaced by their corresponding $v$ variables. By correspondence between the rays in the fan and $v_\alpha$ variables, there is a correspondence between rays and $u_\alpha$ variables. Each of the fourteen terms above thus corresponds to the $14$ cluster quivers of $\text{Gr}(2,6)$ cluster algebra. Indeed the $u$ variables above facilitate a push forward of the kinematic associahedron to the worldsheet associahedron, leading to the CHY integral on the worldsheet as found in \cite{Arkani-Hamed:2017mur}. The corresponding generalisation for $\text{Gr}(k,n)$ cluster algebra can be found in \cite{Arkani-Hamed:2019mrd}. In the following sections we will use the $u$ variables in $\text{Gr}(3,6)$ to find a mapping between the four-point one-loop cluster polytope and the $(3,6)$ CEGM amplitude. 

\subsection{Kinematic associahedron}
\label{sec:kin-assoc}
We discussed earlier that a cluster quiver in $\text{Gr}(2,n)$ corresponds to a complete triangulation of a cyclic polygon with $n$ edges. This is precisely how one obtains the vertices of the kinematic associahedron \cite{Arkani-Hamed:2017mur}. To realise the associahedron in the kinematic space, associate to the $u$ coordinates in the previous section corresponding kinematic coordinates $X$ with the same index structure. Equivalently, we can associate to each unfrozen $\mathcal{A}$ coordinate an $X$ variable with the same indices albeit now the indices are symmetric. Thus corresponding to, 
\begin{align}
	\{u_{13}, u_{14},u_{15},u_{24}, u_{25}, u_{26}, u_{35}, u_{36}, u_{46}\},
\end{align}
we have,
\begin{align}
	\{X_{13}, X_{14},X_{15},X_{24}, X_{25}, X_{26}, X_{35}, X_{36}, X_{46}\}.
\end{align}
However, $u_\alpha$ coordinates above are related to $v_\alpha$ coordinates,
\begin{align}
	v_\alpha\equiv \{s_{12},s_{123},s_{56},s_{23},s_{234},s_{16},s_{34},s_{345},s_{45}\}.
\end{align} 
This induces a mapping between the $X$ and $v$ coordinates. Thus the $X$ coordinates can be interpreted via Mandelstam variables as,
\begin{align}
	X_{ij}=s_{ii+1\cdots j-1}.
\end{align}
Note that the above mapping is independent of any gauge choice as the mapping between the $u_\alpha$ and $v_\alpha$ coordinates were obtained in a gauge invariant manner by using the scattering potential, as we have seen previously.

Now that we have the space of kinematic variables $X_{ij}$, we can ask how to realise the kinematic associahedron in this space. There are $\frac{n(n-3)}{2}$ number of kinematic variables whereas the kinematic space for scattering in cubic theories is $n-3$ dimensional. Therefore we need $\frac{(n-2)(n-3)}{2}$ number of constraints in this space to etch out a polytope. To find the constraints for $\text{Gr}(2,6)$, remember that the $v_\alpha$ variables are associated with the rays,
\begin{align}
	\{\mathbf{e}_1,\; \mathbf{e}_2, \; \mathbf{e}_3, \; -\mathbf{e}_1, \; -\mathbf{e}_2, \; -\mathbf{e}_3, \; \mathbf{e}_1-\mathbf{e}_2, \; \mathbf{e}_1-\mathbf{e}_3, \; \mathbf{e}_2-\mathbf{e}_3\}.
\end{align}
Therefore $X_{13},X_{14},X_{15}$ are associated with the basis rays $\mathbf{e}_1,\mathbf{e}_2$ and $\mathbf{e}_3$. Assign these kinematic variables to be the basis for the three dimensional kinematic space for the six-point amplitude. Define,
\begin{align}
	Y=(1,X_{13},X_{14},X_{15}).
\end{align} 
The constraints are then given as,
\begin{align}
	Y\cdot W_{ij}=0,
\end{align}
where the $W_{ij}$ is given by the association of $X_{ij}$ to the corresponding rays in the fan. For eg., $X_{46}$ is associated with the ray $\mathbf{e}_2-\mathbf{e}_3=(0,1,-1)$. Then $W_{46}$ is given by,
\begin{align}
W_{46}=(\mathcal{C}_{46},0,1,-1).
\end{align}
The constant $\mathcal{C}_{46}$ is found by,
\begin{align}
Y\cdot W_{46}=X_{46}.
\end{align}
The constraints to etch out the positive geometry is then given by,
\begin{align}
Y\cdot W_{ij}=0.
\end{align}
The full set of constraints can be compactly written as,
\begin{align}
X_{ij}+X_{i+1j+1}-X_{i+1j}-X_{ij+1}=-c_{ij}, \quad 2\leq i < j \leq n,
\end{align}
for non-adjacent $i,j$ and for $\text{Gr}(2,6)$, we have $n=6$. The constants $\mathcal{C}_{ij}$ and $c_{ij}$ are linear combinations of each other. Note that to define $W_{ij}$ we used the association between the rays in the fan and the Mandelstam variables. This, as we have discussed earlier, depends on our choice of the initial cluster or equivalently, on the choice of which punctures to gauge fix under $\text{SL}(2,\mathbb{C})$ for the web matrix. The initial cluster for the above constraints is Eq.(\ref{26initialcluster}), and the corresponding gauge fixed web matrix is Eq.(\ref{26web}). 

The constraints in \cite{Arkani-Hamed:2017mur} are,
\begin{align}
	X_{ij}+X_{i+1j+1}-X_{i+1j}-X_{ij+1}=-c_{ij}, \quad 1\leq i < j \leq n-1,
\end{align}
for non-adjacent $i,j$. To obtain the same fan as above for these constraints, we need to take $X_{26}, X_{36}$ and $X_{46}$ as the basis for our kinematic space. A simple way to diagnose this is to associate $\mathbf{e}_1,\mathbf{e}_2$ and $\mathbf{e}_3$ to those variables that do not have any $c_{ij}$ with the same indices. This choice of the constraints and the independent variables is reflected in our choice of which punctures to gauge fix in Eq.(\ref{26web}) and correspondingly, which initial cluster to begin with. Our initial cluster was given in Eq.(\ref{26initialcluster}), and has the unfrozen Pl\"ucker coordinates $\langle 13 \rangle, \langle 14 \rangle$ and $\langle 15 \rangle$. The initial cluster to obtain the constraints in \cite{Arkani-Hamed:2017mur} has the unfrozen coordinates $\langle 26 \rangle, \langle 36 \rangle$ and $\langle 46 \rangle$. In the triangulation picture this corresponds to choosing the chords $\{13,14,15\}$ or $\{26,36,46\}$. The latter is an anti clockwise rotation of the former. In either case, the respective constraints realise the kinematic associahedron whose canonical form leads to the same amplitude. Alternatively, if we take the constraints of \cite{Arkani-Hamed:2017mur}, with the independent coordinates $X_{13},X_{14},X_{15}$, the rays are related to the above rays by rotation. Hence once realises a rotated version of the same polytope. Thus the two choices correspond to realising the kinematic associahedron in different orientations in the same space. Different realisations of the associahedron are discussed in \cite{Bazier-Matte:2018rat,padrol2019associahedra}.

\section{\texorpdfstring{$\text{Gr}\left(3,6\right)$} - amplitude}
\label{sec:Gr-36}

In the previous section, we reviewed how rays in the tropical fan are associated with basis variables as well as $\mathcal{ A}$ coordinates in $\text{Gr}(2,6)$. These associations hold for all $\text{Gr}(2,n)$. However, in $\text{Gr}(3,6)$, the number of unfrozen coordinates no longer match the number of  unfrozen Pl\"ucker determinants.  There are $16$ rays and $14$ unfrozen Pl\"ucker coordinates. This is also reflected by the fact that when one computes the basis variables of kinematic space associated with the rays, the basis is overcomplete. The kinematic basis variables satisfy relations among them which makes the cluster polytope non-simplicial with bipyramid facets. To remedy this, a more refined fan was proposed in \cite{Drummond:2020kqg} where new generalised Mandelstam variables were added as well as two new quadratic $\mathcal{A}$ coordinates \cite{speyer2005tropical}. The generalisation of scattering potential function for $k=3, n=6$ reads,
\begin{equation}\label{ptential-3,6}
	F = \sum\limits_{1\le i<j<k\le 6} s_{ijk}\log\langle ijk \rangle + s_{q_{1}}\log q_{1} + s_{q_{2}}\log q_{2}.
\end{equation}
Here $\langle ijk \rangle$ are $3\times 3$ determinants of the minors formed out of three punctures labelled by $\sigma_{i}, \sigma_{j}$ and $\sigma_{k}$ on $\mathbb{CP}^{2}$. $\langle ijk \rangle$ is also equal to Plucker coordinates of $\text{Gr}\left(3,6\right)$ cluster algebra. Two additional variables, $q_{1}$ and $q_{2}$, which were originally not present in CEGM description \cite{Cachazo:2019ngv}, have been added here. These new quadratic variables are defined as,
\begin{align}
	q_1&=\langle 12[34]56\rangle=\langle 124 \rangle \langle 356 \rangle - \langle 123 \rangle \langle 456 \rangle,\nonumber\\
	q_1&=\langle 23[45]61\rangle=\langle 235 \rangle \langle 461 \rangle - \langle 234 \rangle \langle 561 \rangle.
\end{align}
Indeed $q_{1}$ and $q_{2}$ appear as the $\mathcal{ A}$ coordinates when one uses the mutation rule defined in Eq.(\ref{amutation}), as the central nodes of the $\mathcal{D}_4$ shaped clusters that appear in the $\text{Gr}(3,6)$ cluster mutations\footnote{As discussed earlier, mutation rule for the quivers can be performed without associating any $\mathcal{ A}$ coordinates to the quiver but only in terms of abstract nodes. The fan can still be determined from the mutation rules given in Eq.(\ref{fanmutation}) which depend only on the adjacency and coefficient matrices for the quiver and not on the association of $\mathcal{A}$ coordinates. This corresponds to the less refined Speyer-Williams fan.}. 

The corresponding over completeness of the basis is resolved by the addition of two new generalised Mandelstam variables, $s_{q_{1}}$ and $s_{q_{2}}$. Conservation of momenta with these two new variables are given by
\begin{equation}
	\sum\limits_{j < k} s_{ijk} + s_{q_{1}} + s_{q_{2}} = 0, \qquad \forall i.
\end{equation}
Now there are $22$ kinematic variables, which are $\{s_{ijk}, s_{q_{1}}, s_{q_{2}}\}, \quad 1\leq i < j < k \leq 6$. But due to conservation of momenta, not all of these variables are independent. For an amplitude in the canonical ordering, the basis variables \cite{Cachazo:2019ngv} are given by,
\begin{eqnarray}\label{basis-3,6}
	v_{a} & = & \{ s_{123}, s_{234}, s_{345,}  s_{456}, s_{156}, s_{126}, \nonumber\\
	&& t_{1234} + s_{q_{1}}, t_{2345} + s_{q_{2}}, t_{3456}+s_{q_{1}}, t_{4561} + s_{q_{2}}, t_{5612} + s_{q_{1}}, t_{6123}+ s_{q_{2}}, \nonumber\\
	&& r_{123456} + s_{q_{1}}, r_{234561} + s_{q_{2}}, r_{341256} + s_{q_{1}}, r_{452361} + s_{q_{2}}\},
\end{eqnarray}
where $t_{ijkl} = s_{ijk} + s_{jkl} + s_{ikl} + s_{ijl}$ and $r_{ijklmn} = t_{ijkl} + s_{klm} + s_{kln}$. It can immediately be checked that all the other kinematic variables appearing in Eq.(\ref{ptential-3,6}) can be expressed in terms of the above basis elements as follows,
\begin{eqnarray}\label{s-eqns}
	s_{124}&=&-s_{123}-(t_{5612}+s_{q_1})+(r_{123456}+s_{q_1}), \nonumber\\
	s_{125}&=&-s_{126}-(t_{1234}+s_{q_1})+(r_{341256}+s_{q_1}), \nonumber\\
	s_{134}&=&-s_{234}-(t_{3456}+s_{q_1})+(r_{341256}+s_{q_1}), \nonumber\\
	s_{135}&=&s_{234}+s_{456}+s_{126}-(r_{234561}+s_{q_2})-(r_{341256}+s_{q_1}), \nonumber\\
	s_{136}&=&-s_{126}-(t_{4561}+s_{q_2})+(r_{234561}+s_{q_2}), \nonumber\\
	s_{145}&=&-s_{456}-(t_{2345}+s_{q_2})+(r_{234561}+s_{q_2}), \nonumber\\
	s_{146}&=&-s_{156}-(t_{6123}+s_{q_2})+(r_{452361}+s_{q_2}), \nonumber\\
	s_{235}&=&-s_{234}-(t_{6123}+s_{q_2})+(r_{234561}+s_{q_2}), \nonumber\\
	s_{236}&=&-s_{123}-(t_{2345}+s_{q_2})+(r_{452361}+s_{q_2}), \nonumber\\
	s_{245}&=&-s_{345}-(t_{4561}+s_{q_2})+(r_{452361}+s_{q_2}), \nonumber\\
	s_{246}&=&s_{123}+s_{345}+s_{156}-(r_{123456}+s_{q_1})-(r_{452361}+s_{q_2}), \nonumber\\
	s_{256}&=&-s_{156}-(t_{3456}+s_{q_1})+(r_{123456}+s_{q_1}), \nonumber\\
	s_{346}&=&-s_{345}-(t_{1234}+s_{q_1})+(r_{123456}+s_{q_1}), \nonumber\\
	s_{356}&=&-s_{456}-(t_{5612}+s_{q_1})+(r_{341256}+s_{q_1}), \nonumber\\
	s_{q_1}&= &(t_{1234}+s_{q_1})+(t_{3456}+s_{q_1})+(t_{5612}+s_{q_1})-(r_{123456}+s_{q_1})-(r_{341256}+s_{q_1}), \nonumber\\
	s_{q_2}&= &(t_{2345}+s_{q_2})+(t_{4561}+s_{q_2})+(t_{6123}+s_{q_2})-(r_{234561}+s_{q_2})-(r_{452361}+s_{q_2}).
\end{eqnarray}
If we set $s_{q_{1}}$ and $s_{q_{2}}$ to zero, the last two equations in (\ref{s-eqns}) furnish constraints between the $t$ and $r$ variables. These constraints describe the non-simplicial bi-pyramid facets \cite{Cachazo:2019ngv, Guevara:2020lek}. As discussed earlier, in this case, we will have an overcomplete basis of $16$ variables where $14$ among them are inearly independent. 

Using Eq.(\ref{s-eqns}), the generalised potential in Eq.(\ref{ptential-3,6}) can be re-expressed in the following form,
\begin{equation}
	F = \sum\limits_{a} v_{a} \log u_{a}, 
\end{equation}
where the variables $u_{a}$ are given below in the same ordering in accordance to Eq.(\ref{basis-3,6}),
\begin{eqnarray}\label{u-3,6}
	u_a &=& \Biggl\{\frac{\langle 123\rangle \langle 246 \rangle}{\langle 124 \rangle \langle 236\rangle },\frac{\langle 234\rangle \langle 135 \rangle}{\langle 134 \rangle \langle 235\rangle }, \frac{\langle 345\rangle \langle 246 \rangle}{\langle 245 \rangle \langle 346\rangle },\frac{\langle 456\rangle \langle 135 \rangle}{\langle 145 \rangle \langle 356\rangle },\frac{\langle 156\rangle \langle 246 \rangle}{\langle 146 \rangle \langle 256\rangle },\frac{\langle 126\rangle \langle 135 \rangle}{\langle 125 \rangle \langle 136\rangle },\nonumber\\
	&&\frac{\langle 12[34]56 \rangle}{\langle 125 \rangle \langle 346\rangle },\frac{\langle 23[45]61 \rangle}{\langle 145 \rangle \langle 236\rangle },\frac{\langle 12[34]56 \rangle}{\langle 134 \rangle \langle 256\rangle },\frac{\langle 23[45]61 \rangle}{\langle 136 \rangle \langle 245\rangle },\frac{\langle 12[34]56 \rangle}{\langle 124 \rangle \langle 356\rangle },\frac{\langle 23[45]61 \rangle}{\langle 146 \rangle \langle 235\rangle },\nonumber\\
	&&\frac{\langle 124\rangle \langle 256 \rangle \langle 346\rangle}{\langle 246 \rangle \langle 12[34]56\rangle },\frac{\langle 136\rangle \langle 145 \rangle \langle 235\rangle}{\langle 135 \rangle \langle 23[45]61\rangle },\frac{\langle 125\rangle \langle 134 \rangle \langle 356\rangle}{\langle 135 \rangle \langle 12[34]56\rangle },\frac{\langle 146\rangle \langle 236 \rangle \langle 245\rangle}{\langle 246 \rangle \langle 23[45]61\rangle }\Biggr\}.
\end{eqnarray}
It is to be noted that $u_{a}$ variables form $\text{SL}\left(3, \mathbb{C}\right)$ invariant cross ratios. Remember that the mapping between the $v_a$ and $u_a$ variables is independent of any choice of gauge in the moduli space or equivalenty the choice of an initial cluster in the cluster agebra. These variables play an important role in defining the worldsheet polytope \cite{Arkani-Hamed:2019mrd} for the $\text{Gr}\left(3,6\right)$ amplitude.

\subsection{\texorpdfstring{$\mathcal{D}_{4}$} - cluster polytope}

Positive part of tropicalised $\text{Gr}\left(3,6\right)$ is spanned by $16$ ray vectors which are in one-to-one correspondence with the unfrozen nodes or $\mathcal{A}$ coordinates of the $\text{Gr}\left(3,6\right)$ cluster algebra. The ray vectors with an associated choice of cluster coordinates are given below\footnote{These mappings depend on a choice of the initial cluster as discussed in the case of $\text{Gr}(2,6)$.}:
\begin{alignat}{4}\label{ray-3,6}
	\langle 124 \rangle & \rightarrow \left(1, 0, 0, 0\right) &  \qquad \langle 125 \rangle &  \rightarrow  \left(0,1,0,0 \right) & \qquad  \langle 134 \rangle & \rightarrow \left( 0, 0, 1, 0\right) & \qquad  \langle 145 \rangle  & \rightarrow  \left(0, 0, 0, 1\right) \nonumber\\
	\langle 356 \rangle  & \rightarrow \left( -1, 0, 0, 0\right) & \qquad \langle 136 \rangle & \rightarrow \left( 0,-1, 0, 0\right) &\qquad \langle 235 \rangle  & \rightarrow \left( 0, 0, -1, 0\right) & \qquad \langle 236\rangle &  \rightarrow \left(0, 0, 0, -1\right) \nonumber\\
	\langle 146 \rangle  & \rightarrow \left(1, -1, 0, 0\right) & \qquad \langle 245\rangle  &  \rightarrow  \left(1, 0, -1, 0\right) & \qquad \langle 246\rangle  & \rightarrow  \left( 1, 0, 0, -1\right) & \qquad \langle 256\rangle  &  \rightarrow  \left( 0, 1, 0, -1\right) \nonumber\\
	\langle 346\rangle & \rightarrow \left(0, 0, 1, -1\right) & \qquad \langle 135\rangle & \rightarrow \left( -1, 0, 0, 1\right) &\qquad \langle 12\left[34\right]56\rangle & \rightarrow \left( 0, 1, 1, -1\right) & \qquad \langle 23\left[ 45\right]61\rangle & \rightarrow \left( 1, -1, -1, 0\right).
\end{alignat}
There are in total $50$ clusters in the $\text{Gr}\left(3,6\right)$ cluster algebra, and these are presented in Appendix (\ref{sec:clusters-3,6}).

As we have seen in the previous section, the ray vectors can be mapped to the basis elements of the kinematic space given in Eq.(\ref{basis-3,6}). We present below the relations between the $\mathcal{A}$ coordinates of $\text{Gr}\left(3,6\right)$ cluster algebra and the basis elements, through the identifications used in Eq.(\ref{ray-3,6}):
\begin{alignat}{4}
	\langle 124 \rangle  & \rightarrow s_{123} &  \qquad \langle 125 \rangle  & \rightarrow  t_{3456} + s_{q_{1}} & \qquad  \langle 134 \rangle & \rightarrow t_{1234} + s_{q_{1}} & \qquad  \langle 145 \rangle & \rightarrow  s_{456} \nonumber\\
	\langle 356 \rangle  & \rightarrow t_{5612} + s_{q_{1}} & \qquad \langle 136 \rangle & \rightarrow s_{234} &\qquad \langle 235 \rangle  & \rightarrow s_{612} & \qquad \langle 236\rangle  & \rightarrow t_{2345} + s_{q_{2}} \nonumber\\
	\langle 146 \rangle & \rightarrow t_{4561} + s_{q_{2}} & \qquad \langle 245\rangle  &  \rightarrow  t_{6123} + s_{q_{2}} & \qquad \langle 246\rangle  & \rightarrow r_{452361} + s_{q_{2}} & \qquad \langle 256\rangle  &  \rightarrow  s_{345} \nonumber\\
	\langle 346\rangle & \rightarrow s_{561} & \qquad \langle 135\rangle & \rightarrow r_{341256} + s_{q_{2}} &\qquad \langle 12\left[34\right]56\rangle & \rightarrow r_{123456} + s_{q_{1}} & \qquad \langle 23\left[ 45\right]61\rangle & \rightarrow r_{234561} + s_{q_{2}}.
\end{alignat}
Although we obtained the above map using Eq.(\ref{ray-3,6}), which is gauge dependent, the above relations between the Pl\"ucker coordinates and the generalised Mandelstam variables can be seen to be gauge independent just as the map between $X$ and $v_\alpha$ variables in the $\text{Gr}(2,6)$ case discussed in the previous section.  

Positive tropical $\text{Gr}\left(3,6\right)$ is related to a four-dimensional polytope known as the $\mathcal{D}_{4}$ cluster polytope. This polytope has $16$  facets of co-dimension one, each of which can  be represented by a ray vector of $\text{Gr}\left(3,6\right)$ cluster algebra, and there are $50$ vertices corresponding to the $50$ clusters. We can think of the $\mathcal{D}_{4}$ polytope as a four-dimensional positive geometry in the kinematic space whose facets are the basis variables of Eq.(\ref{basis-3,6}) and the vertices correspond to the $50$ terms of the $\text{Gr}\left(3,6\right)$ amplitude with the canonical ordering. Singularities of the amplitude are captured in the boundaries of this polytope.

\subsection{Singularities in the one-loop integrand}
\label{sec:1-loop}

Stringy realisation of cluster polytopes for finite dimensional cluster algebras has been introduced in \cite{Arkani-Hamed:2019mrd}. In this work, the authors considered generalisation of the string amplitudes to write them as canonical forms related to positive geometries of cluster algebras of finite type.  Stringy integral for ABHY associahedron \cite{Arkani-Hamed:2017mur} gives the disk integral of open string amplitude and in the $\alpha ' \rightarrow 0$ limit it reduces to field theory amplitude of bi-adjoint $\phi^{3}$ theory with canonical ordering. In \cite{Arkani-Hamed:2019vag} the authors have proposed that $\alpha '\rightarrow 0$ of the $\mathcal{D}_{4}$ cluster string integral, which is related to $\text{Gr}\left(3,6\right)$ cluster algebra, yields the integrand for one-loop  four-point amplitudes in bi-adjoint $\phi^{3}$ theory. Taking this as the motivation, we find a map between the two sets of kinematic variables: one appearing in \cite{Drummond:2020kqg} containing generalised Mandelstam variables and the other presented in \cite{Arkani-Hamed:2019mrd} which contains singularities of the $\mathcal{D}_{4}$ cluster string integral:
\begin{alignat}{4}\label{X-s}
	X_{1} & \leftrightarrow  s_{456} &\qquad   X_{2}  &\leftrightarrow  t_{6123} + s_{q_{2}} & \qquad   X_{3} & \leftrightarrow  s_{123}  & \qquad X_{4} & \leftrightarrow  t_{3456} + s_{q_{1}}  \nonumber\\\tilde{X}_{1} & \leftrightarrow  s_{234} & \qquad \tilde{X}_{2} & \leftrightarrow  t_{2345} + s_{q_{2}} & \qquad  \tilde{X}_{3} & \leftrightarrow  s_{561} & \qquad \tilde{X}_{4} & \leftrightarrow  t_{5612} + s_{q_{1}} \nonumber\\
	X_{12} & \leftrightarrow  r_{234561} + s_{q_{2}} & \qquad X_{13} & \leftrightarrow  t_{4561} + s_{q_{2}}  & \qquad  X_{23} & \leftrightarrow  r_{452361} + s_{q_{2}} & \qquad X_{24} & \leftrightarrow  s_{345}  \nonumber\\
	X_{34} & \leftrightarrow  r_{123456} + s_{q_{1}}  & \qquad  X_{31} & \leftrightarrow  t_{1234} + s_{q_{1}} & \qquad  X_{41} & \leftrightarrow  r_{341256} + s_{q_{1}} & \qquad X_{42} & \leftrightarrow  s_{612}.
\end{alignat}
Eq.(\ref{X-s}) is the main observation of this paper, and most of the subsequent inferences are based on this. Note that the association is based on the association between $v_a$ and $u_a$ variables, which is independent of the choice of the gauge or the initial cluster in the cluster algebra. 

These $X$ variables in the kinematic space are related to the corresponding $u$ variables of Eq.(\ref{u-3,6}) in the worldsheet, which give a binary realisation of the generalised associahedra \cite{Arkani-Hamed:2019plo}. Facets of the polytope in the kinematic space are in one-to-one correspondence with the boundaries of the space of the $u$ variables. These boundaries are realised in terms of the following non-linear constraints,
\begin{equation}\label{compatibity equations}
	u_{a} + \prod\limits_{b} u_{b}^{a\parallel b} = 1, \quad \forall a.
\end{equation}
The non-negative integer $a\parallel b$ is called compatibility degree from $a$ to $b$ \cite{fomin2003systems, Arkani-Hamed:2020tuz}. If two facets, $a$ and $b$ are compatible, and hence share a common boundary, then $a\parallel b = 0$, or otherwise the facets are incompatible. Each $u_{a}$ takes values in $\left[0,1\right]$; at any boundary, the corresponding $u_{a} \rightarrow 0$, and the incompatible variables $u_{b}$ become $1$. The non-linear constraints in Eq.(\ref{compatibity equations}) follow from the Plucker relations of $\text{Gr}\left(3,6\right)$ cluster algebra. As an example, it can be checked that, 
\begin{equation}
	u_{1} + \tilde{u}_{2}\tilde{u}_{3}\tilde{u}_{4}u_{23}u_{24}u_{34} = 1,
\end{equation} 
follows from $\langle 45[6\rangle \langle 135]\rangle  = 0$. There are $12$ such independent constraints.

The degree of incompatibility, $a\parallel b$ for different variables can be expressed in the following matrix form, 
\begin{equation}
	\begin{blockarray}{ccccccccccccccccc}
		& u_{1} & u_{2} & u_{3} & u_{4} & \tilde{u}_{1} & \tilde{u}_{2} & \tilde{u}_{3} & \tilde{u}_{4} & u_{12} & u_{13} & u_{23} & u_{24} & u_{34} & u_{31} & u_{41} & u_{42} \\  
		\begin{block}{c(cccccccccccccccc)}
			u_{1} & 0 & 0 & 0 & 0 & 0 & 1 & 1 & 1 & 0 & 0 & 1 & 1 & 1 & 0 & 0 & 0  \\
			u_{2} & 0 & 0 & 0 & 0 & 1 & 0 & 1 & 1 & 0 & 0 & 0 & 0 & 1 & 1 & 1 & 0 \\
			u_{3} & 0 & 0 & 0 & 0 & 1 & 1 & 0 & 1 & 1 & 0 & 0 & 0 & 0 & 0 & 1 & 1 \\
			u_{4} & 0 & 0 & 0 & 0 & 1 & 1 & 1 & 0 & 1 & 1 & 1 & 0 & 0 & 0 & 0 & 0 \\
			\tilde{u}_{1} &  0 & 1 & 1 & 1 & 0 & 0 & 0 & 0 & 0 & 0 & 1 & 1 & 1 & 0 & 0 & 0 \\
			\tilde{u}_{2} &  1 & 0 & 1 & 1 & 0 & 0 & 0 & 0 & 0 & 0 & 0 & 0 & 1 & 1 & 1 & 0 \\
			\tilde{u}_{3} &  1 & 1 & 0 & 1 & 0 & 0 & 0 & 0 & 1 & 0 & 0 & 0 & 0 & 0 & 1 & 1 \\
			\tilde{u}_{4} &  1 & 1 & 1 & 0 & 0 & 0 & 0 & 0 & 1 & 1 & 1 & 0 & 0 & 0 & 0 & 0 \\
			u_{12} & 0 & 0 & 1 & 1 & 0 & 0 & 1 & 1 & 0 & 0 & 1 & 1 & 2 & 1 & 1 & 0  \\
			u_{13} & 0 & 0 & 0 & 1 & 0 & 0 & 0 & 1 & 0 & 0 & 0 & 1 & 1 & 0 & 1 & 1  \\
			u_{23} & 1 & 0 & 0 & 1 & 1 & 0 & 0 & 1 & 1 & 0 & 0 & 0 & 1 & 1 & 2 & 1  \\
			u_{24} & 1 & 0 & 0 & 0 & 1 & 0 & 0 & 0 & 1 & 1 & 0 & 0 & 0 & 1 & 1 & 0  \\
			u_{34} & 1 & 1 & 0 & 0 & 1 & 1 & 0 & 0 & 2 & 1 & 1 & 0 & 0 & 0 & 1 & 1  \\
			u_{31} & 0 & 1 & 0 & 0 & 0 & 1 & 0 & 0 & 1 & 0 & 1 & 1 & 0 & 0 & 0 & 1  \\
			u_{41} & 0 & 1 & 1 & 0 & 0 & 1 & 1 & 0 & 1 & 1 & 2 & 1 & 1 & 0 & 0 & 0  \\
			u_{42} & 0 & 0 & 1 & 0 & 0 & 0 & 1 & 0 & 0 & 1 & 1 & 0 & 1 & 1 & 0 & 0  \\
		\end{block}
	\end{blockarray}.
\end{equation}
Using the compatibility degrees, we can write down the full list of compatible  sets of $u$ variables\footnote{These compatible sets can also be obtained from pseudo-triangulation method given in \cite{ceballos2015cluster, brodsky2015cluster}.}. There are $50$ such sets, which are presented below,
\begin{eqnarray}\label{compatible-sets}
	&&\{u_{1}, u_{2}, u_{3}, u_{4}\}, \{u_{1}, u_{2}, u_{3}, u_{13}\}, \{u_{1}, u_{2}, u_{4}, u_{42}\}, \{u_{1}, u_{2}, u_{12}, u_{13}\}, \{u_{1}, u_{2}, u_{12}, u_{42}\}, \nonumber\\
	&&\{u_{1}, u_{3}, u_{4}, u_{31}\}, \{u_{1}, u_{3}, u_{13}, u_{31}\}, \{u_{1}, u_{4}, u_{31}, u_{41}\}, \{u_{1}, u_{4}, u_{41}, u_{42}\}, \{u_{1}, \tilde{u}_{1}, u_{12}, u_{13}\}, \nonumber\\
	&&\{u_{1}, \tilde{u}_{1}, u_{12}, u_{42}\}, \{u_{1}, \tilde{u}_{1}, u_{13}, u_{31}\}, \{u_{1}, \tilde{u}_{1}, u_{31}, u_{41}\}, \{u_{1}, \tilde{u}_{1}, u_{41}, u_{42}\}, \{u_{2}, u_{3}, u_{4}, u_{24}\}, \nonumber\\
	&&\{u_{2}, u_{3}, u_{13}, u_{23}\}, \{u_{2}, u_{3}, u_{23}, u_{24}\}, \{u_{2}, u_{4}, u_{24}, u_{42}\}, \{u_{2}, \tilde{u}_{2}, u_{12}, u_{13}\}, \{u_{2}, \tilde{u}_{2}, u_{12}, u_{42}\}, \nonumber\\
	&&\{u_{2},\tilde{u}_{2}, u_{13}, u_{23}\}, \{u_{2}, \tilde{u}_{2}, u_{23}, u_{24}\}, \{u_{2}, \tilde{u}_{2}, u_{24}, u_{42}\}, \{u_{3}, u_{4}, u_{24}, u_{34}\}, \{u_{3}, u_{4}, u_{34}, u_{31}\}, \nonumber\\
	&&\{u_{3}, \tilde{u}_{3}, u_{23}, u_{24}\}, \{u_{3}, \tilde{u}_{3}, u_{24}, u_{34}\}, \{u_{3}, \tilde{u}_{3}, u_{34}, u_{31}\},\{u_{3}, \tilde{u}_{3}, u_{13}, u_{23}\}, \{u_{3}, \tilde{u}_{3}, u_{13}, u_{31}\}, \nonumber\\
	&&\{u_{4}, \tilde{u}_{4}, u_{24}, u_{34}\}, \{u_{4}, \tilde{u}_{4}, u_{24}, u_{42}\}, \{u_{4}, \tilde{u}_{4}, u_{34}, u_{31}\}, \{u_{4}, \tilde{u}_{4}, u_{31}, u_{41}\}, \{u_{4}, \tilde{u}_{4}, u_{41}, u_{42}\}, \nonumber\\
	&&\{\tilde{u}_{1}, \tilde{u}_{2}, \tilde{u}_{3}, \tilde{u}_{4}\}, \{\tilde{u}_{1}, \tilde{u}_{2}, \tilde{u}_{3}, u_{13}\}, \{\tilde{u}_{1}, \tilde{u}_{2}, \tilde{u}_{4}, u_{42}\}, \{\tilde{u}_{1}, \tilde{u}_{2}, u_{12}, u_{13}\}, \{\tilde{u}_{1}, \tilde{u}_{2}, u_{12}, u_{42}\}, \nonumber\\
	&&\{\tilde{u}_{1}, \tilde{u}_{3}, \tilde{u}_{4}, u_{31}\}, \{\tilde{u}_{1}, \tilde{u}_{3}, u_{13}, u_{31}\}, \{\tilde{u}_{1}, \tilde{u}_{4}, u_{31}, u_{41}\}, \{\tilde{u}_{1}, \tilde{u}_{4}, u_{41}, u_{42}\}, \{\tilde{u}_{2}, \tilde{u}_{3}, \tilde{u}_{4}, u_{24}\}, \nonumber\\
	&&\{\tilde{u}_{2}, \tilde{u}_{3}, u_{13}, u_{23}\}, \{\tilde{u}_{2}, \tilde{u}_{3}, u_{23}, u_{24}\}, \{\tilde{u}_{2}, \tilde{u}_{4}, u_{24}, u_{42}\}, \{\tilde{u}_{3}, \tilde{u}_{4}, u_{24}, u_{34}\}, \{\tilde{u}_{3}, \tilde{u}_{4}, u_{34}, u_{31}\}.
\end{eqnarray}
From these compatibility sets one can immediately read off the full amplitude corresponding to $\text{Gr}\left(3,6\right)$ cluster algebra; every set in Eq.(\ref{compatible-sets}) represents a term in the amplitude. For example, the first set gives $\frac{1}{X_{1}X_{2}X_{3}X_{4}}$. This also provides an alternative representation of the clusters presented in Appendix (\ref{sec:clusters-3,6}). 

In \cite{Arkani-Hamed:2019vag}, the authors have developed Feynman diagrams to depict the terms of the $\text{Gr}\left(3,6\right)$ amplitude. There, every Feynman diagram is formed of four propagators, and these diagrams capture all the singularity structures of one-loop four-point amplitudes in a planar ordered $\phi^{3}$ theory. All the $50$ Feynman diagrams are presented in Sec.(\ref{Sec:Feynman-diagrams}). Exploiting the mappings in Eq.(\ref{X-s}) we can associate the Feynman diagrams with the generalised bi-adjoint amplitude for $k=3, n=6$. With a word of caution, we emphasize that the association with Feynman diagrams in the strict sense is possible only in the limit $s_{q_{1}} \rightarrow 0$ and $s_{q_{2}} \rightarrow 0$. Most of our findings in the next section suggest that the existence of such a limit is a reasonable assumption. 

\subsection{Kinematic polytope}
\label{sec:kinematic-polytope}

The $X_{ij}, X_i, \tilde{X}_i$ variables define the kinematic space where the kinematic $\mathcal{D}_4$ polytope of is realised \cite{Arkani-Hamed:2019vag}. The polytope is realised by kinematic constraints. However, as discussed earlier kinematic constraints depend on the gauge choice or the choice of the initial cluster.

For the fan given in Eq.(\ref{ray-3,6}), we obtain the kinematic constraints to be,
\begin{align}
X_1-X_3+X_{41} &= \mathcal{C}_{41},\nonumber\\
X_{42}+X_{31} &= \mathcal{C}_{42},\nonumber\\
X_2-X_4+X_{31} &= \mathcal{C}_2,\nonumber\\ 
X_1-X_4+X_{34}-X_{31} &= \mathcal{C}_{34},\nonumber\\
\tilde{X}_1+X_4 &= \tilde{ \mathcal{C}}_{1},\nonumber\\
X_4-X_3+X_{13} &= \mathcal{C}_{13},\nonumber\\
\tilde{X}_3+X_1-X_{31} &= \tilde{\mathcal{C}}_3,\nonumber\\
X_3+\tilde{X}_4 &= \tilde{\mathcal{C}}_4,\nonumber\\
X_1-X_4+X_{24} &= \mathcal{C}_{24},\nonumber\\
X_{12}+X_{31}-X_3+X_4 &= \mathcal{C}_{12}\nonumber\\
X_{23}-X_4+X_1 &= \mathcal{C}_{23} \nonumber\\
\tilde{X}_2+X_1&=\tilde{\mathcal{C}}_2,
\end{align}
with $X_3,X_4,X_{31}$ and $X_1$ are the independent variables. The above constraints are found analogously to the constraints for the kinematic associahedron as reviewed in Sec.(\ref{sec:kin-assoc}). In our analysis to obtain this realisation of the kinematic polytope, we have used the initial cluster given below.
\begin{eqnarray}
\begin{tikzpicture}
\node at (-1,1) (1)  {$\langle 124\rangle$};
\node at (1,1) (2)  {$\langle 125\rangle$};
\node at (1,-1) (3) {$\langle 145\rangle$};
\node at (-1,-1) (4) {$\langle 134\rangle$};
\draw [->] (1) -- (2);
\draw [->] (2) -- (3);
\draw [->] (1)-- (4);
\draw [->] (4) -- (3);
\draw [->] (3) -- (1);
\end{tikzpicture}
\end{eqnarray}
To obtain the constraints in \cite{Arkani-Hamed:2019vag}, the choice of the initial cluster is as given below.
\begin{eqnarray}
\begin{tikzpicture}
\node at (-1.5,0) (1)  {$\langle 356\rangle$};
\node at (0,0) (2)  {$\langle 135\rangle$};
\node at (0.5,1) (3) {$\langle 125\rangle$};
\node at (0.5,-1) (4) {$\langle 235\rangle$};
\draw [->] (1) -- (2);
\draw [->] (3) -- (2);
\draw [->] (2) -- (4);
\end{tikzpicture} 
\end{eqnarray} 
Thus the choice of $\text{Gr}(3,6)$ initial cluster corresponds to a different realisation of the $\mathcal{D}_4$ cluster polytope in the kinematic space. Different realisations of the kinematic $\mathcal{D}_4$ polytope is discussed in \cite{Jagadale:2020qfa}. Note that for the constraints given above, the variable $X_0$ discussed in \cite{Arkani-Hamed:2019vag} does not exist as $\tilde{X}_i-X_i$ is not independent of $i$ when different constants are identified in the above constraints. Therefore, the $\text{Gr}(3,6)$ initial cluster choice leads to seeing the full $\mathcal{D}_4$ polytope while the natural halving to $\bar{\mathcal{D}}_4$ polytope is not available.    

\section{Factorisations of the \texorpdfstring{$\text{Gr}\left(3,6\right)$} - amplitude}
\label{Sec:factorisations}

Moduli space of $6$ punctures on $\mathbb{CP}^{2}$ is $4$-dimensional and factorisations of $\text{Gr}\left(3,6\right)$ amplitude originate from various boundaries of  this moduli space \cite{Cachazo:2019ngv}. In this section, we revisit the factorisation properties and their relations with the $\mathcal{D}_{4}$ cluster polytope in kinematic space, whose volume gives $\text{Gr}\left(3,6\right)$ amplitude. When restricted to the boundaries of the polytope, this amplitude exhibits various factorisation properties \cite{Arkani-Hamed:2019vag, Yang:2019esm}. 

\subsection{Boundaries of the \texorpdfstring{$\mathcal{D}_{4}$} - polytope}
\label{sec:polytope-boundary}

The $4$-dimensional polytope has $16$ co-dimension one facets. Each of these facets is associated with a ray given in Eq.(\ref{ray-3,6}) and represents a sub-algebra. Going to any boundary of the polytope is equivalent to computing the residue of the amplitude on the zero of the corresponding $X$ variable. 

There are $12$ $\mathcal{A}_{3}$ facets corresponding to the variables
\begin{equation}
	\biggl\{X_{1}, X_{2}, X_{3},  X_{4},  \tilde{X}_{1},  \tilde{X}_{2}, \tilde{X}_{3}, \tilde{X}_{4}, X_{13}, X_{24}, X_{31}, X_{42}\biggr\},
\end{equation}
and $4$ facets are of the form $\mathcal{A}_{1} \times \mathcal{A}_{1} \times \mathcal{A}_{1}$ corresponding to the variables
\begin{equation}
	\biggl\{X_{12}, X_{23}, X_{34}, X_{41}\biggr\}.
\end{equation}
%

Boundaries of the $\mathcal{D}_{4}$ polytope in kinematic space are in one-to-one correspondence with the polytope in the worldsheet, whose boundaries satisfy Eq.(\ref{compatibity equations}). Every $u_{a} \rightarrow 0$ is a co-dimension $1$ boundary of the worldsheet polytope and it is related to a propagator of the amplitude. Here we implicitly assume the limit $s_{q_{1}} \rightarrow 0$ and $s_{q_{2}} \rightarrow 0$. In this case, the bi-pyramid conditions imply that if four of the variables approach $0$, then the corresponding fifth variable also approaches $0$. However, as the $\text{Gr}(3,6)$ amplitude contains four poles in each term, setting $s_{q_1},s_{q_2}\rightarrow 0$ does not affect the factorisation properties. Unlike the punctures on $\mathbb{CP}^{1}$ for $n$-point amplitudes in $k=2$, where $u_{ij} = 0$ implies pinching of the punctures $\{z_{i}, z_{i+1}, \ldots z_{j-1}\}$ on one side and $\{z_{j}, \ldots z_{n}, z_{1}, \ldots z_{i-1}\}$ pinch on the other side, in this case there are multiple ways to reach any particular $u_{a} = 0$ boundary. 

Let us consider the facet $X_{1} = 0$ which corresponds to the propagator $s_{456}$. In the worldsheet the boundary is $u_{1} = 0$. In terms of the punctures, there are two possibilities:
\begin{enumerate}
	\item $\sigma_{1}, \sigma_{2}$ and $\sigma_{3}$ collide together simultaneously. In this case we have,
	\begin{equation}
		\langle 123 \rangle \sim \mathcal{O}\left(\varepsilon^{2}\right), \qquad \langle 12a \rangle \approx \langle 13a \rangle \approx \langle 23a \rangle \sim \mathcal{O}\left(\varepsilon\right), \qquad \langle abc\rangle \sim \mathcal{O}\left(\varepsilon^{0}\right), \qquad a, b, c \in \{4, 5, 6\}
	\end{equation}
	where $\varepsilon$ is the infinitesimal parameter denoting the rate of collision.
	\item $\sigma_{4}, \sigma_{5}$ and $\sigma_{6}$ are collinear to each other at a rate $\varepsilon$. In this case we have $\langle 456 \rangle \sim \mathcal{O}\left(\varepsilon\right)$ and all other determinants are of $\mathcal{O}\left(\varepsilon^{0}\right)$.
\end{enumerate}
It immediately follows that $u_{1} = \frac{\langle 456 \rangle\langle 135 \rangle}{\langle 145\rangle \langle 356 \rangle} \sim \mathcal{O}\left(\varepsilon\right)$ and goes to $0$. It can also be checked that the incompatible variables, $\bigl\{\tilde{u}_{2}, \tilde{u}_{3}, \tilde{u}_{4}, u_{23}, u_{24}, u_{34}\bigr\} \rightarrow 1$. For example, using $\langle 23\left[5\right.\rangle \langle \left.461\right]\rangle = 0$ we get,
\begin{equation}
	\langle 235\rangle\langle 461\rangle - \langle 234\rangle\langle 615\rangle + \langle 236\rangle\langle 154\rangle - \langle 231\rangle\langle 546\rangle = 0.
\end{equation}
The last term is $\mathcal{O}\left(\varepsilon\right)$ compared to the other terms and therefore, $\tilde{u}_{2} = \frac{\langle23[45]61\rangle}{\langle 145\rangle\langle 236 \rangle} = 1 + \mathcal{O}\left(\varepsilon\right)$.

As another example, we take $X_{2} \rightarrow 0$ facet which corresponds to the propagator $t_{6123}$. In this case there are again two possibilities:
\begin{enumerate}
	\item $\sigma_{4}$ and $\sigma_{5}$ collide with each other. Then we have $\langle 45a\rangle \sim \mathcal{O}\left(\varepsilon\right)$ and determinants of all the other minors are of $\mathcal{O}\left(\varepsilon^{0}\right)$.
	\item $\sigma_{1}, \sigma_{2}, \sigma_{3}, \sigma_{6}$ are simultaneously collinear. In this case ,
	\begin{equation}
		\langle 123\rangle \approx \langle 236 \rangle \approx \langle 126\rangle \approx \langle 136\rangle \sim \mathcal{O}\left(\varepsilon\right).
	\end{equation}
\end{enumerate}
From the Plucker relation $\langle 23\left[5\right.\rangle \langle \left. 461\right]\rangle = 0$, we obtain,
\begin{equation}
	\langle 235\rangle\langle 461\rangle - \langle 234\rangle\langle 615\rangle + \langle 236\rangle\langle 154\rangle - \langle 231\rangle\langle 546\rangle = 0.
\end{equation}
The last two terms are of $\mathcal{O}\left(\varepsilon\right)$ which implies $u_{2} = \frac{\langle 23[45]61\rangle}{\langle 146\rangle \langle 235} \rightarrow 0$. Again, using $\langle 12\left[4\right.\rangle \langle \left. 356\right]\rangle = 0$ gives
\begin{equation}
	\langle 124\rangle \langle 356 \rangle - \langle 123\rangle\langle 564\rangle + \langle 125\rangle\langle 643\rangle - \langle 126\rangle\langle 435\rangle = 0.
\end{equation}
The last term is of $\mathcal{O}\left(\varepsilon\right)$ and thus $u_{31} = \frac{\langle 12[34]56\rangle}{\langle 125\rangle\langle 346\rangle} \rightarrow 1$. Similarly it can be checked that other variables in the set $\bigl\{\tilde{u}_{1}, \tilde{u}_{3}, \tilde{u}_{4}, u_{31}, u_{31}, u_{41}\bigr\} \rightarrow 1$.

In the last example, we consider $X_{12} \rightarrow 0$ facet for which the corresponding propagator is $r_{234561}$. In this configuration two punctures, $\sigma_{1}$ and $\sigma_{6}$ collide with each other and at the same rate become collinear with two other punctures, $\sigma_{4}$ and $\sigma_{5}$. In this case, 
\begin{equation}
	\langle 16a\rangle \approx \langle 456\rangle \approx \langle 145\rangle \sim \mathcal{O}\left(\varepsilon\right), \qquad a\ne 1,6.
\end{equation}
Rest other determinants are of $\mathcal{O}\left(\varepsilon\right)$. This implies, $u_{12} = \frac{\langle 136\rangle\langle 145\rangle\langle 235\rangle}{\langle 135\rangle\langle 23[45]61\rangle} \sim \mathcal{O}\left(\varepsilon\right)$ and goes to $0$. Again, from the Plucker relation $\langle 12\left[4\right.\rangle \langle \left. 356\right]\rangle  = 0$, we obtain,
\begin{equation}
	\langle 124\rangle\langle 356\rangle - \langle 123\rangle\langle 564\rangle + \langle 125\rangle\langle 643\rangle - \langle 126\rangle \langle 435\rangle = 0.
\end{equation}
The second and fourth terms are of $\mathcal{O}\left(\varepsilon\right)$ and therefore we get $u_{31}\rightarrow 1$. Similarly, it can be checked that all the non-compatible variables, $\bigl\{u_{3}, u_{4}, \tilde{u}_{3}, \tilde{u}_{4}, u_{23}, u_{24}, u_{31}, u_{34}, u_{41}\bigr\} \rightarrow 1$.

The above three examples exhaust the possible boundary structures and rest of the cases of $X_{a} \rightarrow 0$ fall under one of these types.

\subsubsection{Forward limit}
At the boundaries $X_{i} = 0$ or $\tilde{X}_{i} = 0$, $i = 1,2,3,4$, the amplitude can be expressed as a forward limit of tree-level amplitudes leading to one-loop four-point amplitudes in the bi-adjoint scalar theory. From the perspective of the worldsheet this means that the residues at the co-dimension one boundaries, $u_{i} = 0$ or $\tilde{u}_{i} = 0$ take the form of the CHY integral representation of six-point amplitudes on the moduli space of punctured $\mathbb{CP}^{1}$. A particular case corresponding to $u_{2} = 0$ boundary is worked out in Appendix (\ref{sec:u2-boundary}). In this case the amplitude factorises on the channel $t_{6123}$ and with a particular gauge fixing, is given by
\begin{equation}
	m_{6}^{(3)}\left(\mathbb{I}| \mathbb{I}\right)  =  \frac{1}{t_{6123}} \int \mathrm{d}\alpha\; \mathrm{d}x_{5}\; \mathrm{d}y_{5}\; \delta\left(E_{\alpha}\right) \delta\left(E_{5}^{(x)}\right) \delta\left(E_{5}^{(y)}\right) \left[\frac{1}{\left(\alpha - \alpha_{1}\right) \left(x_{5} - 1\right)\left(x_{5} - y_{5}\right)}\right]^{2}.
\end{equation}
In the kinematic space, this factorisation of the amplitude is expressed as, 
\begin{equation}
	\mathcal{D}_{4} \xrightarrow{\partial_{X_{i}} \: \text{or} \: \partial_{\tilde{X}_{i}}} \mathcal{A}_{3}\left(i, i+1, \ldots , i-1, \pm, \mp\right).
\end{equation}
When the loop with an $X_{i}$ propagator is cut, the Feynman diagram becomes that of a six-point tree level diagram with two external legs, carrying momenta $\pm \ell$, inserted between the legs $i-1$ and $i$. 

We consider here an example when $X_{1} = 0$. At this boundary we have $u_{1} = 0$ and other incompatible variables, $\bigl\{\tilde{u}_{2}, \tilde{u}_{3}, \tilde{u}_{4}, u_{23}, u_{24}, u_{34}\bigr\}$ set to $1$. On this facet, the generalised potential function in Eq.(\ref{ptential-3,6}) becomes,
\begin{eqnarray}
	F  \mid_{u_{1}\rightarrow 0} & = &  X_{2}\log u_{2} + X_{3} \log u_{3} + X_{4} \log u_{4} + \tilde{X}_{1}\log \tilde{u}_{1} + X_{12}\log u_{12} \nonumber\\
	&& +  X_{13} \log u_{13} + X_{31} \log u_{31} + X_{41} \log u_{41} + X_{42} \log u_{42}. 
\end{eqnarray}
There are $14$  Feynman diagrams in the $\text{Gr}\left(3,6\right)$ amplitude which have non-vanishing residues at $X_{1} = 0$. When the $X_{1}$ propagator is cut, these diagrams take the the following configurations:
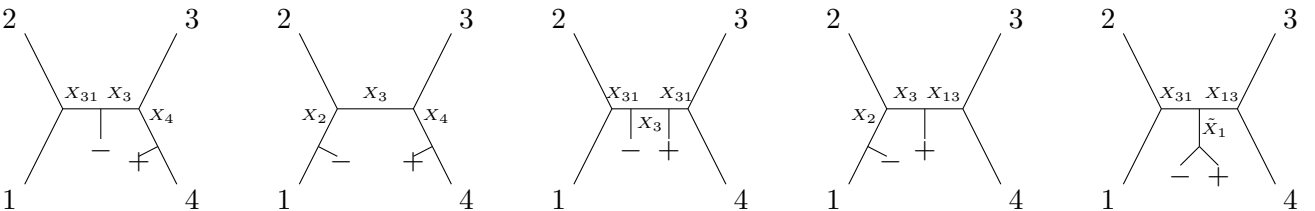
\begin{eqnarray}
	\begin{tikzpicture}                             
		\draw (-1, 1) -- (-0.5,0) -- (-1,-1);
		\draw (1, -1) -- (0.5,0) -- (1,1);
		\draw (0, 0) -- (0,-0.4);
		\draw (0.5, 0) -- (-0.5,0);
		\draw (0.75, -0.5) -- (0.5,-0.63);
		\node at  (-1.2, -1.2) {$1$};
		\node at (-1.2, 1.2) {$2$};
		\node at (1.2, 1.2) {$3$};
		\node at (1.2, -1.2) {$4$};
		\node at (0, -0.55) {$-$};
		\node at (0.5, -0.7) {$+$};
		\node at (-0.25, 0.2) {\tiny{$X_{31}$}};
		\node at (0.8, -0.1) {\tiny{$X_{4}$}};
		\node at (0.25, 0.2) {\tiny{$X_{3}$}};
	\end{tikzpicture} \qquad
	\begin{tikzpicture}                             
		\draw (-1, 1) -- (-0.5,0) -- (-1,-1);
		\draw (1, -1) -- (0.5,0) -- (1,1);
		\draw (-0.75, -0.5) -- (-0.5,-0.63);
		\draw (0.5, 0) -- (-0.5,0);
		\draw (0.75, -0.5) -- (0.5,-0.63);
		\node at  (-1.2, -1.2) {$1$};
		\node at (-1.2, 1.2) {$2$};
		\node at (1.2, 1.2) {$3$};
		\node at (1.2, -1.2) {$4$};
		\node at (-0.45, -0.7) {$-$};
		\node at (0.5, -0.7) {$+$};
		\node at (0.8, -0.1) {\tiny{$X_{4}$}};
		\node at (-0.8, -0.1) {\tiny{$X_{2}$}};
		\node at (0.0, 0.2) {\tiny{$X_{3}$}};
	\end{tikzpicture} \qquad
	\begin{tikzpicture}                             
		\draw (-1, 1) -- (-0.5,0) -- (-1,-1);
		\draw (1, -1) -- (0.5,0) -- (1,1);
		\draw (0.25, 0) -- (0.25,-0.4);
		\draw (0.5, 0) -- (-0.5,0);
		\draw (-0.25, 0) -- (-0.25,-0.4);
		\node at  (-1.2, -1.2) {$1$};
		\node at (-1.2, 1.2) {$2$};
		\node at (1.2, 1.2) {$3$};
		\node at (1.2, -1.2) {$4$};
		\node at (-0.25, -0.55) {$-$};
		\node at (0.25, -0.55) {$+$};
		\node at (-0.35, 0.2) {\tiny{$X_{31}$}};
		\node at (0.35, 0.2) {\tiny{$X_{31}$}};
		\node at (0.0, -0.2) {\tiny{$X_{3}$}};
	\end{tikzpicture} \qquad
	\begin{tikzpicture}                             
		\draw (-1, 1) -- (-0.5,0) -- (-1,-1);
		\draw (1, -1) -- (0.5,0) -- (1,1);
		\draw (0, 0) -- (0,-0.4);
		\draw (0.5, 0) -- (-0.5,0);
		\draw (-0.75, -0.5) -- (-0.5,-0.63);
		\node at  (-1.2, -1.2) {$1$};
		\node at (-1.2, 1.2) {$2$};
		\node at (1.2, 1.2) {$3$};
		\node at (1.2, -1.2) {$4$};
		\node at (0, -0.55) {$+$};
		\node at (-0.45, -0.7) {$-$};
		\node at (0.25, 0.2) {\tiny{$X_{13}$}};
		\node at (-0.8, -0.1) {\tiny{$X_{2}$}};
		\node at (-0.25, 0.2) {\tiny{$X_{3}$}};
	\end{tikzpicture} \qquad
	\begin{tikzpicture}                             
		\draw (-1, 1) -- (-0.5,0) -- (-1,-1);
		\draw (-0.5,0) -- (0.5,0);
		\draw (0, -0.5) -- (0.25,-0.75);
		\draw (0, -0.5) -- (-0.25,-0.75);
		\draw (0,0) -- (0,-0.5);
		\draw (1,1) -- (0.5, 0) -- (1, -1);
		\node at  (-1.2, -1.2) {$1$};
		\node at (-1.2, 1.2) {$2$};
		\node at (1.2, 1.2) {$3$};
		\node at (1.2, -1.2) {$4$};
		\node at (-0.3, 0.2) {\tiny{${X_{31}}$}};
		\node at (0.3, 0.2) {\tiny{$X_{13}$}};
		\node at (0.2, -0.25) {\tiny{$\tilde{X}_{1}$}};
		\node at (-0.25, -0.9) {$-$}; 
		\node at (0.25, -0.9) {$+$}; 
	\end{tikzpicture}  \nonumber\\
	\begin{tikzpicture}                            
		\draw (1,-1) -- (0,-0.5) -- (-1,-1);
		\draw (0,-0.5) -- (0,0.5);
		\draw (1,1) -- (0,0.5) -- (-1,1);
		\draw (-0.5,-0.75) -- (-0.37,-1.0);
		\draw (0.5,-0.75) -- (0.37,-1.0);
		\node at  (-1.2, -1.2) {$1$};
		\node at (-1.2, 1.2) {$2$};
		\node at (1.2, 1.2) {$3$};
		\node at (1.2, -1.2) {$4$};
		\node at (-0.3, -1.05) {$-$}; 
		\node at (0.3, -1.05) {$+$}; 
		\node at (-0.3, -0.4) {\tiny{$X_{2}$}};
		\node at (0.3, -0.4) {\tiny{$X_{4}$}};
		\node at (0.3, 0) {\tiny{$X_{42}$}};
	\end{tikzpicture} \qquad
	\begin{tikzpicture}                             
		\draw (-1, 1) -- (-0.5,0) -- (-1,-1);
		\draw (-0.5,0) -- (0.5,0);
		\draw (1, -1) -- (0.5,0) -- (1,1);
		\draw (0.66,-0.32) -- (0.41,-0.45);
		\draw (0.84,-0.68) -- (0.59,-0.81);
		\node at  (-1.2, -1.2) {$1$};
		\node at (-1.2, 1.2) {$2$};
		\node at (1.2, 1.2) {$3$};
		\node at (1.2, -1.2) {$4$};
		\node at (0.25, -0.5) {$-$};
		\node at (0.5, -0.9) {$+$};
		\node at (0, 0.2) {\tiny{${X_{31}}$}};
		\node at (0.85, -0.1) {\tiny{$X_{41}$}};
		\node at (0.55, -0.6) {\tiny{${X}_{4}$}};
	\end{tikzpicture} \qquad
	\begin{tikzpicture}                             
		\draw (-1, 1) -- (-0.5,0) -- (-1,-1);
		\draw (-0.5,0) -- (0.5,0);
		\draw (1, 1) -- (0.5,0) -- (1,-1);
		\draw (0.75, -0.5) -- (0.5,-0.63);
		\draw (0.5,-0.63) -- (0.17,-0.52);
		\draw (0.5,-0.63) -- (0.39,-0.96);
		\node at  (-1.2, -1.2) {$1$};
		\node at (-1.2, 1.2) {$2$};
		\node at (1.2, 1.2) {$3$};
		\node at (1.2, -1.2) {$4$};
		\node at (0, -0.6) {$-$};
		\node at (0.25, -1.0) {$+$};
		\node at (0, 0.2) {\tiny{${X_{31}}$}};
		\node at (0.65, -0.8) {\tiny{$\tilde{X}_{1}$}};
		\node at (0.9, -0.2) {\tiny{$X_{41}$}};
	\end{tikzpicture} \qquad
	\begin{tikzpicture}                            
		\draw (-1,1) -- (0,0.5) -- (1,1);
		\draw (0,0.5) -- (0,-0.5);
		\draw (-1,-1) -- (0,-0.5) -- (1,-1);
		\draw (0.32,-0.66) -- (0.19,-0.91);
		\draw (0.68,-0.84) -- (0.55,-1.09);
		\node at  (-1.2, -1.2) {$1$};
		\node at (-1.2, 1.2) {$2$};
		\node at (1.2, 1.2) {$3$};
		\node at (1.2, -1.2) {$4$};
		\node at (0, -0.9) {$-$};
		\node at (0.4, -1.2) {$+$};
		\node at (0.3, -0.4) {\tiny{$X_{41}$}};
		\node at (0.4, -0.9) {\tiny{$X_{4}$}};
		\node at (0.3, 0) {\tiny{$X_{42}$}};
	\end{tikzpicture} \qquad
	\begin{tikzpicture}                            
		\draw (-1,1) -- (0,0.5) -- (1,1);
		\draw (0,0.5) -- (0,-0.5);
		\draw (-1,-1) -- (0,-0.5) -- (1,-1);
		\draw (0.5,-0.75) -- (0.37,-1.0);
		\draw (0.37,-1.0) -- (0.04,-1.11);
		\draw (0.37,-1.0) -- (0.48,-1.33);
		\node at  (-1.2, -1.2) {$1$};
		\node at (-1.2, 1.2) {$2$};
		\node at (1.2, 1.2) {$3$};
		\node at (1.2, -1.2) {$4$};
		\node at (0.7, -1.1) {\tiny{$\tilde{X}_{1}$}};
		\node at (0.1, -0.8) {\tiny{$X_{41}$}};
		\node at (-0.15, -1.2) {$-$};
		\node at (0.25, -1.3) {$+$};
		\node at (0.3, 0) {\tiny{$X_{42}$}};
	\end{tikzpicture} \nonumber\\
	\begin{tikzpicture}                             
		\draw (-1, 1) -- (-0.5,0) -- (-1,-1);
		\draw (-0.5,0) -- (0.5,0);
		\draw (1, 1) -- (0.5,0) -- (1,-1);
		\draw (-0.75, -0.5) -- (-0.5,-0.63);
		\draw (-0.5,-0.63) -- (-0.17,-0.52);
		\draw (-0.5,-0.63) -- (-0.39,-0.96);
		\node at  (-1.2, -1.2) {$1$};
		\node at (-1.2, 1.2) {$2$};
		\node at (1.2, 1.2) {$3$};
		\node at (1.2, -1.2) {$4$};
		\node at (0, -0.6) {$+$};
		\node at (-0.25, -1.0) {$-$};
		\node at (0, 0.2) {\tiny{${X_{13}}$}};
		\node at (-0.65, -0.8) {\tiny{$\tilde{X}_{1}$}};
		\node at (-0.9, -0.2) {\tiny{$X_{12}$}};
	\end{tikzpicture} \qquad
	\begin{tikzpicture}                            
		\draw (-1,1) -- (0,0.5) -- (1,1);
		\draw (0,0.5) -- (0,-0.5);
		\draw (-1,-1) -- (0,-0.5) -- (1,-1);
		\draw (-0.5,-0.75) -- (-0.37,-1.0);
		\draw (-.37,-1.0) -- (-0.04,-1.11);
		\draw (-0.37,-1.0) -- (-0.48,-1.33);
		\node at  (-1.2, -1.2) {$1$};
		\node at (-1.2, 1.2) {$2$};
		\node at (1.2, 1.2) {$3$};
		\node at (1.2, -1.2) {$4$};
		\node at (-0.7, -1.1) {\tiny{$\tilde{X}_{1}$}};
		\node at (-0.1, -0.8) {\tiny{$X_{12}$}};
		\node at (0.15, -1.2) {$+$};
		\node at (-0.25, -1.3) {$-$};
		\node at (0.3, 0) {\tiny{$X_{42}$}};
	\end{tikzpicture} \qquad
	\begin{tikzpicture}                             
		\draw (-1, 1) -- (-0.5,0) -- (-1,-1);
		\draw (-0.5,0) -- (0.5,0);
		\draw (1, -1) -- (0.5,0) -- (1,1);
		\draw (-0.66,-0.32) -- (-0.41,-0.45);
		\draw (-0.84,-0.68) -- (-0.59,-0.81);
		\node at  (-1.2, -1.2) {$1$};
		\node at (-1.2, 1.2) {$2$};
		\node at (1.2, 1.2) {$3$};
		\node at (1.2, -1.2) {$4$};
		\node at (-0.25, -0.5) {$+$};
		\node at (-0.5, -0.9) {$-$};
		\node at (0, 0.2) {\tiny{${X_{13}}$}};
		\node at (-0.85, -0.1) {\tiny{$X_{12}$}};
		\node at (-0.55, -0.6) {\tiny{${X}_{2}$}};
	\end{tikzpicture} \qquad
	\begin{tikzpicture}                            
		\draw (-1,1) -- (0,0.5) -- (1,1);
		\draw (0,0.5) -- (0,-0.5);
		\draw (-1,-1) -- (0,-0.5) -- (1,-1);
		\draw (-0.32,-0.66) -- (-0.19,-0.91);
		\draw (-0.68,-0.84) -- (-0.55,-1.09);
		\node at  (-1.2, -1.2) {$1$};
		\node at (-1.2, 1.2) {$2$};
		\node at (1.2, 1.2) {$3$};
		\node at (1.2, -1.2) {$4$};
		\node at (0, -0.9) {$+$};
		\node at (-0.4, -1.2) {$-$};
		\node at (-0.3, -0.4) {\tiny{$X_{12}$}};
		\node at (-0.4, -0.9) {\tiny{$X_{2}$}};
		\node at (0.3, 0) {\tiny{$X_{42}$}};
	\end{tikzpicture}
	\hspace{2cm} 
\end{eqnarray}
The above Feynman diagrams represent a six-point tree-level amplitude in the ordinary ($k=2$) bi-adjoint scalars, $m_{6}^{(2)}\left(-1234+ | -1234+\right)$, with the momenta in the external legs given by $k_{1}, k_{2}, k_{3}, k_{4}, +\ell, -\ell$ respectively. In this limit, we can interpret the generalised Mandelstam variables, written in terms of $X$ variables, as the usual kinematic variables, $s'_{ab} = \left(k_{a} + k_{b}\right)^{2}, \: s'_{abc} = \left(k_{a} + k_{b} + k_{c}\right)^{2}$:
\begin{alignat}{3}\label{X1-set}
	X_{2} & \rightarrow  s'_{-1} & \qquad X_{3} & \rightarrow  s'_{-12}  = s'_{34+} & \qquad  X_{4} & \rightarrow  s'_{4+} \nonumber\\
	X_{12} & \rightarrow  s'_{+-1} = s'_{234} & \qquad X_{13} & \rightarrow  s'_{34} & \qquad X_{31} & \rightarrow  s'_{12} \nonumber\\
	X_{41} & \rightarrow  s'_{4+-} = s'_{123} & \qquad  X_{42} & \rightarrow  s'_{23} & \qquad \tilde{X}_{1} & \rightarrow  s'_{+-}.
\end{alignat}
A six-point tree level amplitude in cubic scalar theory has $9$ independent Mandelstam invariants. Therefore with the identifications in Eq.(\ref{X1-set}) the variables, $\bigl\{X_{2}, X_{3}, X_{4}, \tilde{X}_{1}, X_{12}, X_{13}, X_{31}, X_{41}, X_{42}\bigr\}$ form a basis. 

The expression for one-loop bi-adjoint scalar amplitudes from the forward limit in the CHY formalism \cite{He:2015yua, Cachazo:2015aol} is given by, 
\begin{equation}\label{loop-chy}
	m_{n}^{\text{$1$-loop}}\left(\pi|\rho\right) = \int\frac{d^{D}\ell}{\left(2\pi\right)^{D}} \frac{1}{\ell^{2}} \lim_{k_{\pm}\rightarrow \pm \ell} \sum\limits_{\substack{\alpha\in \text{cyc}\left(\pi\right)\\\beta\in \text{cyc}\left(\rho\right)}} m_{n+2}\left(- \alpha +| - \beta +\right)
\end{equation} 
Therefore adding the contributions coming from the boundaries, $X_{1}, \rightarrow 0$, $X_{2} \rightarrow 0$, $X_{3} \rightarrow 0$ and $X_{4} \rightarrow 0$ or similarly $\tilde{X}_{1} \rightarrow 0$, $\tilde{X}_{2} \rightarrow 0$, $\tilde{X}_{3} \rightarrow 0$ and $\tilde{X}_{4} \rightarrow 0$ gives $m_{4}^{\text{$1$-loop}}\left(1234|1234\right)$ according to Eq.(\ref{loop-chy}). Methods to remove tadpoles and massless bubbles in external legs of the loop amplitude, obtained from forward limit, by projecting out the relevant poles have been studied in \cite{Feng:2016nrf, Feng:2019xiq}.

\subsubsection{Other boundaries}

\paragraph{$\mathcal{A}_{3}$ facets:} First we consider the boundaries when one of $\bigl\{X_{13}, X_{24}, X_{31}, X_{42}\bigr\}$ goes to $0$. These facets are also related to $\mathcal{A}_{3}$ sub-algebra, which also a polytope representation. However, they do not have interpretations of tree-level six-point amplitudes, in fact they take the form of one-loop three-point amplitudes in cubic scalars. But if we go to the boundaries of these boundaries, we find interesting structures. 

$\mathcal{A}_{3}$ polytope has $9$ facets out of which $6$ represent $\mathcal{A}_{2}$ algebras and $3$ facets correspond to $\mathcal{A}_{1} \times \mathcal{A}_{1}$ algebras. When restricted to the $\mathcal{A}_{2}$ facets, the amplitudes become analogous to five-point tree-level amplitudes, which can also be thought of as forward limits to the three-point loop integrand. $\mathcal{A}_{1} \times \mathcal{A}_{1}$ boundaries give rise to one-loop corrections to two-point functions. At the boundaries of $\mathcal{A}_{1} \times \mathcal{A}_{1}$ these amplitudes are equivalent to four-point tree-level amplitudes.

We illustrate the above comments with an example of the boundary $X_{31} \rightarrow 0$. The relevant Feynman diagrams are presented below:
\begin{eqnarray}
	\begin{tikzpicture}                             
		\draw (-0.5,0) -- (0.2,0);
		\draw (0.5, 0) circle (0.3);
		\draw (0.65,0.26) -- (1,1);
		\draw (0.65, -0.26) -- (1,-1);
		\node at (-0.5,0) {$\times$};
		\node at (1.2, 1.2) {$3$};
		\node at (1.2, -1.2) {$4$};
		\node at (0.5, -0.5) {\tiny{$X_{1}$}};
		\node at (1, 0) {\tiny{$X_{4}$}};
		\node at (0.4, 0.5) {\tiny{$X_{3}$}};
	\end{tikzpicture} \qquad
	\begin{tikzpicture}                             
		\draw (-0.5,0) -- (-0.2,0);
		\draw (0, 0) circle (0.2);
		\draw (0.2,0) -- (0.5,0);
		\draw (0.5,0) -- (1,1);
		\draw (0.5, 0) -- (1,-1);
		\node at (-0.5,0) {$\times$};
		\node at (1.2, 1.2) {$3$};
		\node at (1.2, -1.2) {$4$};
		\node at (0, -0.5) {\tiny{$X_{1}$}};
		\node at (0.4, -0.25) {\tiny{$X_{13}$}};
		\node at (0, 0.4) {\tiny{$X_{3}$}};
	\end{tikzpicture} \qquad
	\begin{tikzpicture}                             
		\draw (-0.5,0) -- (0.5,0);
		\draw (0, -0.7) circle (0.2);
		\draw (0,0) -- (0,-0.5);
		\draw (1,1) -- (0.5, 0) -- (1, -1);
		\node at (-0.5,0) {$\times$};
		\node at (1.2, 1.2) {$3$};
		\node at (1.2, -1.2) {$4$};
		\node at (0.3, -0.2) {\tiny{$X_{13}$}};
		\node at (0, -1.1) {\tiny{$X_{1}, \tilde{X}_{1}$}};
	\end{tikzpicture}  \qquad
	\begin{tikzpicture}                             
		\draw (-0.5,0) -- (-0.2,0);
		\draw (0, 0) circle (0.2);
		\draw (0.2,0) -- (0.5,0);
		\draw (0.5,0) -- (1,1);
		\draw (0.5, 0) -- (1,-1);
		\node at (-0.5,0) {$\times$};
		\node at (1.2, 1.2) {$3$};
		\node at (1.2, -1.2) {$4$};
		\node at (0, -0.5) {\tiny{$\tilde{X}_{1}$}};
		\node at (0.4, -0.25) {\tiny{$X_{13}$}};
		\node at (0, 0.4) {\tiny{$\tilde{X}_{3}$}};
	\end{tikzpicture} \qquad
	\begin{tikzpicture}                             
		\draw (-0.5,0) -- (0.5,0);
		\draw (0, 0.7) circle (0.2);
		\draw (0,0) -- (0,0.5);
		\draw (1,1) -- (0.5, 0) -- (1, -1);
		\node at (-0.5,0) {$\times$};
		\node at (1.2, 1.2) {$3$};
		\node at (1.2, -1.2) {$4$};
		\node at (0.3, -0.2) {\tiny{$X_{13}$}};
		\node at (0, 1.1) {\tiny{$X_{3}, \tilde{X}_{3}$}};
	\end{tikzpicture} \nonumber\\
	\begin{tikzpicture}                             
		\draw (-0.5,0) -- (0.2,0);
		\draw (0.5, 0) circle (0.3);
		\draw (0.65,0.26) -- (1,1);
		\draw (0.65, -0.26) -- (1,-1);
		\node at (-0.5,0) {$\times$};
		\node at (1.2, 1.2) {$3$};
		\node at (1.2, -1.2) {$4$};
		\node at (0.5, -0.5) {\tiny{$\tilde{X}_{1}$}};
		\node at (1, 0) {\tiny{$\tilde{X}_{4}$}};
		\node at (0.4, 0.5) {\tiny{$\tilde{X}_{3}$}};
	\end{tikzpicture} \qquad
	\begin{tikzpicture}                             
		\draw (-0.5,0) -- (0.5,0);
		\draw (1, 1) -- (0.5,0) -- (0.66,-0.32);
		\draw (0.75, -0.5) circle (0.2);
		\draw (0.84,-0.68) -- (1,-1);
		\node at (-0.5,0) {$\times$};
		\node at (1.2, 1.2) {$3$};
		\node at (1.2, -1.2) {$4$};
		\node at (0.38, -0.7) {\tiny{${X}_{1}$}};
		\node at (0.9, -0.1) {\tiny{$X_{41}$}};
		\node at (1.1, -0.7) {\tiny{${X}_{4}$}};
	\end{tikzpicture} \qquad
	\begin{tikzpicture}                             
		\draw (-0.5,0) -- (0.5,0);
		\draw (1, 1) -- (0.5,0) -- (1,-1);
		\draw (0.32, -0.72) circle (0.2);
		\draw (0.75, -0.5) -- (0.5,-0.63);
		\node at (-0.5,0) {$\times$};
		\node at (1.2, 1.2) {$3$};
		\node at (1.2, -1.2) {$4$};
		\node at (0.32, -1.1) {\tiny{${X}_{1},\tilde{X}_{1}$}};
		\node at (0.9, -0.2) {\tiny{$X_{41}$}};
	\end{tikzpicture} \qquad
	\begin{tikzpicture}                             
		\draw (-0.5,0) -- (0.5,0);
		\draw (1, 1) -- (0.5,0) -- (1,-1);
		\draw (1.18, -0.28) circle (0.2);
		\draw (0.75, -0.5) -- (1.0,-0.37);
		\node at (-0.5,0) {$\times$};
		\node at (1.2, 1.2) {$3$};
		\node at (1.2, -1.2) {$4$};
		\node at (1.18, 0.1) {\tiny{${X}_{4}$}};
		\node at (1.18, -0.7) {\tiny{$\tilde{X}_{4}$}};
		\node at (0.4, -0.3) {\tiny{$X_{41}$}};
	\end{tikzpicture} \qquad
	\begin{tikzpicture}                             
		\draw (-0.5,0) -- (0.5,0);
		\draw (1, 1) -- (0.5,0) -- (0.66,-0.32);
		\draw (0.75, -0.5) circle (0.2);
		\draw (0.84,-0.68) -- (1,-1);
		\node at (-0.5,0) {$\times$};
		\node at (1.2, 1.2) {$3$};
		\node at (1.2, -1.2) {$4$};
		\node at (0.38, -0.7) {\tiny{$\tilde{X}_{1}$}};
		\node at (0.9, -0.1) {\tiny{$X_{41}$}};
		\node at (1.1, -0.7) {\tiny{$\tilde{X}_{4}$}};
	\end{tikzpicture} \nonumber\\
	\begin{tikzpicture}                             
		\draw (-0.5,0) -- (0.5,0);
		\draw (1, -1) -- (0.5,0) -- (0.66,0.32);
		\draw (0.75, 0.5) circle (0.2);
		\draw (0.84,0.68) -- (1,1);
		\node at (-0.5,0) {$\times$};
		\node at (1.2, 1.2) {$3$};
		\node at (1.2, -1.2) {$4$};
		\node at (0.35, 0.5) {\tiny{${X}_{3}$}};
		\node at (0.9, 0.1) {\tiny{$X_{34}$}};
		\node at (1.15, 0.5) {\tiny{${X}_{4}$}};
	\end{tikzpicture} \qquad
	\begin{tikzpicture}                             
		\draw (-0.5,0) -- (0.5,0);
		\draw (1, -1) -- (0.5,0) -- (1,1);
		\draw (0.32, 0.72) circle (0.2);
		\draw (0.75, 0.5) -- (0.5,0.63);
		\node at (-0.5,0) {$\times$};
		\node at (1.2, 1.2) {$3$};
		\node at (1.2, -1.2) {$4$};
		\node at (-0.32, 0.6) {\tiny{${X}_{3},\tilde{X}_{3}$}};
		\node at (0.9, 0.2) {\tiny{$X_{34}$}};
	\end{tikzpicture} \qquad
	\begin{tikzpicture}                             
		\draw (-0.5,0) -- (0.5,0);
		\draw (1, 1) -- (0.5,0) -- (1,-1);
		\draw (1.18, 0.28) circle (0.2);
		\draw (0.75, 0.5) -- (1.0,0.37);
		\node at (-0.5,0) {$\times$};
		\node at (1.2, 1.2) {$3$};
		\node at (1.2, -1.2) {$4$};
		\node at (1.05, 0.65) {\tiny{${X}_{4}$}};
		\node at (1.05, -0.1) {\tiny{$\tilde{X}_{4}$}};
		\node at (0.35, 0.3) {\tiny{$X_{34}$}};
	\end{tikzpicture} \qquad
	\begin{tikzpicture}                             
		\draw (-0.5,0) -- (0.5,0);
		\draw (1, -1) -- (0.5,0) -- (0.66,0.32);
		\draw (0.75, 0.5) circle (0.2);
		\draw (0.84,0.68) -- (1,1);
		\node at (-0.5,0) {$\times$};
		\node at (1.2, 1.2) {$3$};
		\node at (1.2, -1.2) {$4$};
		\node at (0.35, 0.5) {\tiny{$\tilde{X}_{3}$}};
		\node at (0.9, 0.1) {\tiny{$X_{34}$}};
		\node at (1.15, 0.5) {\tiny{$\tilde{X}_{4}$}};
	\end{tikzpicture} \hspace{1cm}
\end{eqnarray}
The $\times$ in the leg means it is attached to external legs labelled by $1$ and $2$.

The $\mathcal{A}_{2}$ facets are boundaries of the boundaries formed by the intersections of $X_{31}\rightarrow 0$ and one of $\bigl\{X_{1}, X_{3}, X_{4}, \tilde{X}_{1}, \tilde{X}_{3}, \tilde{X}_{4}\bigr\} \rightarrow 0$. For instance, if we go to the boundary with $X_{1}\rightarrow 0$, we have to set $u_{1}\rightarrow 0$, and $\{\tilde{u}_{3}, \tilde{u}_{4}, u_{34}\} \rightarrow 1$. So the generalised potential function of Eq.(\ref{ptential-3,6}) becomes,
\begin{equation}
	F \mid_{u_{31} \rightarrow 0, \; u_{1} \rightarrow 0} = X_{3}\log u_{3} + X_{4} \log u_{4} + \tilde{X}_{1} \log\tilde{u}_{1} + X_{13}\log u_{13} + X_{41}\log u_{41}.
\end{equation}
Thus, the new Feynman diagrams obtained by cutting the propagator $X_{1}$ stand for five-point tree-level amplitude in cubic scalar theory.

The $\mathcal{A}_{1} \times \mathcal{A}_{1}$ facets are boundaries of the boundaries, which are at the intersections of $X_{31}\rightarrow 0$ and one of $\bigl\{X_{13}, X_{34}, X_{41}\bigr\} \rightarrow 0$. As an example, let us go to the boundary where $X_{13} \rightarrow 0$; for this, we have to set $u_{13} \rightarrow 0$ and $\bigl\{u_{4}, \tilde{u}_{4}, u_{41}\bigr\} \rightarrow 1$. Now the boundaries of this $\mathcal{A}_{1}\times \mathcal{ A}_{1}$ facet are the zeoes of $\bigl\{X_{1}, X_{3}, \tilde{X}_{1}, \tilde{X}_{3}\bigr\}$, and each of these represents the $\mathcal{A}_{1}$ polytope. Therefore, if we go to the boundary $X_{1} \rightarrow 0$, then we have to set $u_{1}\rightarrow 0$ and $\tilde{u}_{3}\rightarrow 1$, and the generalised potential function reduces to that of a four-point tree level amplitude.
\begin{equation}
	F \mid_{u_{31}\rightarrow 0, \; u_{13}\rightarrow 0, \; u_{1}\rightarrow 0} = X_{3}\log u_{3} + \tilde{X}_{1}\log \tilde{u}_{1}.
\end{equation}

\paragraph{$\mathcal{A}_{1} \times \mathcal{A}_{1} \times \mathcal{A}_{1}$ facets:} Let us now consider the boundaries when any of  $\bigl\{X_{12}, X_{23}, X_{34}, X_{41}\bigr\}$ goes to zero. These boundaries represent $\mathcal{A}_{1} \times \mathcal{A}_{1} \times \mathcal{A}_{1}$ algebra.  We can illustrate the boundary properties with an example. We take $X_{12} \rightarrow 0$. The residue of the $\text{Gr}\left(3,6\right)$ amplitude on the pole at $X_{12}\rightarrow 0$ can be expressed as,
\begin{equation}
	\left(\frac{1}{X_{1}X_{2}} + \frac{1}{X_{1}\tilde{X}_{1}} + \frac{1}{X_{2}\tilde{X}_{2}} + \frac{1}{\tilde{X}_{1}\tilde{X}_{2}}\right)\left[\frac{1}{X_{13}} + \frac{1}{X_{42}}\right].
\end{equation}
The expression inside the round brackets represents an $\mathcal{A}_{1} \times \mathcal{A}_{1}$ amplitude whereas that in the square brackets is an $\mathcal{A}_{1}$ amplitude. The boundaries of the boundary with $X_{12} \rightarrow 0$ are given by the zeroes of $\bigl\{X_{1}, \tilde{ X}_{1}, X_{2}, \tilde{X}_{2}, X_{13}, X_{42}\bigr\}$. The boundaries, $X_{13} \rightarrow 0$ and $X_{42} \rightarrow 0$ are non-intersecting. Therefore, if we go to the boundary corresponding to $X_{1} \rightarrow 0$, with the condition either $u_{13} \rightarrow 0$ or $u_{42} \rightarrow 0$, we have to set $u_{1}\rightarrow 0$ and $\tilde{u}_{2} \rightarrow 1$. So the generalised potential function becomes
\begin{equation}
	F \mid_{u_{12}\rightarrow 0, \; u_{13} \; \text{or} \; u_{42} \rightarrow 0, \; u_{1}\rightarrow 0} = X_{2} \log u_{2} + \tilde{X}_{1}\log \tilde{u}_{1}.
\end{equation}
This is equivalent to the potential function for four-point tree-level amplitude in cubic scalar theory with $X_{2}$ and $\tilde{X}_{1}$ as two mutable propagators in the appropriate limits.

\subsection{Soft limits}
\label{sec:soft-limits}

In \cite{Sepulveda:2019vrz}, single soft limits of $\text{Gr}\left(k,n\right)$ amplitudes have been studied. In this limit, an $n$-point amplitude factorises on $\left(k-1\right)$ co-dimensional boundaries of $\left(k-1\right)\left(n-k-1\right)$ dimensional moduli space as a product of a soft factor  and an   $\left(n-1\right)$-point amplitude. As an example, we consider the soft limit of the $m_{6}^{(3)}\left(\mathbb{I}|\mathbb{I}\right)$ amplitude when the $6$-th external state is taken to be soft. In this case, the soft factor is given by,
\begin{equation}\label{single-soft-s}
	\mathtt{S}_{6}^{(3)} = \frac{1}{t_{2345}}\left(\frac{1}{s_{612}} + \frac{1}{s_{561}}\right) + \frac{1}{t_{1234}}\left(\frac{1}{s_{561}} + \frac{1}{s_{456}}\right) + \frac{1}{s_{456}s_{612}}.
\end{equation}
Using Eq.(\ref{X-s}), the soft factor can be expressed in terms of the $X$ variables as,
\begin{equation}\label{single-soft-x}
	\mathtt{S}_{6}^{(3)} = \frac{1}{\tilde{X}_{2}}\left(\frac{1}{X_{42}} + \frac{1}{\tilde{X}_{3}}\right) + \frac{1}{X_{31}}\left(\frac{1}{\tilde{X}_{3}} + \frac{1}{X_{1}}\right) + \frac{1}{X_{1}X_{42}}.
\end{equation} 
It was observed that the poles in the single soft factor Eq.(\ref{single-soft-s}) form an $\mathcal{A}_2$ subalgebra. However, as each term appears with different poles in the amplitude, they do not form a facet of the cluster polytope. 

\subsubsection{Double soft theorem}

Double soft limits of $\text{Gr}\left(k,n\right)$ amplitudes have been explored in \cite{Abhishek:2020xfy}. In $\text{Gr}\left(k,n\right)$ amplitudes, the double soft factor contains $4$ propagators and itself is the full amplitude. There are various ways of taking simultaneous double soft limit, depending on the relative positions of the soft external states in a particular planar ordering. The leading order contribution with $\tau^{-6}$ scaling comes from the adjacent double soft factorisations. As we will see, adjacent double soft factors have interesting sub-algebraic properties in cluster algebra. 

Let us consider the $5$-th and $6$-th states to be soft in $m_{6}^{(3)}\left(\mathbb{I}|\mathbb{I}\right)$. Then the double soft factor is, 
\begin{eqnarray}\label{doublesoft}
	\mathtt{S}_{\text{DS}}^{(3)} &=&  \frac{1}{t_{1234}} \left(\frac{1}{s_{156}} + \frac{1}{s_{456}}\right)\Biggl\{\frac{1}{\sum\limits_{a=1}^{3}\left(s_{a45} + s_{a46}\right)}\left(\frac{1}{s_{345} + s_{346}} + \frac{1}{s_{451} + s_{461}}\right)  \nonumber\\
	&&  \hspace{0.5cm}   + \frac{1}{\sum\limits_{a=2}^{4}\left(s_{1a5} + s_{1a6}\right)}\left(\frac{1}{s_{451} + s_{461}} + \frac{1}{s_{512} + s_{612}}\right) + \frac{1}{\left(s_{512} + s_{612}\right) \left(s_{345} + s_{346}\right)}\Biggr\}.
\end{eqnarray}
In this limit, we can identify the propagators appearing in Eq.(\ref{doublesoft}) with that in Eq.(\ref{X-s}) in the following way:
\begin{alignat}{4}
	X_{1} & \rightarrow  s_{456} & \qquad  X_{2} & \rightarrow  t_{6123} & \qquad X_{3} & \rightarrow  s_{123} & \qquad X_{4} & \rightarrow  s_{345} + s_{346} \nonumber\\
	\tilde{X}_{1} & \rightarrow  s_{234} & \qquad \tilde{X}_{2} & \rightarrow  t_{2345} & \qquad \tilde{X}_{3} & \rightarrow  s_{561} & \qquad \tilde{X}_{4} & \rightarrow  s_{125} + s_{126} \nonumber\\
	X_{12} & \rightarrow  r_{234561}& \qquad X_{13} & \rightarrow  s_{145} + s_{146} & \qquad X_{23} & \rightarrow  r_{452361} & \qquad X_{24} & \rightarrow  s_{345} \nonumber\\
	X_{34} & \rightarrow  s_{345} + s_{346} & \qquad X_{31} & \rightarrow  t_{1234} & \qquad  X_{41} & \rightarrow  s_{125} + s_{126} & \qquad  X_{42} & \rightarrow  s_{612}.
\end{alignat}
Here $X_{1}, \tilde{X}_{3}$ and $X_{31}$ scale as $\tau^{2}$ while rest other variables\footnote{Momentum conservations imply $\sum\limits_{a=1}^{3}\left(s_{a45} + s_{a46}\right) = s_{123}$ and $\sum\limits_{a=2}^{4}\left(s_{1a5} + s_{1a6}\right) = s_{234}$. These poles are obtained from the degenerate configurations where the punctures $\sigma_{5}$ and $\sigma_{6}$ simultaneously collide with $\sigma_{4}$ and $\sigma_{1}$ respectively.} scale as $\tau$. Although multiple $X$ variables have the same limits expressed in terms of the generalised Mandelstam variables, compatibility sets in Eq.(\ref{compatible-sets}) determine the appropriate propagator, which should appear in the double soft factor. Therefore Eq.(\ref{doublesoft}) can be expressed as,
\begin{eqnarray}\label{DS-inX}
	\mathtt{S}_{\text{DS}}^{(3)} &=& \frac{1}{X_{31}}\Biggl\{\frac{1}{\tilde{X}_{3}X_{3}}\left(\frac{1}{X_{34}} + \frac{1}{X_{13}}\right) + \frac{1}{X_{1}X_{3}}\left(\frac{1}{X_{4}} + \frac{1}{X_{13}}\right) + \frac{1}{\tilde{X}_{3}\tilde{X}_{1}}\left(\frac{1}{X_{13}} + \frac{1}{\tilde{X}_{4}}\right)   \nonumber\\
	&& \phantom{\frac{1}{X_{31}}\Biggl\{} + \frac{1}{X_{1}\tilde{X}_{1}}\left(\frac{1}{X_{13}} + \frac{1}{X_{41}}\right) + \frac{1}{\tilde{X}_{3}\tilde{X}_{4}X_{34}} + \frac{1}{X_{1}X_{41}X_{4}}\Biggr\}.
\end{eqnarray}
Feynman diagrams leading to the above double soft limit of the amplitude are given below:
\begin{eqnarray}
	\begin{tikzpicture}                             
		\draw (-1, 1) -- (-0.5,0) -- (-1,-1);
		\draw (-0.5,0) -- (0.2,0);
		\draw (0.5, 0) circle (0.3);
		\draw (0.65,0.26) -- (1,1);
		\draw (0.65, -0.26) -- (1,-1);
		\node at  (-1.2, -1.2) {$1$};
		\node at (-1.2, 1.2) {$2$};
		\node at (1.2, 1.2) {$3$};
		\node at (1.2, -1.2) {$4$};
		\node at (-0.2, -0.2) {\tiny{$X_{31}$}};
		\node at (0.5, -0.5) {\tiny{$X_{1}$}};
		\node at (1, 0) {\tiny{$X_{4}$}};
		\node at (0.4, 0.5) {\tiny{$X_{3}$}};
	\end{tikzpicture} \qquad
	\begin{tikzpicture}                             
		\draw (-1, 1) -- (-0.5,0) -- (-1,-1);
		\draw (-0.5,0) -- (-0.2,0);
		\draw (0, 0) circle (0.2);
		\draw (0.2,0) -- (0.5,0);
		\draw (0.5,0) -- (1,1);
		\draw (0.5, 0) -- (1,-1);
		\node at  (-1.2, -1.2) {$1$};
		\node at (-1.2, 1.2) {$2$};
		\node at (1.2, 1.2) {$3$};
		\node at (1.2, -1.2) {$4$};
		\node at (-0.3, -0.25) {\tiny{${X_{31}}$}};
		\node at (0, -0.5) {\tiny{$X_{1}$}};
		\node at (0.4, -0.25) {\tiny{$X_{13}$}};
		\node at (0, 0.4) {\tiny{$X_{3}$}};
	\end{tikzpicture} \qquad
	\begin{tikzpicture}                             
		\draw (-1, 1) -- (-0.5,0) -- (-1,-1);
		\draw (-0.5,0) -- (0.5,0);
		\draw (0, -0.7) circle (0.2);
		\draw (0,0) -- (0,-0.5);
		\draw (1,1) -- (0.5, 0) -- (1, -1);
		\node at  (-1.2, -1.2) {$1$};
		\node at (-1.2, 1.2) {$2$};
		\node at (1.2, 1.2) {$3$};
		\node at (1.2, -1.2) {$4$};
		\node at (-0.3, -0.2) {\tiny{${X_{31}}$}};
		\node at (0.3, -0.2) {\tiny{$X_{13}$}};
		\node at (0, -1.1) {\tiny{$X_{1}, \tilde{X}_{1}$}};
	\end{tikzpicture}  \qquad
	\begin{tikzpicture}                             
		\draw (-1, 1) -- (-0.5,0) -- (-1,-1);
		\draw (-0.5,0) -- (0.5,0);
		\draw (1, 1) -- (0.5,0) -- (0.66,-0.32);
		\draw (0.75, -0.5) circle (0.2);
		\draw (0.84,-0.68) -- (1,-1);
		\node at  (-1.2, -1.2) {$1$};
		\node at (-1.2, 1.2) {$2$};
		\node at (1.2, 1.2) {$3$};
		\node at (1.2, -1.2) {$4$};
		\node at (0, -0.2) {\tiny{${X_{31}}$}};
		\node at (0.38, -0.7) {\tiny{${X}_{1}$}};
		\node at (0.9, -0.1) {\tiny{$X_{41}$}};
		\node at (1.1, -0.7) {\tiny{${X}_{4}$}};
	\end{tikzpicture} \qquad
	\begin{tikzpicture}                             
		\draw (-1, 1) -- (-0.5,0) -- (-1,-1);
		\draw (-0.5,0) -- (0.5,0);
		\draw (1, 1) -- (0.5,0) -- (1,-1);
		\draw (0.32, -0.72) circle (0.2);
		\draw (0.75, -0.5) -- (0.5,-0.63);
		\node at  (-1.2, -1.2) {$1$};
		\node at (-1.2, 1.2) {$2$};
		\node at (1.2, 1.2) {$3$};
		\node at (1.2, -1.2) {$4$};
		\node at (0, -0.2) {\tiny{${X_{31}}$}};
		\node at (0.32, -1.1) {\tiny{${X}_{1},\tilde{X}_{1}$}};
		\node at (0.9, -0.2) {\tiny{$X_{41}$}};
	\end{tikzpicture}  \nonumber\\
	\begin{tikzpicture}                             
		\draw (-1, 1) -- (-0.5,0) -- (-1,-1);
		\draw (-0.5,0) -- (0.2,0);
		\draw (0.5, 0) circle (0.3);
		\draw (0.65,0.26) -- (1,1);
		\draw (0.65, -0.26) -- (1,-1);
		\node at  (-1.2, -1.2) {$1$};
		\node at (-1.2, 1.2) {$2$};
		\node at (1.2, 1.2) {$3$};
		\node at (1.2, -1.2) {$4$};
		\node at (-0.2, -0.2) {\tiny{$X_{31}$}};
		\node at (0.5, -0.5) {\tiny{$\tilde{X}_{1}$}};
		\node at (1, 0) {\tiny{$\tilde{X}_{4}$}};
		\node at (0.4, 0.5) {\tiny{$\tilde{X}_{3}$}};
	\end{tikzpicture} \qquad
	\begin{tikzpicture}                             
		\draw (-1, 1) -- (-0.5,0) -- (-1,-1);
		\draw (-0.5,0) -- (-0.2,0);
		\draw (0, 0) circle (0.2);
		\draw (0.2,0) -- (0.5,0);
		\draw (0.5,0) -- (1,1);
		\draw (0.5, 0) -- (1,-1);
		\node at  (-1.2, -1.2) {$1$};
		\node at (-1.2, 1.2) {$2$};
		\node at (1.2, 1.2) {$3$};
		\node at (1.2, -1.2) {$4$};
		\node at (-0.3, -0.25) {\tiny{${X_{31}}$}};
		\node at (0, -0.5) {\tiny{$\tilde{X}_{1}$}};
		\node at (0.4, -0.25) {\tiny{$X_{13}$}};
		\node at (0, 0.4) {\tiny{$\tilde{X}_{3}$}};
	\end{tikzpicture} \qquad
	\begin{tikzpicture}                             
		\draw (-1, 1) -- (-0.5,0) -- (-1,-1);
		\draw (-0.5,0) -- (0.5,0);
		\draw (0, 0.7) circle (0.2);
		\draw (0,0) -- (0,0.5);
		\draw (1,1) -- (0.5, 0) -- (1, -1);
		\node at  (-1.2, -1.2) {$1$};
		\node at (-1.2, 1.2) {$2$};
		\node at (1.2, 1.2) {$3$};
		\node at (1.2, -1.2) {$4$};
		\node at (-0.3, -0.2) {\tiny{${X_{31}}$}};
		\node at (0.3, -0.2) {\tiny{$X_{13}$}};
		\node at (0, 1.1) {\tiny{$X_{3}, \tilde{X}_{3}$}};
	\end{tikzpicture} \qquad
	\begin{tikzpicture}                             
		\draw (1, 1) -- (0.5,0) -- (1,-1);
		\draw (-0.5,0) -- (0.5,0);
		\draw (-1, 1) -- (-0.5,0) -- (-1,-1);
		\draw (0.32, 0.72) circle (0.2);
		\draw (0.75, 0.5) -- (0.5,0.63);
		\node at  (-1.2, -1.2) {$1$};
		\node at (-1.2, 1.2) {$2$};
		\node at (1.2, 1.2) {$3$};
		\node at (1.2, -1.2) {$4$};
		\node at (0, -0.2) {\tiny{${X_{31}}$}};
		\node at (-0.32, 0.8) {\tiny{${X}_{3},\tilde{X}_{3}$}};
		\node at (0.9, 0.2) {\tiny{$X_{34}$}};
	\end{tikzpicture} \qquad
	\begin{tikzpicture}                             
		\draw (-1, -1) -- (-0.5,0) -- (-1,1);
		\draw (-0.5,0) -- (0.5,0);
		\draw (1, -1) -- (0.5,0) -- (0.66,0.32);
		\draw (0.75, 0.5) circle (0.2);
		\draw (0.84,0.68) -- (1,1);
		\node at  (-1.2, -1.2) {$1$};
		\node at (-1.2, 1.2) {$2$};
		\node at (1.2, 1.2) {$3$};
		\node at (1.2, -1.2) {$4$};
		\node at (0, -0.2) {\tiny{${X_{31}}$}};
		\node at (0.35, 0.5) {\tiny{$\tilde{X}_{3}$}};
		\node at (0.9, 0.1) {\tiny{$X_{34}$}};
		\node at (1.15, 0.5) {\tiny{$\tilde{X}_{4}$}};
	\end{tikzpicture}
\nonumber\\
\end{eqnarray}

Adjacent double soft factors in a planar ordered $\text{Gr}\left(2,n\right)$ amplitude is associated to $\mathcal{A}_{n-4}$ sub-algebras of the $\mathcal{A}_{n-3}$ cluster algebra. However, in the double soft limit only some terms of these $\mathcal{A}_{n-4}$ amplitudes contribute to the leading order in the soft momenta. In the $\text{Gr}\left(3,6\right)$ amplitude, we see the double soft factor in Eq.(\ref{DS-inX}) picks out the leading terms in the $\mathcal{A}_{3}$ sub-algebra corresponding to $X_{31}$ facet of the $\mathcal{D}_{4}$ polytope. Similarly  other $\mathcal{A}_{3}$ sub-algebras represented by $X_{2}, X_{4}, \tilde{X}_{2}, \tilde{X}_{4}$ and $X_{13}$ will be related to the other adjacent double soft factorisations of the $\text{Gr}\left(3,6\right)$ amplitude.

Notice that in Eq.(\ref{doublesoft}), the term inside the second bracket is the single soft factor in Eq.(\ref{single-soft-s}), with shifted propagators $s_{ab6}\rightarrow s_{ab5}+s_{ab6}$. Contrary to the single soft case, here the single soft factor with shifted propagators realise the $\mathcal{A}_2$ subalgebra as a facet of the cluster polytope. This $\mathcal{A}_2$ subalgebra can be understood from the mutation of the six-point tree amplitude obtained in the forward limit corresponding to the variables $X_1,\tilde{ X}_3$, which are the variables that scale as $\tau^2$ and correspond to the single soft factor in $k=2$ with a composite label $m=\{56\}$ going soft as discussed in \cite{Abhishek:2020xfy}.

\section{Conclusion}
\label{sec:conclusion}

CEGM formalism is a novel extension of the CHY formalism from the moduli space of $n$-punctured $\mathbb{CP}^1$ to that of $n$-punctured $\mathbb{CP}^{k-1}$. However, a field theory formulation of these generalised amplitudes is unclear although they exhibit interesting factorisation properties at various boundaries of the moduli space. They also contain an interesting and intricate connection with $\text{Gr}(k,n)$ cluster algebras, which was exploited to compute certain CEGM amplitudes in \cite{Drummond:2019qjk,Drummond:2020kqg}. In this paper, we have used the equivalence between $\text{Gr}(3,6)$ and $\mathcal{D}_4$ cluster algebras to relate the $\left(k=3, n=6\right)$ CEGM amplitude to the cluster polytope for four-point one-loop integrand described in \cite{Arkani-Hamed:2019vag}. We have found a mapping between the generalised Mandelstam variables and the kinematic variables for the one-loop cluster polytope. With the $\text{Gr}(3,6)$ initial cluster, we identify the constraints in the kinematic space, which etch out the polytope in a different realisation than the one in \cite{Arkani-Hamed:2019vag}. We have shown how various factorisations of the one-loop polytope can be interpreted in the moduli space of $6$-punctured $\mathbb{CP}^2$ by using the classification of the CEGM factorisations considered in \cite{Cachazo:2019ngv}. For a particular factorisation leading to a forward limit, we show how the $\mathbb{CP}^2$ CEGM integral reduces to an integral on the moduli space of $\mathbb{CP}^1$ with $6$ punctures as expected from the one-loop CHY integral prescription given in \cite{He:2015yua,Cachazo:2015aol}.

There is a doubling in the number of Feynman diagrams considered in our analysis, due to the presence of both $X_{i}$ and $\tilde{X}_{i}$ variables. In \cite{Arkani-Hamed:2019vag}, it was shown that by choosing $X_{i} - \tilde{X}_{i} = X_{0}$ one can get rid of the redundancies in the Feynman diagrams and in that case $\mathcal{D}_{4}$ polytope is restricted to lesser dimensional $\bar{\mathcal{D}}_{4}$ polytope. It will be interesting to see if $\bar{\mathcal{D}}_{4}$ polytope can be directly related to CEGM amplitudes.

Explicit evaluation of the amplitude for $\left(k=3, n=6\right)$ helps us validate the double soft factor obtained in \cite{Abhishek:2020xfy}. The adjacent double soft theorem for arbitrary $k$ had an intriguing structure with the appearance of single soft factors for $k-1$ and $k$, with modified propagators. This helps us identify the $\mathcal{A}_{2}$ sub-algebra, that was observed for the single soft factor from $\left(k=3, n=6\right)$ amplitude to $\left(k=3, n=5\right)$ amplitude in \cite{Sepulveda:2019vrz}, as a facet of the polytope, when one considers them with the modified propagators. Further, in the forward limit, this sub-algebra has a nice interpretation in terms of the mutations of the six-point tree amplitude. 

Study of subalgebras of $\text{Gr}\left(k,n\right)$ is an interesting mathematical problem, and of relevance to physics due to the relation between $\text{Gr}(4,n)$ cluster algebras and SYM amplitudes\cite{Arkani-Hamed:2019rds,Golden:2013xva,Drummond:2017ssj, Golden:2018gtk}. While we have studied the first non trivial CEGM amplitude with $k=3,n=6$ in the context of the one-loop polytope, it will be interesting to extend this to higher $k$ and $n$. In particular, $\mathcal{D}_n$ cluster polytopes are shown to be related to $n$-point one-loop integrand in \cite{Arkani-Hamed:2019vag}. For the next CEGM amplitude with $k=3,n=7$, $\mathcal{D}_{5}$ forms a sub-algebra. It will be interesting to extend the analysis of this paper, to study how one can obtain the one-loop polytope $\mathcal{D}_{5}$ within the $\text{Gr}(3,7)$ amplitude by analysing the boundaries of the moduli space. This may give further hints for a full interpretation of the $\text{Gr}(3,7)$ CEGM amplitude. 

Using integration rules \cite{Baadsgaard:2015voa, Baadsgaard:2015ifa} in CHY formalism integrands for various theories like $\phi^{4}$ and others can be constructed. The Pfaffian corresponding to $\phi^{4}$ integrand \cite{Cachazo:2014xea} can be expressed as different combinations of Parke-Taylor factors. We leave it as future work to explore the possibility of constructing one-loop integrands from $\text{Gr}\left(3,6\right)$ amplitude for other theories.

\section*{Acknowledgements}
We thank Dileep P. Jatkar for collaboration during initial stages of this work. We are grateful to Anirban Basu, Dileep P. Jatkar, Alok Laddha and Ashoke Sen for several important and illuminating discussions.

We thank the maintenance staff at HRI for their relentless work during the pandemic making the campus a safe and a comfortable space.

\appendix

\section{\texorpdfstring{Factorisation at $u_{2} = 0$} - boundary}
\label{sec:u2-boundary}

At the $u_{2} = 0$ boundary of the worldsheet, the amplitude factorises on the propagator $t_{6123}$. One of the ways to approach this boundary is when the two punctures, $\sigma_{4}$ and $\sigma_{5}$ collide with each other.  

Let us begin with the six-point bi-adjoint amplitude in $k=3$,
\begin{eqnarray}\label{u6-integral}
m_{6}^{(3)}\left(\mathbb{I}| \mathbb{I}\right) & = & \int \mathrm{d}\mu_{6} \left[\frac{1}{|123||234||345||456||561||612|}\right]^{2} \nonumber\\
& = & \int \mathrm{d}^{2}\sigma_{4}\; \mathrm{d}^{2}\sigma_{5}\; \delta^{(2)}\left(E_{4}\right) \delta^{(2)}\left(E_{5}\right)\left[\frac{|236||361|}{|234||345||456||561|}\right]^{2},
\end{eqnarray}
where we have chosen the volume of the $\text{SL}\left(3, \mathbb{C}\right)$ gauge group to be $|123||236||361||612|$, and we have removed the $\delta$-functions for the scattering equations, $E_{1}, E_{2}, E_{3}$ and $E_{6}$. 

We now choose the following parametrisation, 
\begin{equation}
x_{4} = x_{5} + \varepsilon, \qquad y_{4} = y_{5} = \varepsilon\alpha,
\end{equation}
with $\varepsilon \rightarrow 0$. Then the integration measure in Eq.(\ref{u6-integral}) becomes,
\begin{equation}
\mathrm{d}^{2}\sigma_{4}\; \mathrm{d}^{2}\sigma_{5} = \varepsilon\; \mathrm{d}\varepsilon\; \mathrm{d}\alpha\; \mathrm{d}x_{5}\; \mathrm{d}y_{5}.
\end{equation}
From the generalised potential, $\mathcal{S}^{(3)} = \sum\limits_{1\le i < j < k\le 6} s_{ijk} \log |ijk|$, we obtain,
\begin{equation}
\frac{\partial \mathcal{S}^{(3)}}{\partial\varepsilon} = \frac{t_{6123}}{\varepsilon} , \qquad \frac{\partial \mathcal{S}^{(3)}}{\partial\alpha} =  \sum\limits_{b\ne 4, 5} \frac{s_{b45}}{\alpha - \alpha_{b}}, \quad \text{where} \: \alpha_{b} = \frac{y_{b} - y_{5}}{x_{b} - x_{5}}.
\end{equation}
The delta function containing the $E_{4} $ scattering equation transforms as,
\begin{equation}
\delta^{(2)}\left(\partial_{x_{4}}\mathcal{S}^{(3)}, \partial_{y_{4}}\mathcal{S}^{(3)}\right) = \varepsilon\; \delta^{(2)}\left(\partial_{\varepsilon}\mathcal{S}^{(3)}, \partial_{\alpha}\mathcal{S}^{(3)}\right).
\end{equation} 
Therefore, Eq.(\ref{u6-integral}) can be expressed as,
\begin{eqnarray}\label{u6-integral2}
m_{6}^{(3)}\left(\mathbb{I}| \mathbb{I}\right) & = & \int \varepsilon^{2} \; \mathrm{d}\varepsilon\; \mathrm{d}\alpha\; \mathrm{d}x_{5}\; \mathrm{d}y_{5}\; \delta\left(\frac{t_{6123}}{\varepsilon}\right) \delta\left(\sum\limits_{b\ne 4, 5} \frac{s_{b45}}{\alpha - \alpha_{b}}\right) \delta\left(\frac{\partial\mathcal{S}^{(3)}}{\partial x_{5}}\right) \delta\left(\frac{\partial\mathcal{S}^{(3)}}{\partial y_{5}}\right)\nonumber \\
&& \phantom{\int \varepsilon^{2} \; \mathrm{d}\varepsilon\; \mathrm{d}\alpha\; \mathrm{d}x_{5}\; \mathrm{d}y_{5}\; \delta\left(\frac{t_{6123}}{\varepsilon}\right) \delta\left(\sum\limits_{b\ne 4, 5} \frac{s_{b45}}{\alpha - \alpha_{b}}\right)} \times \left[\frac{|236||361|}{|234||345||456||561|}\right]^{2}.
\end{eqnarray}
We can convert the first delta function to a contour integration over $\varepsilon$ around $\varepsilon \rightarrow 0$ pole. A factor $\varepsilon^{-4}$ comes from the terms in the square brackets. Then Eq.(\ref{u6-integral2}) becomes,
\begin{eqnarray}\label{u6-integral3}
m_{6}^{(3)}\left(\mathbb{I}| \mathbb{I}\right) & = & \frac{1}{t_{6123}} \int \mathrm{d}\alpha\; \mathrm{d}x_{5}\; \mathrm{d}y_{5}\;  \delta\left(\sum\limits_{b\ne 4, 5} \frac{s_{b45}}{\alpha - \alpha_{b}}\right) \delta\left(\frac{\partial\mathcal{S}^{(3)}}{\partial x_{5}}\right) \delta\left(\frac{\partial\mathcal{S}^{(3)}}{\partial y_{5}}\right) \nonumber\\
&& \phantom{\frac{1}{t_{6123}} \int \mathrm{d}\alpha\; \mathrm{d}x_{5}\; \mathrm{d}y_{5}\;} \times \left[\frac{|236||361|}{\left(\alpha - \alpha_{3}\right)\left(\alpha - \alpha_{6}\right)\left(x_{3} - x_{5}\right)\left(x_{5} - x_{6}\right)|234||561|}\right]^{2}.
\end{eqnarray}
Therefore, the residue on the factorisation channel $t_{6123}$ is equivalent to the CHY integral representation over the moduli space of $6$-punctured Riemann sphere, $\mathfrak{M}_{0,6}$. This becomes manifest if we choose a gauge where we fix, 
\begin{equation}
\left(\sigma_{1}, \sigma_{2}, \sigma_{3}, \sigma_{6}\right) \rightarrow \begin{pmatrix}
1 & 0 & 0 & 1\\
0 & 1 & 0 & 1\\
0 & 0 & 1 & 1
\end{pmatrix}.
\end{equation}
Then Eq.(\ref{u6-integral3}) can be written as, 
\begin{equation}
m_{6}^{(3)}\left(\mathbb{I}| \mathbb{I}\right)  =  \frac{1}{t_{6123}} \int \mathrm{d}\alpha\; \mathrm{d}x_{5}\; \mathrm{d}y_{5}\; \delta\left(E_{\alpha}\right) \delta\left(E_{5}^{(x)}\right) \delta\left(E_{5}^{(y)}\right) \left[\frac{1}{\left(\alpha - \alpha_{1}\right) \left(x_{5} - 1\right)\left(x_{5} - y_{5}\right)}\right]^{2},
\end{equation}
where, 
\begin{eqnarray}
E_{\alpha} & = & \sum\limits_{a \ne 4,5} \frac{s_{a45}}{\alpha - \alpha_{a}}, \qquad \text{with} \quad \alpha_{1} = \frac{y_{5}}{x_{5}}, \; \alpha_{2} = 0, \; \alpha_{3} = \infty,\;  \alpha_{6} = \frac{1-y_{5}}{1-x_{5}}, \nonumber\\
E_{5}^{(x)} & = & \frac{s_{135}}{x_{5}} + \frac{s_{145}}{x_{5} - \frac{y_{5}}{\alpha}} + \frac{s_{156}}{x_{5} - y_{5}} + \frac{s_{356}}{x_{5} - 1} + \frac{s_{456}}{x_{5} - 1 - \frac{y_{5} - 1}{\alpha}}, \nonumber\\
E_{5}^{(y)} & =& \frac{s_{125}}{y_{5}} + \frac{s_{145}}{y_{5} - \alpha x_{5}} + \frac{s_{156}}{y_{5} - x_{5}} + \frac{s_{256}}{y_{5} - 1} + \frac{s_{456}}{y_{5} - 1 - \alpha\left(x_{5} - 1\right)}.
\end{eqnarray}

\section{Adjacent double soft theorem for arbitrary \texorpdfstring{$k$}- }
\label{sec:doublesoft-appendix}

Double soft limits of CEGM amplitudes have been studied in \cite{Abhishek:2020xfy}. In this section, we present a summary of the double soft theorem in arbitrary $\left(k,n\right)$ amplitudes when the two adjacent external states are taken to be soft simultaneously. For arbitrary $k$, the leading order double soft factor comes from the degenerate configuration. The degenerate solution of the scattering equation receives contributions  from two different situations,
\begin{enumerate}
  \item  when two adjacent punctures, say $n$-th and $(n-1)$-th, on $\mathbb{CP}^{(k-1)}$ approach infinitesimally close to each other,
  \item when two adjacent soft punctures and $(k-2)$ number of hard punctures lie in a codimension one subspace.
\end{enumerate}
For both the cases above, the determinant $|\sigma_{a_{1}}\; \sigma_{a_{1}}\cdots\sigma_{a_{k-2}}\;\sigma_{n-1}\;\sigma_{n}|\sim \mathcal{O}(\tau )$, where the parameter $\tau$ defines the soft limit in terms of the generalized Mandelstam variables given below,
\begin{eqnarray}
s_{a_1\;a_2\cdots\;a_{k-1}\;n}&=&\tau\;\hat{s}_{a_1\;a_2\cdots a_{k-1}\;n},\nonumber\\
s_{a_1\;a_2\cdots\;a_{k-1}\;n-1}&=&\tau\;\hat{s}_{a_1\;a_2\cdots a_{k-1}\;n-1},\nonumber\\
s_{a_1\;a_2\cdots\;a_{k-2}\;n-1\;n}&=&\tau^2\;\hat{s}_{a_1\;a_2\cdots a_{k-2}\;n-1\;n}.
\end{eqnarray}
At the leading order, the configuration (1) dominates over the other for the degenerate adjacent case. The non-degenerate solutions contribute at a further lower order in the adjacent double soft factor.
After choosing appropriate parametrisations for general $k$,  the simultaneous double soft factor for the adjacent soft external states $n$ and $(n-1)$ is,
\begin{equation}
  \label{eq:mainresult}
  \mathtt{S}_{\text{DS}}^{(k)}  =
  \frac{1}{\sum\limits_{1\le a_{1}\cdots <a_{k-2}\le n-2}
    s_{a_{1}\cdots a_{k-2}\; n-1\; n}} \;\mathtt{S}^{(k-1)}
  \left(s_{a_{1}\cdots a_{k-2}\; m} \to s_{a_{1}\cdots a_{k-2}\; n-1\; n} \right)
  \; \mathtt{S}^{(k)}\ ,
\end{equation}
where the single soft factor $\mathtt{S}^{(k-1)}$ for $k-1$ is defined with $m$ as the composite level for $\{n-1\; n\}$, and $\mathtt{S}^{(k)}$ is the single soft factor for $k$, but with the shifted generalized Mandelstam variables $(s_{a_{1}\cdots a_{k-1}\; n} + s_{a_{1}\cdots a_{k-1}\; n-1})$. The leading simultaneous double soft factor for the adjacent case scales as $\tau^{-3(k-1)}$ as $\tau \to 0$, for arbitrary $k$-value. We can also check for general $k$, that the non-adjacent double soft factor contributes in the sub-leading order. 

\section{\texorpdfstring{$\text{Gr}\left(3,6\right)$} - cluster algebra}
\label{sec:Gr-36-appendix}

In this appendix, we give some computational details of Sec.(\ref{sec:Gr-36}).

\subsection{\texorpdfstring{$u$}- equations for \texorpdfstring{$\mathcal{D}_4$} - cluster}

We briefly describe the method of finding degree of compatibility between $u_{a}$ and $u_{b}$ as explained in \cite{Arkani-Hamed:2020tuz}. For $\mathcal{D}_4$ type cluster, we define a $4$-gon $PP_4$ with vertices  $\{1,2,3,4\}$ (in clockwise order) and an additional point $0$ at the center of the usual $4$-gon $P_4$. Also define a set $\{a\}$ of certain arcs denoted by $a$, inside the $PP_4$, which join the vertices and the point $0$.  There are two types of arcs,
\begin{enumerate}
  \item  for $1 \leq i \neq j \leq 4$ and $i \neq j+1$ mod $4$ we have an arc $(i,j)$ counterclockwise connecting  $i$ and $j$ surrounding the point $0$,
  \item for each $1 \leq i \leq 4 $ we have two types of arcs, $[i]$ and $[\tilde{i}]$ connecting $i$ and $0$.
\end{enumerate}
The corresponding $u$-variables are denoted by $u_{ij}$, $u_i$ and $\tilde{u}_{i}$ respectively. See the  Fig.(\ref{fig:u-labels}) below, where one can easily point out that the variables like $u_{21},u_{32},u_{43},u_{14}$ are not possible because the corresponding arcs either lie outside the $PP_4$ or become clockwise. But whereas, for $u_{13}$ with $u_{31}$ and $u_{24}$ with $u_{42}$ both the pairs are possible.

The $u$-equation for the $\mathcal{D}_4$ cluster is given as,
 
 \begin{equation}
    u_a = 1 - \prod_{\{b\}} u_{b}^{a\parallel b},\quad \forall a\in\{a\},
 \end{equation}
 where the compatibility degree $a\parallel b$, for $a,b\in\{a\} $, is equal to the minimum number of points where the arc $b$ intersect with the arc $a$, if either $a$ or $b$ are not connected to $0$. If both $a$ and $b$ are connected to $0$, we have $[i]\parallel [j]=[\tilde{i}]\parallel [\tilde{j}]=[\tilde{i}]\parallel [i]=0$ but $[i]\parallel[\tilde{j}]=[\tilde{j}]\parallel [i]=1$ if $i\neq j$. The representative equations for $\mathcal{D}_4$ are,
 \begin{eqnarray}
 u_{12} &=& 1-u_3\tilde{u}_3u_4\tilde{u}_4u_{23}u_{24}u_{31}u_{34}^2u_{41}, \quad\text{4 equations},\nonumber\\
 u_{13} &=& 1-u_4\tilde{u}_4u_{24}u_{34}u_{41}u_{42}, \quad\text{4 equations},\nonumber\\
 u_{1} &=& 1- \tilde{u}_2\tilde{u}_3\tilde{u}_4u_{23}u_{24}u_{34},\quad\text{8 equations},
 \end{eqnarray}
 So, for $\mathcal{D}_4$ cluster, totally we have 16 equations.
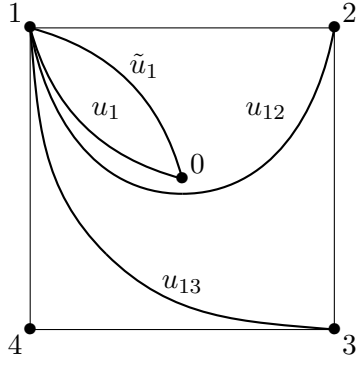
\begin{figure}[H]\label{fig:u-labels}
\begin{center}
\begin{tikzpicture}                         
\draw (-2, -2) -- (-2,2) -- (2,2) -- (2,-2) -- (-2,-2);
\node at  (-2.2, -2.2) {$4$};
\node at (-2.2, 2.2) {$1$};
\node at (2.2, 2.2) {$2$};
\node at (2.2, -2.2) {$3$};
\node at (0.2, 0.2) {$0$};
\node at (-2, -2)(4) {$\bullet$};
\node at (-2, 2) (1) {$\bullet$};
\node at (2, 2) (2) {$\bullet$};
\node at (2, -2) (3) {$\bullet$};
\node at (0, 0) (0) {$\bullet$};
\node at (-1, 0.9)  {$u_1$};
\node at (-0.5, 1.5)  {$\tilde{u}_1$};
\node at (0, -1.4)  {$u_{13}$};
\node at (1.1, 0.9)  {${u}_{12}$};
\draw [ thick] (-2,2) to[bend left] node {} (0,0);
\draw [ thick] (-2,2) to[bend right] node {} (0,0);
\draw [ thick] (-2,2) to[out=-80, in=180] node {} (0,-0.2);
\draw [ thick] (0,-0.2) to [out=0, in=-100] node {} (2,2);
\draw [ thick] (-2,2) to[out=-85, in=135] node {} (-1,-1);
\draw [ thick] (-1,-1) to[out=-45, in=175] node {} (2,-2);
\end{tikzpicture}
\end{center}
\caption{A representative example showing the conventions for labelling $u$ variables}
\end{figure}

\subsection{Cluster quivers}
\label{sec:clusters-3,6}
 
The $50$ clusters for the $\text{Gr}\left(3,6\right)$ algebra are given below:

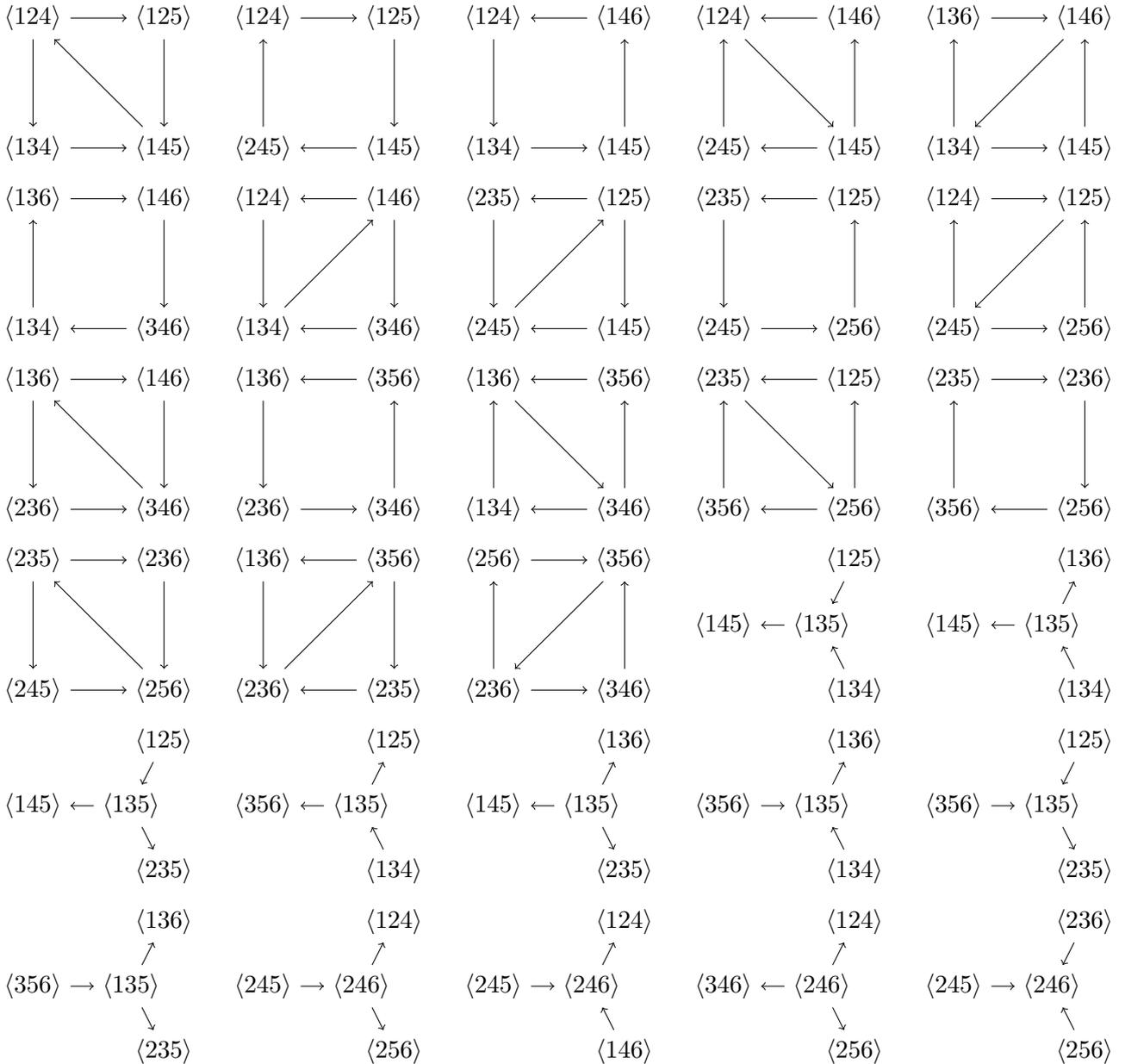
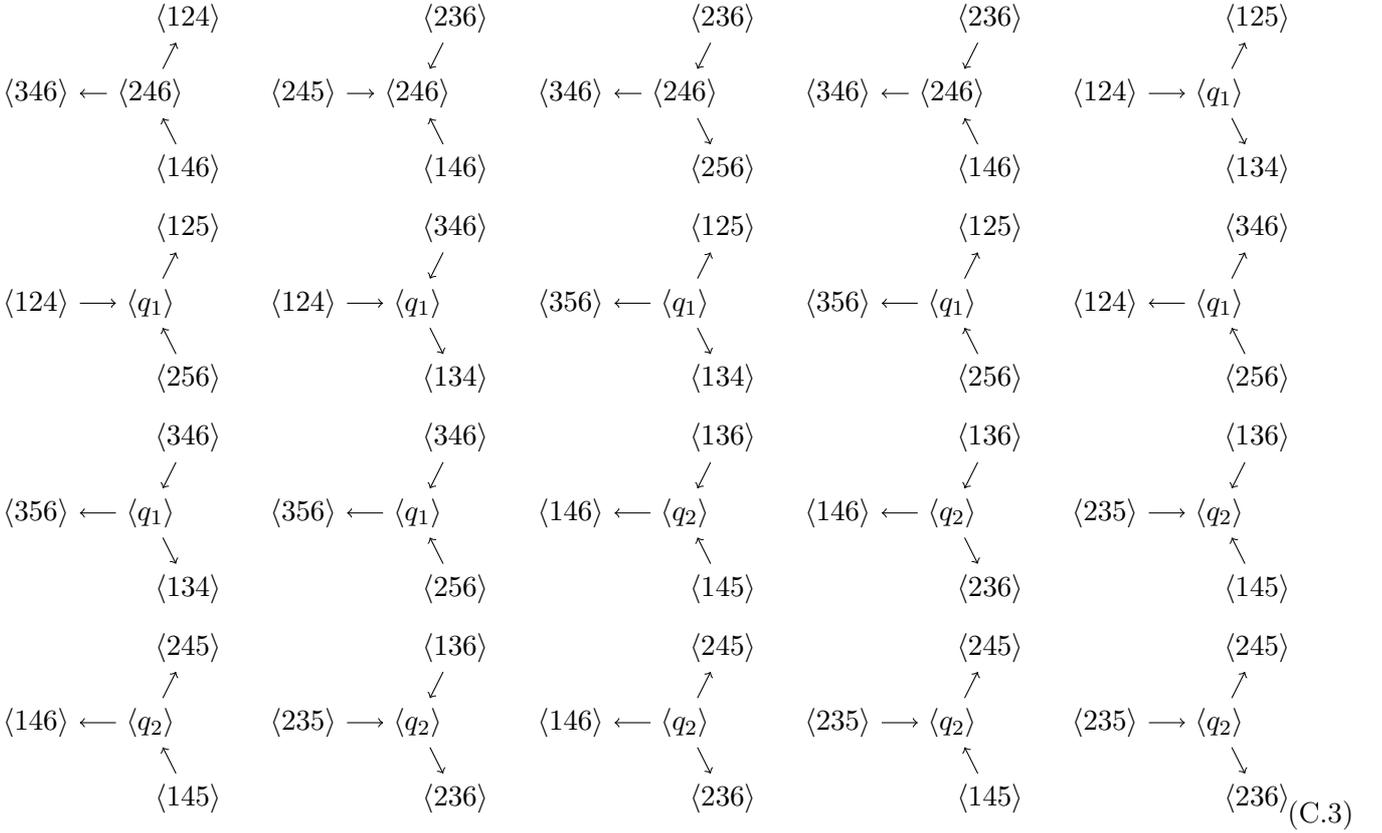
\begin{eqnarray}
\begin{tikzpicture}
\node at (-1,1) (1)  {$\langle 124\rangle$};
\node at (1,1) (2)  {$\langle 125\rangle$};
\node at (1,-1) (3) {$\langle 145\rangle$};
\node at (-1,-1) (4) {$\langle 134\rangle$};
\draw [->] (1) -- (2);
\draw [->] (2) -- (3);
\draw [->] (1)-- (4);
\draw [->] (4) -- (3);
\draw [->] (3) -- (1);
\end{tikzpicture} \quad
\begin{tikzpicture}
\node at (-1,1) (1)  {$\langle 124\rangle$};
\node at (1,1) (2)  {$\langle 125\rangle$};
\node at (1,-1) (3) {$\langle 145\rangle$};
\node at (-1,-1) (4) {$\langle 245\rangle$};
\draw [->] (1) -- (2);
\draw [->] (2) -- (3);
\draw [->] (3)-- (4);
\draw [->] (4) -- (1);
\end{tikzpicture} \quad 
\begin{tikzpicture}
\node at (-1,1) (1)  {$\langle 124\rangle$};
\node at (1,1) (2)  {$\langle 146\rangle$};
\node at (1,-1) (3) {$\langle 145\rangle$};
\node at (-1,-1) (4) {$\langle 134\rangle$};
\draw [->] (2) -- (1);
\draw [->] (1) -- (4);
\draw [->] (4)-- (3);
\draw [->] (3) -- (2);
\end{tikzpicture} \quad 
\begin{tikzpicture}
\node at (-1,1) (1)  {$\langle 124\rangle$};
\node at (1,1) (2)  {$\langle 146\rangle$};
\node at (1,-1) (3) {$\langle 145\rangle$};
\node at (-1,-1) (4) {$\langle 245\rangle$};
\draw [->] (1) -- (3);
\draw [->] (2) -- (1);
\draw [->] (4)-- (1);
\draw [->] (3) -- (4);
\draw [->] (3) -- (2);
\end{tikzpicture} \quad
\begin{tikzpicture}
\node at (-1,1) (1)  {$\langle 136\rangle$};
\node at (1,1) (2)  {$\langle 146\rangle$};
\node at (1,-1) (3) {$\langle 145\rangle$};
\node at (-1,-1) (4) {$\langle 134\rangle$};
\draw [->] (1) -- (2);
\draw [->] (2) -- (4);
\draw [->] (4)-- (1);
\draw [->] (4) -- (3);
\draw [->] (3) -- (2);
\end{tikzpicture} \nonumber\\
\begin{tikzpicture}
\node at (-1,1) (1)  {$\langle 136\rangle$};
\node at (1,1) (2)  {$\langle 146\rangle$};
\node at (1,-1) (3) {$\langle 346\rangle$};
\node at (-1,-1) (4) {$\langle 134\rangle$};
\draw [->] (1) -- (2);
\draw [->] (2) -- (3);
\draw [->] (3)-- (4);
\draw [->] (4) -- (1);
\end{tikzpicture} \quad 
\begin{tikzpicture}
\node at (-1,1) (1)  {$\langle 124\rangle$};
\node at (1,1) (2)  {$\langle 146\rangle$};
\node at (1,-1) (3) {$\langle 346\rangle$};
\node at (-1,-1) (4) {$\langle 134\rangle$};
\draw [->] (2) -- (1);
\draw [->] (2) -- (3);
\draw [->] (3)-- (4);
\draw [->] (1) -- (4);
\draw [->] (4) -- (2);
\end{tikzpicture} \quad 
\begin{tikzpicture}
\node at (-1,1) (1)  {$\langle 235\rangle$};
\node at (1,1) (2)  {$\langle 125\rangle$};
\node at (1,-1) (3) {$\langle 145\rangle$};
\node at (-1,-1) (4) {$\langle 245\rangle$};
\draw [->] (2) -- (1);
\draw [->] (2) -- (3);
\draw [->] (3)-- (4);
\draw [->] (1) -- (4);
\draw [->] (4) -- (2);
\end{tikzpicture} \quad 
\begin{tikzpicture}
\node at (-1,1) (1)  {$\langle 235\rangle$};
\node at (1,1) (2)  {$\langle 125\rangle$};
\node at (1,-1) (3) {$\langle 256\rangle$};
\node at (-1,-1) (4) {$\langle 245\rangle$};
\draw [->] (2) -- (1);
\draw [->] (1) -- (4);
\draw [->] (4)-- (3);
\draw [->] (3) -- (2);
\end{tikzpicture} \quad 
\begin{tikzpicture}
\node at (-1,1) (1)  {$\langle 124\rangle$};
\node at (1,1) (2)  {$\langle 125\rangle$};
\node at (1,-1) (3) {$\langle 256\rangle$};
\node at (-1,-1) (4) {$\langle 245\rangle$};
\draw [->] (1) -- (2);
\draw [->] (3) -- (2);
\draw [->] (4)-- (3);
\draw [->] (4) -- (1);
\draw [->] (2) -- (4);
\end{tikzpicture} \nonumber\\
\begin{tikzpicture}
\node at (-1,1) (1)  {$\langle 136\rangle$};
\node at (1,1) (2)  {$\langle 146\rangle$};
\node at (1,-1) (3) {$\langle 346\rangle$};
\node at (-1,-1) (4) {$\langle 236\rangle$};
\draw [->] (1) -- (2);
\draw [->] (1) -- (4);
\draw [->] (4)-- (3);
\draw [->] (2) -- (3);
\draw [->] (3) -- (1);
\end{tikzpicture} \quad 
\begin{tikzpicture}
\node at (-1,1) (1)  {$\langle 136\rangle$};
\node at (1,1) (2)  {$\langle 356\rangle$};
\node at (1,-1) (3) {$\langle 346\rangle$};
\node at (-1,-1) (4) {$\langle 236\rangle$};
\draw [->] (2) -- (1);
\draw [->] (1) -- (4);
\draw [->] (4)-- (3);
\draw [->] (3) -- (2);
\end{tikzpicture} \quad 
\begin{tikzpicture}
\node at (-1,1) (1)  {$\langle 136\rangle$};
\node at (1,1) (2)  {$\langle 356\rangle$};
\node at (1,-1) (3) {$\langle 346\rangle$};
\node at (-1,-1) (4) {$\langle 134\rangle$};
\draw [->] (2) -- (1);
\draw [->] (4) -- (1);
\draw [->] (3)-- (4);
\draw [->] (3) -- (2);
\draw [->] (1) -- (3);
\end{tikzpicture} \quad 
\begin{tikzpicture}
\node at (-1,1) (1)  {$\langle 235\rangle$};
\node at (1,1) (2)  {$\langle 125\rangle$};
\node at (1,-1) (3) {$\langle 256\rangle$};
\node at (-1,-1) (4) {$\langle 356\rangle$};
\draw [->] (2) -- (1);
\draw [->] (4) -- (1);
\draw [->] (3)-- (4);
\draw [->] (3) -- (2);
\draw [->] (1) -- (3);
\end{tikzpicture} \quad 
\begin{tikzpicture}
\node at (-1,1) (1)  {$\langle 235\rangle$};
\node at (1,1) (2)  {$\langle 236\rangle$};
\node at (1,-1) (3) {$\langle 256\rangle$};
\node at (-1,-1) (4) {$\langle 356\rangle$};
\draw [->] (1) -- (2);
\draw [->] (2) -- (3);
\draw [->] (3)-- (4);
\draw [->] (4) -- (1);
\end{tikzpicture} \nonumber\\
\begin{tikzpicture}
\node at (-1,1) (1)  {$\langle 235\rangle$};
\node at (1,1) (2)  {$\langle 236\rangle$};
\node at (1,-1) (3) {$\langle 256\rangle$};
\node at (-1,-1) (4) {$\langle 245\rangle$};
\draw [->] (1) -- (2);
\draw [->] (1) -- (4);
\draw [->] (4)-- (3);
\draw [->] (2) -- (3);
\draw [->] (3) -- (1);
\end{tikzpicture} \quad 
\begin{tikzpicture}
\node at (-1,1) (1)  {$\langle 136\rangle$};
\node at (1,1) (2)  {$\langle 356\rangle$};
\node at (1,-1) (3) {$\langle 235\rangle$};
\node at (-1,-1) (4) {$\langle 236\rangle$};
\draw [->] (2) -- (1);
\draw [->] (1) -- (4);
\draw [->] (3)-- (4);
\draw [->] (2) -- (3);
\draw [->] (4) -- (2);
\end{tikzpicture} \quad 
\begin{tikzpicture}
\node at (-1,1) (1)  {$\langle 256\rangle$};
\node at (1,1) (2)  {$\langle 356\rangle$};
\node at (1,-1) (3) {$\langle 346\rangle$};
\node at (-1,-1) (4) {$\langle 236\rangle$};
\draw [->] (1) -- (2);
\draw [->] (4) -- (1);
\draw [->] (4)-- (3);
\draw [->] (3) -- (2);
\draw [->] (2) -- (4);
\end{tikzpicture} \quad 
\begin{tikzpicture}
\node at (-1.5,0) (1)  {$\langle 145\rangle$};
\node at (0,0) (2)  {$\langle 135\rangle$};
\node at (0.5,1) (3) {$\langle 125\rangle$};
\node at (0.5,-1) (4) {$\langle 134\rangle$};
\draw [->] (2) -- (1);
\draw [->] (3) -- (2);
\draw [->] (4) -- (2);
\end{tikzpicture} \quad 
\begin{tikzpicture}
\node at (-1.5,0) (1)  {$\langle 145\rangle$};
\node at (0,0) (2)  {$\langle 135\rangle$};
\node at (0.5,1) (3) {$\langle 136\rangle$};
\node at (0.5,-1) (4) {$\langle 134\rangle$};
\draw [->] (2) -- (1);
\draw [->] (2) -- (3);
\draw [->] (4) -- (2);
\end{tikzpicture} \nonumber\\
\begin{tikzpicture}
\node at (-1.5,0) (1)  {$\langle 145\rangle$};
\node at (0,0) (2)  {$\langle 135\rangle$};
\node at (0.5,1) (3) {$\langle 125\rangle$};
\node at (0.5,-1) (4) {$\langle 235\rangle$};
\draw [->] (2) -- (1);
\draw [->] (3) -- (2);
\draw [->] (2) -- (4);
\end{tikzpicture} \quad 
\begin{tikzpicture}
\node at (-1.5,0) (1)  {$\langle 356\rangle$};
\node at (0,0) (2)  {$\langle 135\rangle$};
\node at (0.5,1) (3) {$\langle 125\rangle$};
\node at (0.5,-1) (4) {$\langle 134\rangle$};
\draw [->] (2) -- (1);
\draw [->] (2) -- (3);
\draw [->] (4) -- (2);
\end{tikzpicture} \quad 
\begin{tikzpicture}
\node at (-1.5,0) (1)  {$\langle 145\rangle$};
\node at (0,0) (2)  {$\langle 135\rangle$};
\node at (0.5,1) (3) {$\langle 136\rangle$};
\node at (0.5,-1) (4) {$\langle 235\rangle$};
\draw [->] (2) -- (1);
\draw [->] (2) -- (3);
\draw [->] (2) -- (4);
\end{tikzpicture} \quad 
\begin{tikzpicture}
\node at (-1.5,0) (1)  {$\langle 356\rangle$};
\node at (0,0) (2)  {$\langle 135\rangle$};
\node at (0.5,1) (3) {$\langle 136\rangle$};
\node at (0.5,-1) (4) {$\langle 134\rangle$};
\draw [->] (1) -- (2);
\draw [->] (2) -- (3);
\draw [->] (4) -- (2);
\end{tikzpicture} \quad 
\begin{tikzpicture}
\node at (-1.5,0) (1)  {$\langle 356\rangle$};
\node at (0,0) (2)  {$\langle 135\rangle$};
\node at (0.5,1) (3) {$\langle 125\rangle$};
\node at (0.5,-1) (4) {$\langle 235\rangle$};
\draw [->] (1) -- (2);
\draw [->] (3) -- (2);
\draw [->] (2) -- (4);
\end{tikzpicture} \nonumber\\
\begin{tikzpicture}
\node at (-1.5,0) (1)  {$\langle 356\rangle$};
\node at (0,0) (2)  {$\langle 135\rangle$};
\node at (0.5,1) (3) {$\langle 136\rangle$};
\node at (0.5,-1) (4) {$\langle 235\rangle$};
\draw [->] (1) -- (2);
\draw [->] (2) -- (3);
\draw [->] (2) -- (4);
\end{tikzpicture} \quad 
\begin{tikzpicture}
\node at (-1.5,0) (1)  {$\langle 245\rangle$};
\node at (0,0) (2)  {$\langle 246\rangle$};
\node at (0.5,1) (3) {$\langle 124\rangle$};
\node at (0.5,-1) (4) {$\langle 256\rangle$};
\draw [->] (1) -- (2);
\draw [->] (2) -- (3);
\draw [->] (2) -- (4);
\end{tikzpicture} \quad
\begin{tikzpicture}
\node at (-1.5,0) (1)  {$\langle 245\rangle$};
\node at (0,0) (2)  {$\langle 246\rangle$};
\node at (0.5,1) (3) {$\langle 124\rangle$};
\node at (0.5,-1) (4) {$\langle 146\rangle$};
\draw [->] (1) -- (2);
\draw [->] (2) -- (3);
\draw [->] (4) -- (2);
\end{tikzpicture} \quad
\begin{tikzpicture}
\node at (-1.5,0) (1)  {$\langle 346\rangle$};
\node at (0,0) (2)  {$\langle 246\rangle$};
\node at (0.5,1) (3) {$\langle 124\rangle$};
\node at (0.5,-1) (4) {$\langle 256\rangle$};
\draw [->] (2) -- (1);
\draw [->] (2) -- (3);
\draw [->] (2) -- (4);
\end{tikzpicture} \quad
\begin{tikzpicture}
\node at (-1.5,0) (1)  {$\langle 245\rangle$};
\node at (0,0) (2)  {$\langle 246\rangle$};
\node at (0.5,1) (3) {$\langle 236\rangle$};
\node at (0.5,-1) (4) {$\langle 256\rangle$};
\draw [->] (1) -- (2);
\draw [->] (3) -- (2);
\draw [->] (4) -- (2);
\end{tikzpicture} \nonumber\\
\begin{tikzpicture}
\node at (-1.5,0) (1)  {$\langle 346\rangle$};
\node at (0,0) (2)  {$\langle 246\rangle$};
\node at (0.5,1) (3) {$\langle 124\rangle$};
\node at (0.5,-1) (4) {$\langle 146\rangle$};
\draw [->] (2) -- (1);
\draw [->] (2) -- (3);
\draw [->] (4) -- (2);
\end{tikzpicture} \quad
\begin{tikzpicture}
\node at (-1.5,0) (1)  {$\langle 245\rangle$};
\node at (0,0) (2)  {$\langle 246\rangle$};
\node at (0.5,1) (3) {$\langle 236\rangle$};
\node at (0.5,-1) (4) {$\langle 146\rangle$};
\draw [->] (1) -- (2);
\draw [->] (3) -- (2);
\draw [->] (4) -- (2);
\end{tikzpicture} \quad
\begin{tikzpicture}
\node at (-1.5,0) (1)  {$\langle 346\rangle$};
\node at (0,0) (2)  {$\langle 246\rangle$};
\node at (0.5,1) (3) {$\langle 236\rangle$};
\node at (0.5,-1) (4) {$\langle 256\rangle$};
\draw [->] (2) -- (1);
\draw [->] (3) -- (2);
\draw [->] (2) -- (4);
\end{tikzpicture} \quad
\begin{tikzpicture}
\node at (-1.5,0) (1)  {$\langle 346\rangle$};
\node at (0,0) (2)  {$\langle 246\rangle$};
\node at (0.5,1) (3) {$\langle 236\rangle$};
\node at (0.5,-1) (4) {$\langle 146\rangle$};
\draw [->] (2) -- (1);
\draw [->] (3) -- (2);
\draw [->] (4) -- (2);
\end{tikzpicture} \quad
\begin{tikzpicture}
\node at (-1.5,0) (1)  {$\langle 124\rangle$};
\node at (0,0) (2)  {$\langle q_1\rangle$};
\node at (0.5,1) (3) {$\langle 125\rangle$};
\node at (0.5,-1) (4) {$\langle 134\rangle$};
\draw [->] (1) -- (2);
\draw [->] (2) -- (3);
\draw [->] (2) -- (4);
\end{tikzpicture} \nonumber\\
\begin{tikzpicture}
\node at (-1.5,0) (1)  {$\langle 124\rangle$};
\node at (0,0) (2)  {$\langle q_1\rangle$};
\node at (0.5,1) (3) {$\langle 125\rangle$};
\node at (0.5,-1) (4) {$\langle 256\rangle$};
\draw [->] (1) -- (2);
\draw [->] (2) -- (3);
\draw [->] (4) -- (2);
\end{tikzpicture} \quad
\begin{tikzpicture}
\node at (-1.5,0) (1)  {$\langle 124\rangle$};
\node at (0,0) (2)  {$\langle q_1\rangle$};
\node at (0.5,1) (3) {$\langle 346\rangle$};
\node at (0.5,-1) (4) {$\langle 134\rangle$};
\draw [->] (1) -- (2);
\draw [->] (3) -- (2);
\draw [->] (2) -- (4);
\end{tikzpicture} \quad
\begin{tikzpicture}
\node at (-1.5,0) (1)  {$\langle 356\rangle$};
\node at (0,0) (2)  {$\langle q_1\rangle$};
\node at (0.5,1) (3) {$\langle 125\rangle$};
\node at (0.5,-1) (4) {$\langle 134\rangle$};
\draw [->] (2) -- (1);
\draw [->] (2) -- (3);
\draw [->] (2) -- (4);
\end{tikzpicture} \quad
\begin{tikzpicture}
\node at (-1.5,0) (1)  {$\langle 356\rangle$};
\node at (0,0) (2)  {$\langle q_1\rangle$};
\node at (0.5,1) (3) {$\langle 125\rangle$};
\node at (0.5,-1) (4) {$\langle 256\rangle$};
\draw [->] (2) -- (1);
\draw [->] (2) -- (3);
\draw [->] (4) -- (2);
\end{tikzpicture} \quad
\begin{tikzpicture}
\node at (-1.5,0) (1)  {$\langle 124\rangle$};
\node at (0,0) (2)  {$\langle q_1\rangle$};
\node at (0.5,1) (3) {$\langle 346\rangle$};
\node at (0.5,-1) (4) {$\langle 256\rangle$};
\draw [->] (2) -- (1);
\draw [->] (2) -- (3);
\draw [->] (4) -- (2);
\end{tikzpicture} \nonumber\\
\begin{tikzpicture}
\node at (-1.5,0) (1)  {$\langle 356\rangle$};
\node at (0,0) (2)  {$\langle q_1\rangle$};
\node at (0.5,1) (3) {$\langle 346\rangle$};
\node at (0.5,-1) (4) {$\langle 134\rangle$};
\draw [->] (2) -- (1);
\draw [->] (3) -- (2);
\draw [->] (2) -- (4);
\end{tikzpicture} \quad
\begin{tikzpicture}
\node at (-1.5,0) (1)  {$\langle 356\rangle$};
\node at (0,0) (2)  {$\langle q_1\rangle$};
\node at (0.5,1) (3) {$\langle 346\rangle$};
\node at (0.5,-1) (4) {$\langle 256\rangle$};
\draw [->] (2) -- (1);
\draw [->] (3) -- (2);
\draw [->] (4) -- (2);
\end{tikzpicture} \quad
\begin{tikzpicture}
\node at (-1.5,0) (1)  {$\langle 146\rangle$};
\node at (0,0) (2)  {$\langle q_2\rangle$};
\node at (0.5,1) (3) {$\langle 136\rangle$};
\node at (0.5,-1) (4) {$\langle 145\rangle$};
\draw [->] (2) -- (1);
\draw [->] (3) -- (2);
\draw [->] (4) -- (2);
\end{tikzpicture} \quad
\begin{tikzpicture}
\node at (-1.5,0) (1)  {$\langle 146\rangle$};
\node at (0,0) (2)  {$\langle q_2\rangle$};
\node at (0.5,1) (3) {$\langle 136\rangle$};
\node at (0.5,-1) (4) {$\langle 236\rangle$};
\draw [->] (2) -- (1);
\draw [->] (3) -- (2);
\draw [->] (2) -- (4);
\end{tikzpicture} \quad
\begin{tikzpicture}
\node at (-1.5,0) (1)  {$\langle 235\rangle$};
\node at (0,0) (2)  {$\langle q_2\rangle$};
\node at (0.5,1) (3) {$\langle 136\rangle$};
\node at (0.5,-1) (4) {$\langle 145\rangle$};
\draw [->] (1) -- (2);
\draw [->] (3) -- (2);
\draw [->] (4) -- (2);
\end{tikzpicture} \nonumber\\
\begin{tikzpicture}
\node at (-1.5,0) (1)  {$\langle 146\rangle$};
\node at (0,0) (2)  {$\langle q_2\rangle$};
\node at (0.5,1) (3) {$\langle 245\rangle$};
\node at (0.5,-1) (4) {$\langle 145\rangle$};
\draw [->] (2) -- (1);
\draw [->] (2) -- (3);
\draw [->] (4) -- (2);
\end{tikzpicture} \quad
\begin{tikzpicture}
\node at (-1.5,0) (1)  {$\langle 235\rangle$};
\node at (0,0) (2)  {$\langle q_2\rangle$};
\node at (0.5,1) (3) {$\langle 136\rangle$};
\node at (0.5,-1) (4) {$\langle 236\rangle$};
\draw [->] (1) -- (2);
\draw [->] (3) -- (2);
\draw [->] (2) -- (4);
\end{tikzpicture} \quad
\begin{tikzpicture}
\node at (-1.5,0) (1)  {$\langle 146\rangle$};
\node at (0,0) (2)  {$\langle q_2\rangle$};
\node at (0.5,1) (3) {$\langle 245\rangle$};
\node at (0.5,-1) (4) {$\langle 236\rangle$};
\draw [->] (2) -- (1);
\draw [->] (2) -- (3);
\draw [->] (2) -- (4);
\end{tikzpicture} \quad
\begin{tikzpicture}
\node at (-1.5,0) (1)  {$\langle 235\rangle$};
\node at (0,0) (2)  {$\langle q_2\rangle$};
\node at (0.5,1) (3) {$\langle 245\rangle$};
\node at (0.5,-1) (4) {$\langle 145\rangle$};
\draw [->] (1) -- (2);
\draw [->] (2) -- (3);
\draw [->] (4) -- (2);
\end{tikzpicture} \quad
\begin{tikzpicture}
\node at (-1.5,0) (1)  {$\langle 235\rangle$};
\node at (0,0) (2)  {$\langle q_2\rangle$};
\node at (0.5,1) (3) {$\langle 245\rangle$};
\node at (0.5,-1) (4) {$\langle 236\rangle$};
\draw [->] (1) -- (2);
\draw [->] (2) -- (3);
\draw [->] (2) -- (4);
\end{tikzpicture} 
\end{eqnarray}
where $\langle q_1\rangle:=\langle 12[34]56\rangle \qquad \langle q_2\rangle:=\langle 23[45]61\rangle$.

\subsection{Feynman diagrams}
\label{Sec:Feynman-diagrams}
Feynman diagrams in the same ordering as the clusters in Appendix (\ref{sec:clusters-3,6}) are given:

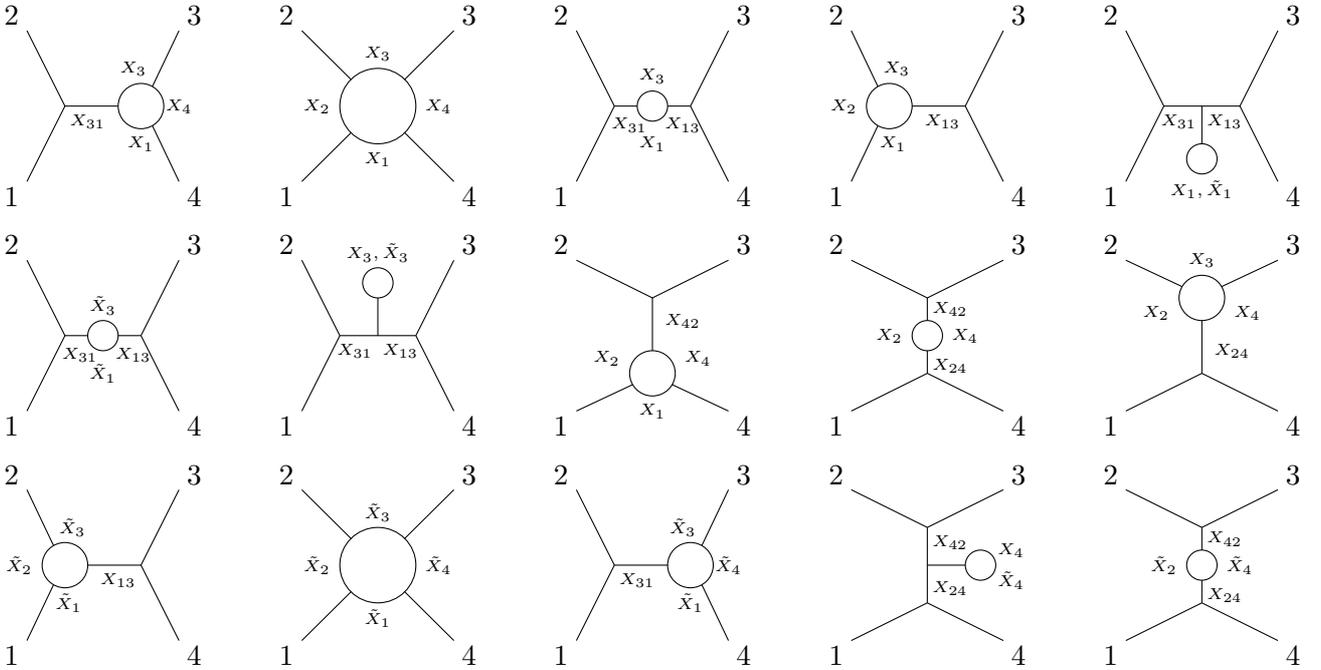
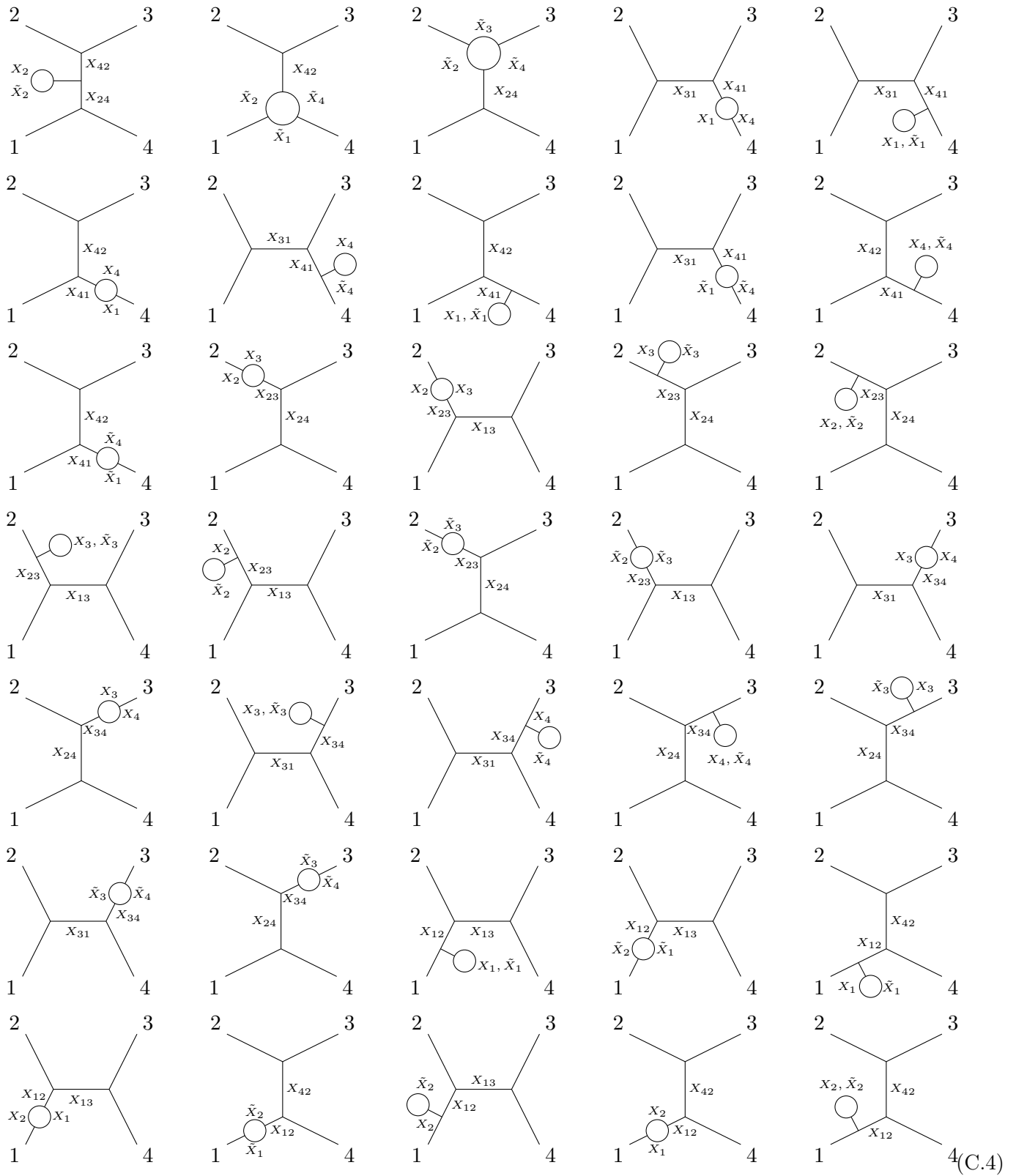
\begin{eqnarray}
\begin{tikzpicture}                             
\draw (-1, 1) -- (-0.5,0) -- (-1,-1);
\draw (-0.5,0) -- (0.2,0);
\draw (0.5, 0) circle (0.3);
\draw (0.65,0.26) -- (1,1);
\draw (0.65, -0.26) -- (1,-1);
\node at  (-1.2, -1.2) {$1$};
\node at (-1.2, 1.2) {$2$};
\node at (1.2, 1.2) {$3$};
\node at (1.2, -1.2) {$4$};
\node at (-0.2, -0.2) {\tiny{$X_{31}$}};
\node at (0.5, -0.5) {\tiny{$X_{1}$}};
\node at (1, 0) {\tiny{$X_{4}$}};
\node at (0.4, 0.5) {\tiny{$X_{3}$}};
\end{tikzpicture} \qquad
\begin{tikzpicture}                             
\draw (0,0) circle(0.5);
\draw (-1, 1) -- (-0.35, 0.35) ;
\draw (-1, -1) -- (-0.35, -0.35);
\draw (0.35, 0.35) -- (1,1);
\draw (0.35, -0.35) -- (1, -1);
\node at  (-1.2, -1.2) {$1$};
\node at (-1.2, 1.2) {$2$};
\node at (1.2, 1.2) {$3$};
\node at (1.2, -1.2) {$4$}; 
\node at (0, -0.7) {\tiny{$X_{1}$}};
\node at (-0.8, 0) {\tiny{$X_{2}$}};
\node at (0, 0.7) {\tiny{$X_{3}$}};
\node at (0.8, 0) {\tiny{$X_{4}$}};
\end{tikzpicture} \qquad 
\begin{tikzpicture}                             
\draw (-1, 1) -- (-0.5,0) -- (-1,-1);
\draw (-0.5,0) -- (-0.2,0);
\draw (0, 0) circle (0.2);
\draw (0.2,0) -- (0.5,0);
\draw (0.5,0) -- (1,1);
\draw (0.5, 0) -- (1,-1);
\node at  (-1.2, -1.2) {$1$};
\node at (-1.2, 1.2) {$2$};
\node at (1.2, 1.2) {$3$};
\node at (1.2, -1.2) {$4$};
\node at (-0.3, -0.25) {\tiny{${X_{31}}$}};
\node at (0, -0.5) {\tiny{$X_{1}$}};
\node at (0.4, -0.25) {\tiny{$X_{13}$}};
\node at (0, 0.4) {\tiny{$X_{3}$}};
\end{tikzpicture} \qquad
\begin{tikzpicture}                             
\draw (1, 1) -- (0.5,0) -- (1,-1);
\draw (0.5,0) -- (-0.2,0);
\draw (-0.5, 0) circle (0.3);
\draw (-0.65,0.26) -- (-1,1);
\draw (-0.65, -0.26) -- (-1,-1);
\node at  (-1.2, -1.2) {$1$};
\node at (-1.2, 1.2) {$2$};
\node at (1.2, 1.2) {$3$};
\node at (1.2, -1.2) {$4$};
\node at (0.2, -0.2) {\tiny{$X_{13}$}};
\node at (-0.45, -0.5) {\tiny{$X_{1}$}};
\node at (-1.1, 0) {\tiny{$X_{2}$}};
\node at (-0.4, 0.5) {\tiny{$X_{3}$}};
\end{tikzpicture} \qquad
\begin{tikzpicture}                             
\draw (-1, 1) -- (-0.5,0) -- (-1,-1);
\draw (-0.5,0) -- (0.5,0);
\draw (0, -0.7) circle (0.2);
\draw (0,0) -- (0,-0.5);
\draw (1,1) -- (0.5, 0) -- (1, -1);
\node at  (-1.2, -1.2) {$1$};
\node at (-1.2, 1.2) {$2$};
\node at (1.2, 1.2) {$3$};
\node at (1.2, -1.2) {$4$};
\node at (-0.3, -0.2) {\tiny{${X_{31}}$}};
\node at (0.3, -0.2) {\tiny{$X_{13}$}};
\node at (0, -1.1) {\tiny{$X_{1}, \tilde{X}_{1}$}};
\end{tikzpicture}  \nonumber\\
\begin{tikzpicture}                             
\draw (-1, 1) -- (-0.5,0) -- (-1,-1);
\draw (-0.5,0) -- (-0.2,0);
\draw (0, 0) circle (0.2);
\draw (0.2,0) -- (0.5,0);
\draw (0.5,0) -- (1,1);
\draw (0.5, 0) -- (1,-1);
\node at  (-1.2, -1.2) {$1$};
\node at (-1.2, 1.2) {$2$};
\node at (1.2, 1.2) {$3$};
\node at (1.2, -1.2) {$4$};
\node at (-0.3, -0.25) {\tiny{${X_{31}}$}};
\node at (0, -0.5) {\tiny{$\tilde{X}_{1}$}};
\node at (0.4, -0.25) {\tiny{$X_{13}$}};
\node at (0, 0.4) {\tiny{$\tilde{X}_{3}$}};
\end{tikzpicture} \qquad
\begin{tikzpicture}                             
\draw (-1, 1) -- (-0.5,0) -- (-1,-1);
\draw (-0.5,0) -- (0.5,0);
\draw (0, 0.7) circle (0.2);
\draw (0,0) -- (0,0.5);
\draw (1,1) -- (0.5, 0) -- (1, -1);
\node at  (-1.2, -1.2) {$1$};
\node at (-1.2, 1.2) {$2$};
\node at (1.2, 1.2) {$3$};
\node at (1.2, -1.2) {$4$};
\node at (-0.3, -0.2) {\tiny{${X_{31}}$}};
\node at (0.3, -0.2) {\tiny{$X_{13}$}};
\node at (0, 1.1) {\tiny{$X_{3}, \tilde{X}_{3}$}};
\end{tikzpicture} \qquad
\begin{tikzpicture}                            
\draw (-1,1) -- (0,0.5) -- (1,1);
\draw (0,0.5) -- (-0,-0.2);
\draw (0,-0.5) circle (0.3);
\draw (-1,-1) -- (-0.26, -0.65);
\draw (1,-1) -- (0.26, -0.65);
\node at  (-1.2, -1.2) {$1$};
\node at (-1.2, 1.2) {$2$};
\node at (1.2, 1.2) {$3$};
\node at (1.2, -1.2) {$4$};
\node at (0, -1) {\tiny{$X_{1}$}};
\node at (-0.6, -0.3) {\tiny{$X_{2}$}};
\node at (0.6, -0.3) {\tiny{$X_{4}$}};
\node at (0.4, 0.2) {\tiny{$X_{42}$}};
\end{tikzpicture} \qquad
\begin{tikzpicture}                            
\draw (-1,1) -- (0,0.5) -- (1,1);
\draw (0,0.5) -- (0,0.2);
\draw (0,0) circle (0.2);
\draw (-1,-1) -- (0,-0.5) -- (1,-1);
\draw (0,-0.5) -- (0,-0.2);
\node at  (-1.2, -1.2) {$1$};
\node at (-1.2, 1.2) {$2$};
\node at (1.2, 1.2) {$3$};
\node at (1.2, -1.2) {$4$};
\node at (0.3, -0.4) {\tiny{$X_{24}$}};
\node at (-0.5, 0) {\tiny{$X_{2}$}};
\node at (0.5, 0) {\tiny{$X_{4}$}};
\node at (0.3, 0.35) {\tiny{$X_{42}$}};
\end{tikzpicture} \qquad
\begin{tikzpicture}                            
\draw (-1,-1) -- (0,-0.5) -- (1,-1);
\draw (0,-0.5) -- (0,0.2);
\draw (0,0.5) circle (0.3);
\draw (-1,1) -- (-0.26, 0.65);
\draw (1,1) -- (0.26, 0.65);
\node at  (-1.2, -1.2) {$1$};
\node at (-1.2, 1.2) {$2$};
\node at (1.2, 1.2) {$3$};
\node at (1.2, -1.2) {$4$};
\node at (0, 1) {\tiny{$X_{3}$}};
\node at (-0.6, 0.3) {\tiny{$X_{2}$}};
\node at (0.6, 0.3) {\tiny{$X_{4}$}};
\node at (0.4, -0.2) {\tiny{$X_{24}$}};
\end{tikzpicture} \nonumber\\
\begin{tikzpicture}                             
\draw (1, 1) -- (0.5,0) -- (1,-1);
\draw (0.5,0) -- (-0.2,0);
\draw (-0.5, 0) circle (0.3);
\draw (-0.65,0.26) -- (-1,1);
\draw (-0.65, -0.26) -- (-1,-1);
\node at  (-1.2, -1.2) {$1$};
\node at (-1.2, 1.2) {$2$};
\node at (1.2, 1.2) {$3$};
\node at (1.2, -1.2) {$4$};
\node at (0.2, -0.2) {\tiny{$X_{13}$}};
\node at (-0.45, -0.5) {\tiny{$\tilde{X}_{1}$}};
\node at (-1.1, 0) {\tiny{$\tilde{X}_{2}$}};
\node at (-0.4, 0.5) {\tiny{$\tilde{X}_{3}$}};
\end{tikzpicture} \qquad
\begin{tikzpicture}                             
\draw (0,0) circle(0.5);
\draw (-1, 1) -- (-0.35, 0.35) ;
\draw (-1, -1) -- (-0.35, -0.35);
\draw (0.35, 0.35) -- (1,1);
\draw (0.35, -0.35) -- (1, -1);
\node at  (-1.2, -1.2) {$1$};
\node at (-1.2, 1.2) {$2$};
\node at (1.2, 1.2) {$3$};
\node at (1.2, -1.2) {$4$}; 
\node at (0, -0.7) {\tiny{$\tilde{X}_{1}$}};
\node at (-0.8, 0) {\tiny{$\tilde{X}_{2}$}};
\node at (0, 0.7) {\tiny{$\tilde{X}_{3}$}};
\node at (0.8, 0) {\tiny{$\tilde{X}_{4}$}};
\end{tikzpicture} \qquad
\begin{tikzpicture}                             
\draw (-1, 1) -- (-0.5,0) -- (-1,-1);
\draw (-0.5,0) -- (0.2,0);
\draw (0.5, 0) circle (0.3);
\draw (0.65,0.26) -- (1,1);
\draw (0.65, -0.26) -- (1,-1);
\node at  (-1.2, -1.2) {$1$};
\node at (-1.2, 1.2) {$2$};
\node at (1.2, 1.2) {$3$};
\node at (1.2, -1.2) {$4$};
\node at (-0.2, -0.2) {\tiny{$X_{31}$}};
\node at (0.5, -0.5) {\tiny{$\tilde{X}_{1}$}};
\node at (1, 0) {\tiny{$\tilde{X}_{4}$}};
\node at (0.4, 0.5) {\tiny{$\tilde{X}_{3}$}};
\end{tikzpicture} \qquad
\begin{tikzpicture}                            
\draw (-1,1) -- (0,0.5) -- (1,1);
\draw (0,0.5) -- (0,-0.5);
\draw (0,0) -- (0.5,0);
\draw (0.7, 0) circle (0.2);
\draw (-1,-1) -- (0,-0.5) -- (1,-1);
\node at  (-1.2, -1.2) {$1$};
\node at (-1.2, 1.2) {$2$};
\node at (1.2, 1.2) {$3$};
\node at (1.2, -1.2) {$4$};
\node at (0.3, -0.3) {\tiny{$X_{24}$}};
\node at (1.1, -0.2) {\tiny{$\tilde{X}_{4}$}};
\node at (1.1, 0.2) {\tiny{$X_{4}$}};
\node at (0.3, 0.3) {\tiny{$X_{42}$}};
\end{tikzpicture} \qquad
\begin{tikzpicture}                            
\draw (-1,1) -- (0,0.5) -- (1,1);
\draw (0,0.5) -- (0,0.2);
\draw (0,0) circle (0.2);
\draw (-1,-1) -- (0,-0.5) -- (1,-1);
\draw (0,-0.5) -- (0,-0.2);
\node at  (-1.2, -1.2) {$1$};
\node at (-1.2, 1.2) {$2$};
\node at (1.2, 1.2) {$3$};
\node at (1.2, -1.2) {$4$};
\node at (0.3, -0.4) {\tiny{$X_{24}$}};
\node at (-0.5, 0) {\tiny{$\tilde{X}_{2}$}};
\node at (0.5, 0) {\tiny{$\tilde{X}_{4}$}};
\node at (0.3, 0.35) {\tiny{$X_{42}$}};
\end{tikzpicture} \nonumber\\
\begin{tikzpicture}                            
\draw (-1,1) -- (0,0.5) -- (1,1);
\draw (0,0.5) -- (0,-0.5);
\draw (0,0) -- (-0.5,0);
\draw (-0.7, 0) circle (0.2);
\draw (-1,-1) -- (0,-0.5) -- (1,-1);
\node at  (-1.2, -1.2) {$1$};
\node at (-1.2, 1.2) {$2$};
\node at (1.2, 1.2) {$3$};
\node at (1.2, -1.2) {$4$};
\node at (0.3, -0.3) {\tiny{$X_{24}$}};
\node at (-1.1, -0.2) {\tiny{$\tilde{X}_{2}$}};
\node at (-1.1, 0.2) {\tiny{$X_{2}$}};
\node at (0.3, 0.3) {\tiny{$X_{42}$}};
\end{tikzpicture} \qquad
\begin{tikzpicture}                            
\draw (-1,1) -- (0,0.5) -- (1,1);
\draw (0,0.5) -- (-0,-0.2);
\draw (0,-0.5) circle (0.3);
\draw (-1,-1) -- (-0.26, -0.65);
\draw (1,-1) -- (0.26, -0.65);
\node at  (-1.2, -1.2) {$1$};
\node at (-1.2, 1.2) {$2$};
\node at (1.2, 1.2) {$3$};
\node at (1.2, -1.2) {$4$};
\node at (0, -1) {\tiny{$\tilde{X}_{1}$}};
\node at (-0.6, -0.3) {\tiny{$\tilde{X}_{2}$}};
\node at (0.6, -0.3) {\tiny{$\tilde{X}_{4}$}};
\node at (0.4, 0.2) {\tiny{$X_{42}$}};
\end{tikzpicture} \qquad
\begin{tikzpicture}                            
\draw (-1,-1) -- (0,-0.5) -- (1,-1);
\draw (0,-0.5) -- (0,0.2);
\draw (0,0.5) circle (0.3);
\draw (-1,1) -- (-0.26, 0.65);
\draw (1,1) -- (0.26, 0.65);
\node at  (-1.2, -1.2) {$1$};
\node at (-1.2, 1.2) {$2$};
\node at (1.2, 1.2) {$3$};
\node at (1.2, -1.2) {$4$};
\node at (0, 1) {\tiny{$\tilde{X}_{3}$}};
\node at (-0.6, 0.3) {\tiny{$\tilde{X}_{2}$}};
\node at (0.6, 0.3) {\tiny{$\tilde{X}_{4}$}};
\node at (0.4, -0.2) {\tiny{$X_{24}$}};
\end{tikzpicture} \qquad
\begin{tikzpicture}                             
\draw (-1, 1) -- (-0.5,0) -- (-1,-1);
\draw (-0.5,0) -- (0.5,0);
\draw (1, 1) -- (0.5,0) -- (0.66,-0.32);
\draw (0.75, -0.5) circle (0.2);
\draw (0.84,-0.68) -- (1,-1);
\node at  (-1.2, -1.2) {$1$};
\node at (-1.2, 1.2) {$2$};
\node at (1.2, 1.2) {$3$};
\node at (1.2, -1.2) {$4$};
\node at (0, -0.2) {\tiny{${X_{31}}$}};
\node at (0.38, -0.7) {\tiny{${X}_{1}$}};
\node at (0.9, -0.1) {\tiny{$X_{41}$}};
\node at (1.1, -0.7) {\tiny{${X}_{4}$}};
\end{tikzpicture} \qquad
\begin{tikzpicture}                             
\draw (-1, 1) -- (-0.5,0) -- (-1,-1);
\draw (-0.5,0) -- (0.5,0);
\draw (1, 1) -- (0.5,0) -- (1,-1);
\draw (0.32, -0.72) circle (0.2);
\draw (0.75, -0.5) -- (0.5,-0.63);
\node at  (-1.2, -1.2) {$1$};
\node at (-1.2, 1.2) {$2$};
\node at (1.2, 1.2) {$3$};
\node at (1.2, -1.2) {$4$};
\node at (0, -0.2) {\tiny{${X_{31}}$}};
\node at (0.32, -1.1) {\tiny{${X}_{1},\tilde{X}_{1}$}};
\node at (0.9, -0.2) {\tiny{$X_{41}$}};
\end{tikzpicture} \nonumber\\
\begin{tikzpicture}                            
\draw (-1,1) -- (0,0.5) -- (1,1);
\draw (0,0.5) -- (0,-0.5);
\draw (0.5,-0.75) circle (0.2);
\draw (-1,-1) -- (0,-0.5) -- (0.32,-0.66);
\draw (0.68,-0.84) -- (1,-1);
\node at  (-1.2, -1.2) {$1$};
\node at (-1.2, 1.2) {$2$};
\node at (1.2, 1.2) {$3$};
\node at (1.2, -1.2) {$4$};
\node at (0, -0.8) {\tiny{$X_{41}$}};
\node at (0.6, -1.1) {\tiny{$X_{1}$}};
\node at (0.6, -0.4) {\tiny{$X_{4}$}};
\node at (0.3, 0) {\tiny{$X_{42}$}};
\end{tikzpicture} \qquad
\begin{tikzpicture}                             
\draw (-1, 1) -- (-0.5,0) -- (-1,-1);
\draw (-0.5,0) -- (0.5,0);
\draw (1, 1) -- (0.5,0) -- (1,-1);
\draw (1.18, -0.28) circle (0.2);
\draw (0.75, -0.5) -- (1.0,-0.37);
\node at  (-1.2, -1.2) {$1$};
\node at (-1.2, 1.2) {$2$};
\node at (1.2, 1.2) {$3$};
\node at (1.2, -1.2) {$4$};
\node at (0, 0.2) {\tiny{${X_{31}}$}};
\node at (1.18, 0.1) {\tiny{${X}_{4}$}};
\node at (1.18, -0.7) {\tiny{$\tilde{X}_{4}$}};
\node at (0.4, -0.3) {\tiny{$X_{41}$}};
\end{tikzpicture} \qquad
\begin{tikzpicture}                            
\draw (-1,1) -- (0,0.5) -- (1,1);
\draw (0,0.5) -- (0,-0.5);
\draw (-1,-1) -- (0,-0.5) -- (1,-1);
\draw (0.5,-0.75) -- (0.37,-1.0);
\draw (0.28,-1.18) circle (0.2);
\node at  (-1.2, -1.2) {$1$};
\node at (-1.2, 1.2) {$2$};
\node at (1.2, 1.2) {$3$};
\node at (1.2, -1.2) {$4$};
\node at (0.1, -0.8) {\tiny{$X_{41}$}};
\node at (-0.32, -1.2) {\tiny{${X}_{1},\tilde{X}_{1}$}};
\node at (0.3, 0) {\tiny{$X_{42}$}};
\end{tikzpicture} \qquad
\begin{tikzpicture}                             
\draw (-1, 1) -- (-0.5,0) -- (-1,-1);
\draw (-0.5,0) -- (0.5,0);
\draw (1, 1) -- (0.5,0) -- (0.66,-0.32);
\draw (0.75, -0.5) circle (0.2);
\draw (0.84,-0.68) -- (1,-1);
\node at  (-1.2, -1.2) {$1$};
\node at (-1.2, 1.2) {$2$};
\node at (1.2, 1.2) {$3$};
\node at (1.2, -1.2) {$4$};
\node at (0, -0.2) {\tiny{${X_{31}}$}};
\node at (0.38, -0.7) {\tiny{$\tilde{X}_{1}$}};
\node at (0.9, -0.1) {\tiny{$X_{41}$}};
\node at (1.1, -0.7) {\tiny{$\tilde{X}_{4}$}};
\end{tikzpicture} \qquad
\begin{tikzpicture}                             
\draw (1, -1) -- (0,-0.5) -- (-1,-1);
\draw (0,0.5) -- (0,-0.5);
\draw (1, 1) -- (0,0.5) -- (-1,1);
\draw (0.72, -0.32) circle (0.2);
\draw (0.5, -0.75) -- (0.63,-0.5);
\node at  (-1.2, -1.2) {$1$};
\node at (-1.2, 1.2) {$2$};
\node at (1.2, 1.2) {$3$};
\node at (1.2, -1.2) {$4$};
\node at (-0.3, 0) {\tiny{${X_{42}}$}};
\node at (0.8, 0.1) {\tiny{${X}_{4},\tilde{X}_{4}$}};
\node at (0.1, -0.8) {\tiny{$X_{41}$}};
\end{tikzpicture} \nonumber\\
\begin{tikzpicture}                            
\draw (-1,1) -- (0,0.5) -- (1,1);
\draw (0,0.5) -- (0,-0.5);
\draw (0.5,-0.75) circle (0.2);
\draw (-1,-1) -- (0,-0.5) -- (0.32,-0.66);
\draw (0.68,-0.84) -- (1,-1);
\node at  (-1.2, -1.2) {$1$};
\node at (-1.2, 1.2) {$2$};
\node at (1.2, 1.2) {$3$};
\node at (1.2, -1.2) {$4$};
\node at (0, -0.8) {\tiny{$X_{41}$}};
\node at (0.6, -1.1) {\tiny{$\tilde{X}_{1}$}};
\node at (0.6, -0.4) {\tiny{$\tilde{X}_{4}$}};
\node at (0.3, 0) {\tiny{$X_{42}$}};
\end{tikzpicture} \qquad
\begin{tikzpicture}                            
\draw (1,-1) -- (0,-0.5) -- (-1,-1);
\draw (0,0.5) -- (0,-0.5);
\draw (-0.5,0.75) circle (0.2);
\draw (1,1) -- (0,0.5) -- (-0.32,0.66);
\draw (-0.68,0.84) -- (-1,1);
\node at  (-1.2, -1.2) {$1$};
\node at (-1.2, 1.2) {$2$};
\node at (1.2, 1.2) {$3$};
\node at (1.2, -1.2) {$4$};
\node at (-0.25, 0.4) {\tiny{$X_{23}$}};
\node at (-0.9, 0.7) {\tiny{$X_{2}$}};
\node at (-0.5, 1.1) {\tiny{$X_{3}$}};
\node at (0.3, 0) {\tiny{$X_{24}$}};
\end{tikzpicture} \qquad
\begin{tikzpicture}                             
\draw (1, -1) -- (0.5,0) -- (1,1);
\draw (-0.5,0) -- (0.5,0);
\draw (-1, -1) -- (-0.5,0) -- (-0.66,0.32);
\draw (-0.75, 0.5) circle (0.2);
\draw (-0.84,0.68) -- (-1,1);
\node at  (-1.2, -1.2) {$1$};
\node at (-1.2, 1.2) {$2$};
\node at (1.2, 1.2) {$3$};
\node at (1.2, -1.2) {$4$};
\node at (0, -0.2) {\tiny{${X_{13}}$}};
\node at (-0.35, 0.5) {\tiny{${X}_{3}$}};
\node at (-0.8, 0.1) {\tiny{$X_{23}$}};
\node at (-1.15, 0.5) {\tiny{${X}_{2}$}};
\end{tikzpicture} \qquad
\begin{tikzpicture}                            
\draw (1,-1) -- (0,-0.5) -- (-1,-1);
\draw (0,-0.5) -- (0,0.5);
\draw (1,1) -- (0,0.5) -- (-1,1);
\draw (-0.5,0.75) -- (-0.37,1.0);
\draw (-0.28,1.18) circle (0.2);
\node at  (-1.2, -1.2) {$1$};
\node at (-1.2, 1.2) {$2$};
\node at (1.2, 1.2) {$3$};
\node at (1.2, -1.2) {$4$};
\node at (-0.3, 0.4) {\tiny{$X_{23}$}};
\node at (-0.7, 1.2) {\tiny{${X}_{3}$}};
\node at (0.1, 1.2) {\tiny{$\tilde{X}_{3}$}};
\node at (0.3, 0) {\tiny{$X_{24}$}};
\end{tikzpicture} \qquad
\begin{tikzpicture}                             
\draw (-1, 1) -- (0,0.5) -- (1,1);
\draw (0,-0.5) -- (0,0.5);
\draw (-1, -1) -- (0,-0.5) -- (1,-1);
\draw (-0.72, 0.32) circle (0.2);
\draw (-0.5, 0.75) -- (-0.63,0.5);
\node at  (-1.2, -1.2) {$1$};
\node at (-1.2, 1.2) {$2$};
\node at (1.2, 1.2) {$3$};
\node at (1.2, -1.2) {$4$};
\node at (0.3, 0) {\tiny{${X_{24}}$}};
\node at (-0.8, -0.1) {\tiny{${X}_{2},\tilde{X}_{2}$}};
\node at (-0.25, 0.4) {\tiny{$X_{23}$}};
\end{tikzpicture} \nonumber\\
\begin{tikzpicture}                             
\draw (-1, 1) -- (-0.5,0) -- (-1,-1);
\draw (-0.5,0) -- (0.5,0);
\draw (1, 1) -- (0.5,0) -- (1,-1);
\draw (-0.32, 0.72) circle (0.2);
\draw (-0.75, 0.5) -- (-0.5,0.63);
\node at  (-1.2, -1.2) {$1$};
\node at (-1.2, 1.2) {$2$};
\node at (1.2, 1.2) {$3$};
\node at (1.2, -1.2) {$4$};
\node at (0, -0.2) {\tiny{${X_{13}}$}};
\node at (0.32, 0.8) {\tiny{${X}_{3},\tilde{X}_{3}$}};
\node at (-0.9, 0.2) {\tiny{$X_{23}$}};
\end{tikzpicture} \qquad
\begin{tikzpicture}                             
\draw (-1, 1) -- (-0.5,0) -- (-1,-1);
\draw (-0.5,0) -- (0.5,0);
\draw (1, 1) -- (0.5,0) -- (1,-1);
\draw (-1.18, 0.28) circle (0.2);
\draw (-0.75, 0.5) -- (-1.0,0.37);
\node at  (-1.2, -1.2) {$1$};
\node at (-1.2, 1.2) {$2$};
\node at (1.2, 1.2) {$3$};
\node at (1.2, -1.2) {$4$};
\node at (0, -0.15) {\tiny{${X_{13}}$}};
\node at (-1.05, 0.65) {\tiny{${X}_{2}$}};
\node at (-1.05, -0.1) {\tiny{$\tilde{X}_{2}$}};
\node at (-0.35, 0.3) {\tiny{$X_{23}$}};
\end{tikzpicture} \qquad
\begin{tikzpicture}                            
\draw (1,-1) -- (0,-0.5) -- (-1,-1);
\draw (0,0.5) -- (0,-0.5);
\draw (-0.5,0.75) circle (0.2);
\draw (1,1) -- (0,0.5) -- (-0.32,0.66);
\draw (-0.68,0.84) -- (-1,1);
\node at  (-1.2, -1.2) {$1$};
\node at (-1.2, 1.2) {$2$};
\node at (1.2, 1.2) {$3$};
\node at (1.2, -1.2) {$4$};
\node at (-0.25, 0.4) {\tiny{$X_{23}$}};
\node at (-0.9, 0.7) {\tiny{$\tilde{X}_{2}$}};
\node at (-0.5, 1.1) {\tiny{$\tilde{X}_{3}$}};
\node at (0.3, 0) {\tiny{$X_{24}$}};
\end{tikzpicture} \qquad
\begin{tikzpicture}                             
\draw (1, -1) -- (0.5,0) -- (1,1);
\draw (-0.5,0) -- (0.5,0);
\draw (-1, -1) -- (-0.5,0) -- (-0.66,0.32);
\draw (-0.75, 0.5) circle (0.2);
\draw (-0.84,0.68) -- (-1,1);
\node at  (-1.2, -1.2) {$1$};
\node at (-1.2, 1.2) {$2$};
\node at (1.2, 1.2) {$3$};
\node at (1.2, -1.2) {$4$};
\node at (0, -0.2) {\tiny{${X_{13}}$}};
\node at (-0.35, 0.5) {\tiny{$\tilde{X}_{3}$}};
\node at (-0.8, 0.1) {\tiny{$X_{23}$}};
\node at (-1.15, 0.5) {\tiny{$\tilde{X}_{2}$}};
\end{tikzpicture} \qquad
\begin{tikzpicture}                             
\draw (-1, -1) -- (-0.5,0) -- (-1,1);
\draw (-0.5,0) -- (0.5,0);
\draw (1, -1) -- (0.5,0) -- (0.66,0.32);
\draw (0.75, 0.5) circle (0.2);
\draw (0.84,0.68) -- (1,1);
\node at  (-1.2, -1.2) {$1$};
\node at (-1.2, 1.2) {$2$};
\node at (1.2, 1.2) {$3$};
\node at (1.2, -1.2) {$4$};
\node at (0, -0.2) {\tiny{${X_{31}}$}};
\node at (0.35, 0.5) {\tiny{${X}_{3}$}};
\node at (0.9, 0.1) {\tiny{$X_{34}$}};
\node at (1.15, 0.5) {\tiny{${X}_{4}$}};
\end{tikzpicture} \nonumber\\
\begin{tikzpicture}                            
\draw (-1,-1) -- (0,-0.5) -- (1,-1);
\draw (0,0.5) -- (0,-0.5);
\draw (0.5,0.75) circle (0.2);
\draw (-1,1) -- (0,0.5) -- (0.32,0.66);
\draw (0.68,0.84) -- (1,1);
\node at  (-1.2, -1.2) {$1$};
\node at (-1.2, 1.2) {$2$};
\node at (1.2, 1.2) {$3$};
\node at (1.2, -1.2) {$4$};
\node at (0.25, 0.4) {\tiny{$X_{34}$}};
\node at (0.9, 0.7) {\tiny{${X}_{4}$}};
\node at (0.5, 1.1) {\tiny{${X}_{3}$}};
\node at (-0.3, 0) {\tiny{$X_{24}$}};
\end{tikzpicture} \qquad
\begin{tikzpicture}                             
\draw (1, 1) -- (0.5,0) -- (1,-1);
\draw (-0.5,0) -- (0.5,0);
\draw (-1, 1) -- (-0.5,0) -- (-1,-1);
\draw (0.32, 0.72) circle (0.2);
\draw (0.75, 0.5) -- (0.5,0.63);
\node at  (-1.2, -1.2) {$1$};
\node at (-1.2, 1.2) {$2$};
\node at (1.2, 1.2) {$3$};
\node at (1.2, -1.2) {$4$};
\node at (0, -0.2) {\tiny{${X_{31}}$}};
\node at (-0.32, 0.8) {\tiny{${X}_{3},\tilde{X}_{3}$}};
\node at (0.9, 0.2) {\tiny{$X_{34}$}};
\end{tikzpicture} \qquad
\begin{tikzpicture}                             
\draw (-1, 1) -- (-0.5,0) -- (-1,-1);
\draw (-0.5,0) -- (0.5,0);
\draw (1, 1) -- (0.5,0) -- (1,-1);
\draw (1.18, 0.28) circle (0.2);
\draw (0.75, 0.5) -- (1.0,0.37);
\node at  (-1.2, -1.2) {$1$};
\node at (-1.2, 1.2) {$2$};
\node at (1.2, 1.2) {$3$};
\node at (1.2, -1.2) {$4$};
\node at (0, -0.15) {\tiny{${X_{31}}$}};
\node at (1.05, 0.65) {\tiny{${X}_{4}$}};
\node at (1.05, -0.1) {\tiny{$\tilde{X}_{4}$}};
\node at (0.35, 0.3) {\tiny{$X_{34}$}};
\end{tikzpicture} \qquad
\begin{tikzpicture}                             
\draw (-1, 1) -- (0,0.5) -- (1,1);
\draw (0,-0.5) -- (0,0.5);
\draw (-1, -1) -- (0,-0.5) -- (1,-1);
\draw (0.72, 0.32) circle (0.2);
\draw (0.5, 0.75) -- (0.63,0.5);
\node at  (-1.2, -1.2) {$1$};
\node at (-1.2, 1.2) {$2$};
\node at (1.2, 1.2) {$3$};
\node at (1.2, -1.2) {$4$};
\node at (-0.3, 0) {\tiny{${X_{24}}$}};
\node at (0.8, -0.1) {\tiny{${X}_{4},\tilde{X}_{4}$}};
\node at (0.25, 0.4) {\tiny{$X_{34}$}};
\end{tikzpicture} \qquad
\begin{tikzpicture}                            
\draw (1,-1) -- (0,-0.5) -- (-1,-1);
\draw (0,-0.5) -- (0,0.5);
\draw (1,1) -- (0,0.5) -- (-1,1);
\draw (0.5,0.75) -- (0.37,1.0);
\draw (0.28,1.18) circle (0.2);
\node at  (-1.2, -1.2) {$1$};
\node at (-1.2, 1.2) {$2$};
\node at (1.2, 1.2) {$3$};
\node at (1.2, -1.2) {$4$};
\node at (0.3, 0.4) {\tiny{$X_{34}$}};
\node at (0.7, 1.2) {\tiny{${X}_{3}$}};
\node at (-0.1, 1.2) {\tiny{$\tilde{X}_{3}$}};
\node at (-0.3, 0) {\tiny{$X_{24}$}};
\end{tikzpicture} \nonumber\\
\begin{tikzpicture}                             
\draw (-1, -1) -- (-0.5,0) -- (-1,1);
\draw (-0.5,0) -- (0.5,0);
\draw (1, -1) -- (0.5,0) -- (0.66,0.32);
\draw (0.75, 0.5) circle (0.2);
\draw (0.84,0.68) -- (1,1);
\node at  (-1.2, -1.2) {$1$};
\node at (-1.2, 1.2) {$2$};
\node at (1.2, 1.2) {$3$};
\node at (1.2, -1.2) {$4$};
\node at (0, -0.2) {\tiny{${X_{31}}$}};
\node at (0.35, 0.5) {\tiny{$\tilde{X}_{3}$}};
\node at (0.9, 0.1) {\tiny{$X_{34}$}};
\node at (1.15, 0.5) {\tiny{$\tilde{X}_{4}$}};
\end{tikzpicture} \qquad
\begin{tikzpicture}                            
\draw (-1,-1) -- (0,-0.5) -- (1,-1);
\draw (0,0.5) -- (0,-0.5);
\draw (0.5,0.75) circle (0.2);
\draw (-1,1) -- (0,0.5) -- (0.32,0.66);
\draw (0.68,0.84) -- (1,1);
\node at  (-1.2, -1.2) {$1$};
\node at (-1.2, 1.2) {$2$};
\node at (1.2, 1.2) {$3$};
\node at (1.2, -1.2) {$4$};
\node at (0.25, 0.4) {\tiny{$X_{34}$}};
\node at (0.9, 0.7) {\tiny{$\tilde{X}_{4}$}};
\node at (0.5, 1.1) {\tiny{$\tilde{X}_{3}$}};
\node at (-0.3, 0) {\tiny{$X_{24}$}};
\end{tikzpicture} \qquad
\begin{tikzpicture}                             
\draw (-1, 1) -- (-0.5,0) -- (-1,-1);
\draw (-0.5,0) -- (0.5,0);
\draw (1, 1) -- (0.5,0) -- (1,-1);
\draw (-0.32, -0.72) circle (0.2);
\draw (-0.75, -0.5) -- (-0.5,-0.63);
\node at  (-1.2, -1.2) {$1$};
\node at (-1.2, 1.2) {$2$};
\node at (1.2, 1.2) {$3$};
\node at (1.2, -1.2) {$4$};
\node at (0, -0.2) {\tiny{${X_{13}}$}};
\node at (0.32, -0.8) {\tiny{${X}_{1},\tilde{X}_{1}$}};
\node at (-0.9, -0.2) {\tiny{$X_{12}$}};
\end{tikzpicture} \qquad
\begin{tikzpicture}                             
\draw (1, -1) -- (0.5,0) -- (1,1);
\draw (-0.5,0) -- (0.5,0);
\draw (-1, 1) -- (-0.5,0) -- (-0.66,-0.32);
\draw (-0.75, -0.5) circle (0.2);
\draw (-0.84,-0.68) -- (-1,-1);
\node at  (-1.2, -1.2) {$1$};
\node at (-1.2, 1.2) {$2$};
\node at (1.2, 1.2) {$3$};
\node at (1.2, -1.2) {$4$};
\node at (0, -0.2) {\tiny{${X_{13}}$}};
\node at (-0.35, -0.5) {\tiny{$\tilde{X}_{1}$}};
\node at (-0.85, -0.1) {\tiny{$X_{12}$}};
\node at (-1.15, -0.5) {\tiny{$\tilde{X}_{2}$}};
\end{tikzpicture} \qquad
\begin{tikzpicture}                            
\draw (1,-1) -- (0,-0.5) -- (-1,-1);
\draw (0,-0.5) -- (0,0.5);
\draw (1,1) -- (0,0.5) -- (-1,1);
\draw (-0.5,-0.75) -- (-0.37,-1.0);
\draw (-0.28,-1.18) circle (0.2);
\node at  (-1.2, -1.2) {$1$};
\node at (-1.2, 1.2) {$2$};
\node at (1.2, 1.2) {$3$};
\node at (1.2, -1.2) {$4$};
\node at (-0.3, -0.4) {\tiny{$X_{12}$}};
\node at (-0.7, -1.2) {\tiny{${X}_{1}$}};
\node at (0.1, -1.2) {\tiny{$\tilde{X}_{1}$}};
\node at (0.3, 0) {\tiny{$X_{42}$}};
\end{tikzpicture} \nonumber\\
\begin{tikzpicture}                             
\draw (1, -1) -- (0.5,0) -- (1,1);
\draw (-0.5,0) -- (0.5,0);
\draw (-1, 1) -- (-0.5,0) -- (-0.66,-0.32);
\draw (-0.75, -0.5) circle (0.2);
\draw (-0.84,-0.68) -- (-1,-1);
\node at  (-1.2, -1.2) {$1$};
\node at (-1.2, 1.2) {$2$};
\node at (1.2, 1.2) {$3$};
\node at (1.2, -1.2) {$4$};
\node at (0, -0.2) {\tiny{${X_{13}}$}};
\node at (-0.35, -0.5) {\tiny{${X}_{1}$}};
\node at (-0.85, -0.1) {\tiny{$X_{12}$}};
\node at (-1.15, -0.5) {\tiny{${X}_{2}$}};
\end{tikzpicture} \qquad
\begin{tikzpicture}                            
\draw (1,1) -- (0,0.5) -- (-1,1);
\draw (0,0.5) -- (0,-0.5);
\draw (-0.5,-0.75) circle (0.2);
\draw (1,-1) -- (0,-0.5) -- (-0.32,-0.66);
\draw (-0.68,-0.84) -- (-1,-1);
\node at  (-1.2, -1.2) {$1$};
\node at (-1.2, 1.2) {$2$};
\node at (1.2, 1.2) {$3$};
\node at (1.2, -1.2) {$4$};
\node at (0, -0.75) {\tiny{$X_{12}$}};
\node at (-0.5, -0.4) {\tiny{$\tilde{X}_{2}$}};
\node at (-0.5, -1.1) {\tiny{$\tilde{X}_{1}$}};
\node at (0.3, 0) {\tiny{$X_{42}$}};
\end{tikzpicture} \qquad
\begin{tikzpicture}                             
\draw (-1, 1) -- (-0.5,0) -- (-1,-1);
\draw (-0.5,0) -- (0.5,0);
\draw (1, 1) -- (0.5,0) -- (1,-1);
\draw (-1.18, -0.28) circle (0.2);
\draw (-0.75, -0.5) -- (-1.0,-0.37);
\node at  (-1.2, -1.2) {$1$};
\node at (-1.2, 1.2) {$2$};
\node at (1.2, 1.2) {$3$};
\node at (1.2, -1.2) {$4$};
\node at (0, 0.15) {\tiny{${X_{13}}$}};
\node at (-1.05, -0.65) {\tiny{${X}_{2}$}};
\node at (-1.05, 0.1) {\tiny{$\tilde{X}_{2}$}};
\node at (-0.35, -0.3) {\tiny{$X_{12}$}};
\end{tikzpicture} \qquad
\begin{tikzpicture}                            
\draw (1,1) -- (0,0.5) -- (-1,1);
\draw (0,0.5) -- (0,-0.5);
\draw (-0.5,-0.75) circle (0.2);
\draw (1,-1) -- (0,-0.5) -- (-0.32,-0.66);
\draw (-0.68,-0.84) -- (-1,-1);
\node at  (-1.2, -1.2) {$1$};
\node at (-1.2, 1.2) {$2$};
\node at (1.2, 1.2) {$3$};
\node at (1.2, -1.2) {$4$};
\node at (0, -0.75) {\tiny{$X_{12}$}};
\node at (-0.5, -0.4) {\tiny{${X}_{2}$}};
\node at (-0.5, -1.1) {\tiny{${X}_{1}$}};
\node at (0.3, 0) {\tiny{$X_{42}$}};
\end{tikzpicture} \qquad
\begin{tikzpicture}                             
\draw (-1, 1) -- (0,0.5) -- (1,1);
\draw (0,-0.5) -- (0,0.5);
\draw (-1, -1) -- (0,-0.5) -- (1,-1);
\draw (-0.72, -0.32) circle (0.2);
\draw (-0.5, -0.75) -- (-0.63,-0.5);
\node at  (-1.2, -1.2) {$1$};
\node at (-1.2, 1.2) {$2$};
\node at (1.2, 1.2) {$3$};
\node at (1.2, -1.2) {$4$};
\node at (0.3, 0) {\tiny{${X_{42}}$}};
\node at (-0.8, 0.1) {\tiny{${X}_{2},\tilde{X}_{2}$}};
\node at (-0.1, -0.8) {\tiny{$X_{12}$}};
\end{tikzpicture} 
\end{eqnarray}

\bibliographystyle{utphys}
\bibliography{grkn}

\providecommand{\href}[2]{#2}\begingroup\raggedright\begin{thebibliography}{10}

\bibitem{Parke:1986gb}
S.~J. Parke and T.~Taylor, ``{An Amplitude for $n$ Gluon Scattering},''
  \href{http://dx.doi.org/10.1103/PhysRevLett.56.2459}{{\em Phys. Rev. Lett.}
  {\bfseries 56} (1986) 2459}.

\bibitem{Witten:2003nn}
E.~Witten, ``{Perturbative gauge theory as a string theory in twistor space},''
  \href{http://dx.doi.org/10.1007/s00220-004-1187-3}{{\em Commun. Math. Phys.}
  {\bfseries 252} (2004) 189--258},
  \href{http://arxiv.org/abs/hep-th/0312171}{{\ttfamily arXiv:hep-th/0312171}}.

\bibitem{Roiban:2004yf}
R.~Roiban, M.~Spradlin, and A.~Volovich, ``{On the tree level S matrix of
  Yang-Mills theory},''
  \href{http://dx.doi.org/10.1103/PhysRevD.70.026009}{{\em Phys. Rev. D}
  {\bfseries 70} (2004) 026009},
  \href{http://arxiv.org/abs/hep-th/0403190}{{\ttfamily arXiv:hep-th/0403190}}.

\bibitem{Cachazo:2004kj}
F.~Cachazo, P.~Svrcek, and E.~Witten, ``{MHV vertices and tree amplitudes in
  gauge theory},'' \href{http://dx.doi.org/10.1088/1126-6708/2004/09/006}{{\em
  JHEP} {\bfseries 09} (2004) 006},
  \href{http://arxiv.org/abs/hep-th/0403047}{{\ttfamily arXiv:hep-th/0403047}}.

\bibitem{Britto:2005fq}
R.~Britto, F.~Cachazo, B.~Feng, and E.~Witten, ``{Direct proof of tree-level
  recursion relation in Yang-Mills theory},''
  \href{http://dx.doi.org/10.1103/PhysRevLett.94.181602}{{\em Phys. Rev. Lett.}
  {\bfseries 94} (2005) 181602},
  \href{http://arxiv.org/abs/hep-th/0501052}{{\ttfamily arXiv:hep-th/0501052}}.

\bibitem{ArkaniHamed:2008yf}
N.~Arkani-Hamed and J.~Kaplan, ``{On Tree Amplitudes in Gauge Theory and
  Gravity},'' \href{http://dx.doi.org/10.1088/1126-6708/2008/04/076}{{\em JHEP}
  {\bfseries 04} (2008) 076}, \href{http://arxiv.org/abs/0801.2385}{{\ttfamily
  arXiv:0801.2385 [hep-th]}}.

\bibitem{Mason:2013sva}
L.~Mason and D.~Skinner, ``{Ambitwistor strings and the scattering
  equations},'' \href{http://dx.doi.org/10.1007/JHEP07(2014)048}{{\em JHEP}
  {\bfseries 07} (2014) 048}, \href{http://arxiv.org/abs/1311.2564}{{\ttfamily
  arXiv:1311.2564 [hep-th]}}.

\bibitem{Geyer:2014fka}
Y.~Geyer, A.~E. Lipstein, and L.~J. Mason, ``{Ambitwistor Strings in Four
  Dimensions},'' \href{http://dx.doi.org/10.1103/PhysRevLett.113.081602}{{\em
  Phys. Rev. Lett.} {\bfseries 113} no.~8, (2014) 081602},
  \href{http://arxiv.org/abs/1404.6219}{{\ttfamily arXiv:1404.6219 [hep-th]}}.

\bibitem{Casali:2015vta}
E.~Casali, Y.~Geyer, L.~Mason, R.~Monteiro, and K.~A. Roehrig, ``{New
  Ambitwistor String Theories},''
  \href{http://dx.doi.org/10.1007/JHEP11(2015)038}{{\em JHEP} {\bfseries 11}
  (2015) 038}, \href{http://arxiv.org/abs/1506.08771}{{\ttfamily
  arXiv:1506.08771 [hep-th]}}.

\bibitem{Geyer:2015bja}
Y.~Geyer, L.~Mason, R.~Monteiro, and P.~Tourkine, ``{Loop Integrands for
  Scattering Amplitudes from the Riemann Sphere},''
  \href{http://dx.doi.org/10.1103/PhysRevLett.115.121603}{{\em Phys. Rev.
  Lett.} {\bfseries 115} no.~12, (2015) 121603},
  \href{http://arxiv.org/abs/1507.00321}{{\ttfamily arXiv:1507.00321
  [hep-th]}}.

\bibitem{Geyer:2015jch}
Y.~Geyer, L.~Mason, R.~Monteiro, and P.~Tourkine, ``{One-loop amplitudes on the
  Riemann sphere},'' \href{http://dx.doi.org/10.1007/JHEP03(2016)114}{{\em
  JHEP} {\bfseries 03} (2016) 114},
  \href{http://arxiv.org/abs/1511.06315}{{\ttfamily arXiv:1511.06315
  [hep-th]}}.

\bibitem{Geyer:2016wjx}
Y.~Geyer, L.~Mason, R.~Monteiro, and P.~Tourkine, ``{Two-Loop Scattering
  Amplitudes from the Riemann Sphere},''
  \href{http://dx.doi.org/10.1103/PhysRevD.94.125029}{{\em Phys. Rev. D}
  {\bfseries 94} no.~12, (2016) 125029},
  \href{http://arxiv.org/abs/1607.08887}{{\ttfamily arXiv:1607.08887
  [hep-th]}}.

\bibitem{Geyer:2017ela}
Y.~Geyer and R.~Monteiro, ``{Gluons and gravitons at one loop from ambitwistor
  strings},'' \href{http://dx.doi.org/10.1007/JHEP03(2018)068}{{\em JHEP}
  {\bfseries 03} (2018) 068}, \href{http://arxiv.org/abs/1711.09923}{{\ttfamily
  arXiv:1711.09923 [hep-th]}}.

\bibitem{Geyer:2018xwu}
Y.~Geyer and R.~Monteiro, ``{Two-Loop Scattering Amplitudes from Ambitwistor
  Strings: from Genus Two to the Nodal Riemann Sphere},''
  \href{http://dx.doi.org/10.1007/JHEP11(2018)008}{{\em JHEP} {\bfseries 11}
  (2018) 008}, \href{http://arxiv.org/abs/1805.05344}{{\ttfamily
  arXiv:1805.05344 [hep-th]}}.

\bibitem{Berkovits:2019bbx}
N.~Berkovits, M.~Guillen, and L.~Mason, ``{Supertwistor description of
  ambitwistor strings},'' \href{http://dx.doi.org/10.1007/JHEP01(2020)020}{{\em
  JHEP} {\bfseries 01} (2020) 020},
  \href{http://arxiv.org/abs/1908.06899}{{\ttfamily arXiv:1908.06899
  [hep-th]}}.

\bibitem{Cachazo:2013iaa}
F.~Cachazo, S.~He, and E.~Y. Yuan, ``{Scattering in Three Dimensions from
  Rational Maps},'' \href{http://dx.doi.org/10.1007/JHEP10(2013)141}{{\em JHEP}
  {\bfseries 10} (2013) 141}, \href{http://arxiv.org/abs/1306.2962}{{\ttfamily
  arXiv:1306.2962 [hep-th]}}.

\bibitem{Cachazo:2013gna}
F.~Cachazo, S.~He, and E.~Y. Yuan, ``{Scattering equations and
  Kawai-Lewellen-Tye orthogonality},''
  \href{http://dx.doi.org/10.1103/PhysRevD.90.065001}{{\em Phys. Rev. D}
  {\bfseries 90} no.~6, (2014) 065001},
  \href{http://arxiv.org/abs/1306.6575}{{\ttfamily arXiv:1306.6575 [hep-th]}}.

\bibitem{Cachazo:2013hca}
F.~Cachazo, S.~He, and E.~Y. Yuan, ``{Scattering of Massless Particles in
  Arbitrary Dimensions},''
  \href{http://dx.doi.org/10.1103/PhysRevLett.113.171601}{{\em Phys. Rev.
  Lett.} {\bfseries 113} no.~17, (2014) 171601},
  \href{http://arxiv.org/abs/1307.2199}{{\ttfamily arXiv:1307.2199 [hep-th]}}.

\bibitem{Cachazo:2013iea}
F.~Cachazo, S.~He, and E.~Y. Yuan, ``{Scattering of Massless Particles:
  Scalars, Gluons and Gravitons},''
  \href{http://dx.doi.org/10.1007/JHEP07(2014)033}{{\em JHEP} {\bfseries 07}
  (2014) 033}, \href{http://arxiv.org/abs/1309.0885}{{\ttfamily arXiv:1309.0885
  [hep-th]}}.

\bibitem{Cachazo:2014fwa}
F.~Cachazo and A.~Strominger, ``{Evidence for a New Soft Graviton Theorem},''
  \href{http://arxiv.org/abs/1404.4091}{{\ttfamily arXiv:1404.4091 [hep-th]}}.

\bibitem{Cachazo:2014nsa}
F.~Cachazo, S.~He, and E.~Y. Yuan, ``{Einstein-Yang-Mills Scattering Amplitudes
  From Scattering Equations},''
  \href{http://dx.doi.org/10.1007/JHEP01(2015)121}{{\em JHEP} {\bfseries 01}
  (2015) 121}, \href{http://arxiv.org/abs/1409.8256}{{\ttfamily arXiv:1409.8256
  [hep-th]}}.

\bibitem{Cachazo:2014xea}
F.~Cachazo, S.~He, and E.~Y. Yuan, ``{Scattering Equations and Matrices: From
  Einstein To Yang-Mills, DBI and NLSM},''
  \href{http://dx.doi.org/10.1007/JHEP07(2015)149}{{\em JHEP} {\bfseries 07}
  (2015) 149}, \href{http://arxiv.org/abs/1412.3479}{{\ttfamily arXiv:1412.3479
  [hep-th]}}.

\bibitem{Cachazo:2016sdc}
F.~Cachazo and G.~Zhang, ``{Minimal Basis in Four Dimensions and Scalar
  Blocks},'' \href{http://arxiv.org/abs/1601.06305}{{\ttfamily arXiv:1601.06305
  [hep-th]}}.

\bibitem{Naculich_2014}
S.~G. Naculich, ``Scattering equations and bcj relations for gauge and
  gravitational amplitudes with massive scalar particles,''
  \href{http://dx.doi.org/10.1007/jhep09(2014)029}{{\em Journal of High Energy
  Physics} {\bfseries 2014} no.~9, (Sep, 2014) }.
  \url{http://dx.doi.org/10.1007/JHEP09(2014)029}.

\bibitem{Naculich_2015}
S.~G. Naculich, ``Chy representations for gauge theory and gravity amplitudes
  with up to three massive particles,''
  \href{http://dx.doi.org/10.1007/jhep05(2015)050}{{\em Journal of High Energy
  Physics} {\bfseries 2015} no.~5, (May, 2015) }.
  \url{http://dx.doi.org/10.1007/JHEP05(2015)050}.

\bibitem{Weinzierl:2014ava}
S.~Weinzierl, ``{Fermions and the scattering equations},''
  \href{http://dx.doi.org/10.1007/JHEP03(2015)141}{{\em JHEP} {\bfseries 03}
  (2015) 141}, \href{http://arxiv.org/abs/1412.5993}{{\ttfamily arXiv:1412.5993
  [hep-th]}}.

\bibitem{Adamo:2013tsa}
T.~Adamo, E.~Casali, and D.~Skinner, ``{Ambitwistor strings and the scattering
  equations at one loop},''
  \href{http://dx.doi.org/10.1007/JHEP04(2014)104}{{\em JHEP} {\bfseries 04}
  (2014) 104}, \href{http://arxiv.org/abs/1312.3828}{{\ttfamily arXiv:1312.3828
  [hep-th]}}.

\bibitem{Adamo:2015hoa}
T.~Adamo and E.~Casali, ``{Scattering equations, supergravity integrands, and
  pure spinors},'' \href{http://dx.doi.org/10.1007/JHEP05(2015)120}{{\em JHEP}
  {\bfseries 05} (2015) 120}, \href{http://arxiv.org/abs/1502.06826}{{\ttfamily
  arXiv:1502.06826 [hep-th]}}.

\bibitem{He:2015yua}
S.~He and E.~Y. Yuan, ``{One-loop Scattering Equations and Amplitudes from
  Forward Limit},'' \href{http://dx.doi.org/10.1103/PhysRevD.92.105004}{{\em
  Phys. Rev.} {\bfseries D92} no.~10, (2015) 105004},
\href{http://arxiv.org/abs/1508.06027}{{\ttfamily arXiv:1508.06027 [hep-th]}}.

\bibitem{Cachazo:2015aol}
F.~Cachazo, S.~He, and E.~Y. Yuan, ``{One-Loop Corrections from Higher
  Dimensional Tree Amplitudes},''
  \href{http://dx.doi.org/10.1007/JHEP08(2016)008}{{\em JHEP} {\bfseries 08}
  (2016) 008}, \href{http://arxiv.org/abs/1512.05001}{{\ttfamily
  arXiv:1512.05001 [hep-th]}}.

\bibitem{Feng:2016nrf}
B.~Feng, ``{CHY-construction of Planar Loop Integrands of Cubic Scalar
  Theory},'' \href{http://dx.doi.org/10.1007/JHEP05(2016)061}{{\em JHEP}
  {\bfseries 05} (2016) 061}, \href{http://arxiv.org/abs/1601.05864}{{\ttfamily
  arXiv:1601.05864 [hep-th]}}.

\bibitem{Feng:2019xiq}
B.~Feng and C.~Hu, ``{One-loop CHY-Integrand of Bi-adjoint Scalar Theory},''
  \href{http://dx.doi.org/10.1007/JHEP02(2020)187}{{\em JHEP} {\bfseries 02}
  (2020) 187}, \href{http://arxiv.org/abs/1912.12960}{{\ttfamily
  arXiv:1912.12960 [hep-th]}}.

\bibitem{Farrow_2020}
J.~A. Farrow, Y.~Geyer, A.~E. Lipstein, R.~Monteiro, and R.~Stark-Much{\~a}o,
  ``Propagators, bcfw recursion and new scattering equations at one loop,''
  \href{http://dx.doi.org/10.1007/jhep10(2020)074}{{\em Journal of High Energy
  Physics} {\bfseries 2020} no.~10, (Oct, 2020) }.
  \url{http://dx.doi.org/10.1007/JHEP10(2020)074}.

\bibitem{Schwab}
B.~U.~W. Schwab and A.~Volovich, ``{Subleading Soft Theorem in Arbitrary
  Dimensions from Scattering Equations},''
  \href{http://dx.doi.org/10.1103/PhysRevLett.113.101601}{{\em Phys. Rev.
  Lett.} {\bfseries 113} no.~10, (2014) 101601},
\href{http://arxiv.org/abs/1404.7749}{{\ttfamily arXiv:1404.7749 [hep-th]}}.

\bibitem{Afkhami}
N.~Afkhami-Jeddi, ``{Soft Graviton Theorem in Arbitrary Dimensions},''
\href{http://arxiv.org/abs/1405.3533}{{\ttfamily arXiv:1405.3533 [hep-th]}}.

\bibitem{Zlotnikov}
M.~Zlotnikov, ``{Sub-sub-leading soft-graviton theorem in arbitrary
  dimension},'' \href{http://dx.doi.org/10.1007/JHEP10(2014)148}{{\em JHEP}
  {\bfseries 10} (2014) 148},
\href{http://arxiv.org/abs/1407.5936}{{\ttfamily arXiv:1407.5936 [hep-th]}}.

\bibitem{Kalousios}
C.~Kalousios and F.~Rojas, ``{Next to subleading soft-graviton theorem in
  arbitrary dimensions},''
  \href{http://dx.doi.org/10.1007/JHEP01(2015)107}{{\em JHEP} {\bfseries 01}
  (2015) 107},
\href{http://arxiv.org/abs/1407.5982}{{\ttfamily arXiv:1407.5982 [hep-th]}}.

\bibitem{DoubleSoftPRD}
F.~Cachazo, S.~He, and E.~Y. Yuan, ``{New Double Soft Emission Theorems},''
  \href{http://dx.doi.org/10.1103/PhysRevD.92.065030}{{\em Phys. Rev.}
  {\bfseries D92} no.~6, (2015) 065030},
\href{http://arxiv.org/abs/1503.04816}{{\ttfamily arXiv:1503.04816 [hep-th]}}.

\bibitem{VolovichZlotnikov}
A.~Volovich, C.~Wen, and M.~Zlotnikov, ``{Double Soft Theorems in Gauge and
  String Theories},'' \href{http://dx.doi.org/10.1007/JHEP07(2015)095}{{\em
  JHEP} {\bfseries 07} (2015) 095},
\href{http://arxiv.org/abs/1504.05559}{{\ttfamily arXiv:1504.05559 [hep-th]}}.

\bibitem{Saha:2017yqi}
A.~P. Saha, ``{Double soft limit of the graviton amplitude from the
  Cachazo-He-Yuan formalism},''
  \href{http://dx.doi.org/10.1103/PhysRevD.96.045002}{{\em Phys. Rev. D}
  {\bfseries 96} no.~4, (2017) 045002},
  \href{http://arxiv.org/abs/1702.02350}{{\ttfamily arXiv:1702.02350
  [hep-th]}}.

\bibitem{Saha:2016kjr}
A.~P. Saha, ``{Double Soft Theorem for Perturbative Gravity},''
  \href{http://dx.doi.org/10.1007/JHEP09(2016)165}{{\em JHEP} {\bfseries 09}
  (2016) 165}, \href{http://arxiv.org/abs/1607.02700}{{\ttfamily
  arXiv:1607.02700 [hep-th]}}.

\bibitem{Chakrabarti:2017zmh}
S.~Chakrabarti, S.~P. Kashyap, B.~Sahoo, A.~Sen, and M.~Verma, ``{Testing
  Subleading Multiple Soft Graviton Theorem for CHY Prescription},''
  \href{http://dx.doi.org/10.1007/JHEP01(2018)090}{{\em JHEP} {\bfseries 01}
  (2018) 090}, \href{http://arxiv.org/abs/1709.07883}{{\ttfamily
  arXiv:1709.07883 [hep-th]}}.

\bibitem{Nandan_2017}
D.~Nandan, J.~Plefka, and W.~Wormsbecher, ``Collinear limits beyond the leading
  order from the scattering equations,''
  \href{http://dx.doi.org/10.1007/jhep02(2017)038}{{\em Journal of High Energy
  Physics} {\bfseries 2017} no.~2, (Feb, 2017) }.
  \url{http://dx.doi.org/10.1007/JHEP02(2017)038}.

\bibitem{Cachazo:2019ngv}
F.~Cachazo, N.~Early, A.~Guevara, and S.~Mizera, ``{Scattering Equations: From
  Projective Spaces to Tropical Grassmannians},''
  \href{http://dx.doi.org/10.1007/JHEP06(2019)039}{{\em JHEP} {\bfseries 06}
  (2019) 039}, \href{http://arxiv.org/abs/1903.08904}{{\ttfamily
  arXiv:1903.08904 [hep-th]}}.

\bibitem{Cachazo:2016ror}
F.~Cachazo, S.~Mizera, and G.~Zhang, ``{Scattering Equations: Real Solutions
  and Particles on a Line},''
  \href{http://dx.doi.org/10.1007/JHEP03(2017)151}{{\em JHEP} {\bfseries 03}
  (2017) 151}, \href{http://arxiv.org/abs/1609.00008}{{\ttfamily
  arXiv:1609.00008 [hep-th]}}.

\bibitem{Cachazo:2019apa}
F.~Cachazo and J.~M. Rojas, ``{Notes on Biadjoint Amplitudes, ${\rm
  Trop}\,G(3,7)$ and $X(3,7)$ Scattering Equations},''
  \href{http://dx.doi.org/10.1007/JHEP04(2020)176}{{\em JHEP} {\bfseries 04}
  (2020) 176}, \href{http://arxiv.org/abs/1906.05979}{{\ttfamily
  arXiv:1906.05979 [hep-th]}}.

\bibitem{Sepulveda:2019vrz}
D.~Garc{\'\i}a~Sep{\'u}lveda and A.~Guevara, ``{A Soft Theorem for the Tropical
  Grassmannian},'' \href{http://arxiv.org/abs/1909.05291}{{\ttfamily
  arXiv:1909.05291 [hep-th]}}.

\bibitem{Borges:2019csl}
F.~Borges and F.~Cachazo, ``{Generalized Planar Feynman Diagrams:
  Collections},'' \href{http://arxiv.org/abs/1910.10674}{{\ttfamily
  arXiv:1910.10674 [hep-th]}}.

\bibitem{Cachazo:2019ble}
F.~Cachazo, B.~Umbert, and Y.~Zhang, ``{Singular Solutions in Soft Limits},''
  \href{http://dx.doi.org/10.1007/JHEP05(2020)148}{{\em JHEP} {\bfseries 05}
  (2020) 148}, \href{http://arxiv.org/abs/1911.02594}{{\ttfamily
  arXiv:1911.02594 [hep-th]}}.

\bibitem{Cachazo:2019xjx}
F.~Cachazo, A.~Guevara, B.~Umbert, and Y.~Zhang, ``{Planar Matrices and Arrays
  of Feynman Diagrams},'' \href{http://arxiv.org/abs/1912.09422}{{\ttfamily
  arXiv:1912.09422 [hep-th]}}.

\bibitem{Guevara:2020lek}
A.~Guevara and Y.~Zhang, ``{Planar Matrices and Arrays of Feynman Diagrams:
  Poles for Higher $k$},'' \href{http://arxiv.org/abs/2007.15679}{{\ttfamily
  arXiv:2007.15679 [hep-th]}}.

\bibitem{Abhishek:2020xfy}
M.~Abhishek, S.~Hegde, D.~P. Jatkar, and A.~P. Saha, ``{Double Soft Theorem for
  Generalised Biadjoint Scalar Amplitudes},''
  \href{http://dx.doi.org/10.21468/SciPostPhys.10.2.036}{{\em SciPost Phys.}
  {\bfseries 10} (2021) 36}, \href{http://arxiv.org/abs/2008.07271}{{\ttfamily
  arXiv:2008.07271}}. \url{https://scipost.org/10.21468/SciPostPhys.10.2.036}.

\bibitem{Cachazo:2020wgu}
F.~Cachazo and N.~Early, ``{Planar Kinematics: Cyclic Fixed Points, Mirror
  Superpotential, k-Dimensional Catalan Numbers, and Root Polytopes},''
  \href{http://arxiv.org/abs/2010.09708}{{\ttfamily arXiv:2010.09708
  [math.CO]}}.

\bibitem{Drummond:2019qjk}
J.~Drummond, J.~Foster, O.~G\"urdogan, and C.~Kalousios, ``{Tropical
  Grassmannians, cluster algebras and scattering amplitudes},''
  \href{http://dx.doi.org/10.1007/JHEP04(2020)146}{{\em JHEP} {\bfseries 04}
  (2020) 146}, \href{http://arxiv.org/abs/1907.01053}{{\ttfamily
  arXiv:1907.01053 [hep-th]}}.

\bibitem{Drummond:2020kqg}
J.~Drummond, J.~Foster, O.~G\"urdo\u{g}an, and C.~Kalousios, ``{Tropical fans,
  scattering equations and amplitudes},''
  \href{http://arxiv.org/abs/2002.04624}{{\ttfamily arXiv:2002.04624
  [hep-th]}}.

\bibitem{fomin2002cluster}
S.~Fomin and A.~Zelevinsky, ``Cluster algebras i: foundations,'' {\em Journal
  of the American Mathematical Society} {\bfseries 15} no.~2, (2002) 497--529.

\bibitem{Fomin_2003}
S.~Fomin and A.~Zelevinsky, ``Cluster algebras ii: Finite type
  classification,'' \href{http://dx.doi.org/10.1007/s00222-003-0302-y}{{\em
  Inventiones mathematicae} {\bfseries 154} no.~1, (May, 2003) 63--121}.
  \url{http://dx.doi.org/10.1007/s00222-003-0302-y}.

\bibitem{2006math......2259F}
S.~{Fomin} and A.~{Zelevinsky}, ``{Cluster algebras IV: Coefficients},'' {\em
  arXiv Mathematics e-prints} (Feb., 2006) math/0602259,
  \href{http://arxiv.org/abs/math/0602259}{{\ttfamily arXiv:math/0602259
  [math.RA]}}.

\bibitem{fomin2003systems}
S.~Fomin and A.~Zelevinsky, ``Y-systems and generalized associahedra,'' {\em
  Annals of Mathematics} {\bfseries 158} no.~3, (2003) 977--1018.

\bibitem{Arkani-Hamed:2013jha}
N.~Arkani-Hamed and J.~Trnka, ``{The Amplituhedron},''
  \href{http://dx.doi.org/10.1007/JHEP10(2014)030}{{\em JHEP} {\bfseries 10}
  (2014) 030}, \href{http://arxiv.org/abs/1312.2007}{{\ttfamily arXiv:1312.2007
  [hep-th]}}.

\bibitem{Arkani-Hamed:2016byb}
N.~Arkani-Hamed, J.~L. Bourjaily, F.~Cachazo, A.~B. Goncharov, A.~Postnikov,
  and J.~Trnka, \href{http://dx.doi.org/10.1017/CBO9781316091548}{{\em
  {Grassmannian Geometry of Scattering Amplitudes}}}.
\newblock Cambridge University Press, 4, 2016.
\newblock \href{http://arxiv.org/abs/1212.5605}{{\ttfamily arXiv:1212.5605
  [hep-th]}}.

\bibitem{Arkani-Hamed:2017tmz}
N.~Arkani-Hamed, Y.~Bai, and T.~Lam, ``{Positive Geometries and Canonical
  Forms},'' \href{http://dx.doi.org/10.1007/JHEP11(2017)039}{{\em JHEP}
  {\bfseries 11} (2017) 039}, \href{http://arxiv.org/abs/1703.04541}{{\ttfamily
  arXiv:1703.04541 [hep-th]}}.

\bibitem{Arkani-Hamed:2017mur}
N.~Arkani-Hamed, Y.~Bai, S.~He, and G.~Yan, ``{Scattering Forms and the
  Positive Geometry of Kinematics, Color and the Worldsheet},''
  \href{http://dx.doi.org/10.1007/JHEP05(2018)096}{{\em JHEP} {\bfseries 05}
  (2018) 096}, \href{http://arxiv.org/abs/1711.09102}{{\ttfamily
  arXiv:1711.09102 [hep-th]}}.

\bibitem{He:2018pue}
S.~He, G.~Yan, C.~Zhang, and Y.~Zhang, ``{Scattering Forms, Worldsheet Forms
  and Amplitudes from Subspaces},''
  \href{http://dx.doi.org/10.1007/JHEP08(2018)040}{{\em JHEP} {\bfseries 08}
  (2018) 040}, \href{http://arxiv.org/abs/1803.11302}{{\ttfamily
  arXiv:1803.11302 [hep-th]}}.

\bibitem{Banerjee:2018tun}
P.~Banerjee, A.~Laddha, and P.~Raman, ``{Stokes polytopes: the positive
  geometry for $\phi^{4}$ interactions},''
  \href{http://dx.doi.org/10.1007/JHEP08(2019)067}{{\em JHEP} {\bfseries 08}
  (2019) 067}, \href{http://arxiv.org/abs/1811.05904}{{\ttfamily
  arXiv:1811.05904 [hep-th]}}.

\bibitem{Raman:2019utu}
P.~Raman, ``{The positive geometry for $\phi^{p}$ interactions},''
  \href{http://dx.doi.org/10.1007/JHEP10(2019)271}{{\em JHEP} {\bfseries 10}
  (2019) 271}, \href{http://arxiv.org/abs/1906.02985}{{\ttfamily
  arXiv:1906.02985 [hep-th]}}.

\bibitem{Aneesh:2019ddi}
P.~B. Aneesh, M.~Jagadale, and N.~Kalyanapuram, ``{Accordiohedra as positive
  geometries for generic scalar field theories},''
  \href{http://dx.doi.org/10.1103/PhysRevD.100.106013}{{\em Phys. Rev. D}
  {\bfseries 100} no.~10, (2019) 106013},
  \href{http://arxiv.org/abs/1906.12148}{{\ttfamily arXiv:1906.12148
  [hep-th]}}.

\bibitem{Kalyanapuram:2019nnf}
N.~Kalyanapuram, ``{Stokes Polytopes and Intersection Theory},''
  \href{http://dx.doi.org/10.1103/PhysRevD.101.105010}{{\em Phys. Rev. D}
  {\bfseries 101} no.~10, (2020) 105010},
  \href{http://arxiv.org/abs/1910.12195}{{\ttfamily arXiv:1910.12195
  [hep-th]}}.

\bibitem{Aneesh:2019cvt}
P.~B. Aneesh, P.~Banerjee, M.~Jagadale, R.~R. John, A.~Laddha, and S.~Mahato,
  ``{On positive geometries of quartic interactions: Stokes polytopes, lower
  forms on associahedra and world-sheet forms},''
  \href{http://dx.doi.org/10.1007/JHEP04(2020)149}{{\em JHEP} {\bfseries 04}
  (2020) 149}, \href{http://arxiv.org/abs/1911.06008}{{\ttfamily
  arXiv:1911.06008 [hep-th]}}.

\bibitem{Salvatori:2019phs}
G.~Salvatori and S.~Stanojevic, ``{Scattering Amplitudes and Simple Canonical
  Forms for Simple Polytopes},''
  \href{http://arxiv.org/abs/1912.06125}{{\ttfamily arXiv:1912.06125
  [hep-th]}}.

\bibitem{Arkani-Hamed:2019plo}
N.~Arkani-Hamed, S.~He, T.~Lam, and H.~Thomas, ``{Binary Geometries,
  Generalized Particles and Strings, and Cluster Algebras},''
  \href{http://arxiv.org/abs/1912.11764}{{\ttfamily arXiv:1912.11764
  [hep-th]}}.

\bibitem{Arkani-Hamed:2019vag}
N.~Arkani-Hamed, S.~He, G.~Salvatori, and H.~Thomas, ``{Causal Diamonds,
  Cluster Polytopes and Scattering Amplitudes},''
  \href{http://arxiv.org/abs/1912.12948}{{\ttfamily arXiv:1912.12948
  [hep-th]}}.

\bibitem{He:2020ray}
S.~He, L.~Ren, and Y.~Zhang, ``{Notes on polytopes, amplitudes and boundary
  configurations for Grassmannian string integrals},''
  \href{http://dx.doi.org/10.1007/JHEP04(2020)140}{{\em JHEP} {\bfseries 04}
  (2020) 140}, \href{http://arxiv.org/abs/2001.09603}{{\ttfamily
  arXiv:2001.09603 [hep-th]}}.

\bibitem{Arkani-Hamed:2020cig}
N.~Arkani-Hamed, T.~Lam, and M.~Spradlin, ``{Positive configuration space},''
  \href{http://arxiv.org/abs/2003.03904}{{\ttfamily arXiv:2003.03904
  [math.CO]}}.

\bibitem{He:2020onr}
S.~He, Z.~Li, P.~Raman, and C.~Zhang, ``{Stringy canonical forms and binary
  geometries from associahedra, cyclohedra and generalized permutohedra},''
  \href{http://arxiv.org/abs/2005.07395}{{\ttfamily arXiv:2005.07395
  [hep-th]}}.

\bibitem{Kalyanapuram:2020vil}
N.~Kalyanapuram and R.~G. Jha, ``{Positive Geometries for all Scalar Theories
  from Twisted Intersection Theory},''
  \href{http://dx.doi.org/10.1103/PhysRevResearch.2.033119}{{\em Phys. Rev.
  Res.} {\bfseries 2} no.~3, (2020) 033119},
  \href{http://arxiv.org/abs/2006.15359}{{\ttfamily arXiv:2006.15359
  [hep-th]}}.

\bibitem{Kalyanapuram:2020tsr}
N.~Kalyanapuram, ``{Geometric Recursion from Polytope Triangulations and
  Twisted Homology},''
  \href{http://dx.doi.org/10.1103/PhysRevD.102.125027}{{\em Phys. Rev. D}
  {\bfseries 102} (2020) 125027},
  \href{http://arxiv.org/abs/2008.06956}{{\ttfamily arXiv:2008.06956
  [hep-th]}}.

\bibitem{Kalyanapuram:2020axt}
N.~Kalyanapuram, ``{On Polytopes and Generalizations of the KLT Relations},''
  \href{http://dx.doi.org/10.1007/JHEP12(2020)057}{{\em JHEP} {\bfseries 12}
  (2020) 057}, \href{http://arxiv.org/abs/2009.10114}{{\ttfamily
  arXiv:2009.10114 [hep-th]}}.

\bibitem{Arkani-Hamed:2019mrd}
N.~Arkani-Hamed, S.~He, and T.~Lam, ``{Stringy Canonical Forms},''
  \href{http://arxiv.org/abs/1912.08707}{{\ttfamily arXiv:1912.08707
  [hep-th]}}.

\bibitem{Jagadale:2020qfa}
M.~Jagadale and A.~Laddha, ``{On the Positive Geometry of Quartic Interactions
  III : One Loop Integrands from Polytopes},''
  \href{http://arxiv.org/abs/2007.12145}{{\ttfamily arXiv:2007.12145
  [hep-th]}}.

\bibitem{Salvatori:2018aha}
G.~Salvatori, ``{1-loop Amplitudes from the Halohedron},''
  \href{http://dx.doi.org/10.1007/JHEP12(2019)074}{{\em JHEP} {\bfseries 12}
  (2019) 074}, \href{http://arxiv.org/abs/1806.01842}{{\ttfamily
  arXiv:1806.01842 [hep-th]}}.

\bibitem{Bazier-Matte:2018rat}
V.~Bazier-Matte, G.~Douville, K.~Mousavand, H.~Thomas, and
  E.~Y\i{}ld\i{}r\i{}m, ``{ABHY Associahedra and Newton polytopes of
  $F$-polynomials for finite type cluster algebras},''
  \href{http://arxiv.org/abs/1808.09986}{{\ttfamily arXiv:1808.09986
  [math.RT]}}.

\bibitem{padrol2019associahedra}
A.~Padrol, Y.~Palu, V.~Pilaud, and P.-G. Plamondon, ``Associahedra for finite
  type cluster algebras and minimal relations between $\mathbf{g}$-vectors,''
  2019.

\bibitem{speyer2005tropical}
D.~Speyer and L.~Williams, ``The tropical totally positive grassmannian,'' {\em
  Journal of Algebraic Combinatorics} {\bfseries 22} no.~2, (2005) 189--210.

\bibitem{Arkani-Hamed:2020tuz}
N.~Arkani-Hamed, S.~He, and T.~Lam, ``{Cluster configuration spaces of finite
  type},'' \href{http://arxiv.org/abs/2005.11419}{{\ttfamily arXiv:2005.11419
  [math.AG]}}.

\bibitem{ceballos2015cluster}
C.~Ceballos and V.~Pilaud, ``Cluster algebras of type d: pseudotriangulations
  approach,'' 2015.

\bibitem{brodsky2015cluster}
S.~B. Brodsky, C.~Ceballos, and J.-P. Labb{\'e}, ``Cluster algebras of type
  $d_4$, tropical planes, and the positive tropical grassmannian,'' 2015.

\bibitem{Yang:2019esm}
Q.~Yang, ``{Triangulations for ABHY Polytopes and Recursions for Tree and Loop
  Amplitudes},'' \href{http://arxiv.org/abs/1912.09163}{{\ttfamily
  arXiv:1912.09163 [hep-th]}}.

\bibitem{Arkani-Hamed:2019rds}
N.~Arkani-Hamed, T.~Lam, and M.~Spradlin, ``{Non-perturbative geometries for
  planar $\mathcal{N}=4$ SYM amplitudes},''
  \href{http://arxiv.org/abs/1912.08222}{{\ttfamily arXiv:1912.08222
  [hep-th]}}.

\bibitem{Golden:2013xva}
J.~Golden, A.~B. Goncharov, M.~Spradlin, C.~Vergu, and A.~Volovich, ``{Motivic
  Amplitudes and Cluster Coordinates},''
  \href{http://dx.doi.org/10.1007/JHEP01(2014)091}{{\em JHEP} {\bfseries 01}
  (2014) 091}, \href{http://arxiv.org/abs/1305.1617}{{\ttfamily arXiv:1305.1617
  [hep-th]}}.

\bibitem{Drummond:2017ssj}
J.~Drummond, J.~Foster, and O.~G\"urdo\u{g}an, ``{Cluster Adjacency Properties
  of Scattering Amplitudes in $N=4$ Supersymmetric Yang-Mills Theory},''
  \href{http://dx.doi.org/10.1103/PhysRevLett.120.161601}{{\em Phys. Rev.
  Lett.} {\bfseries 120} no.~16, (2018) 161601},
  \href{http://arxiv.org/abs/1710.10953}{{\ttfamily arXiv:1710.10953
  [hep-th]}}.

\bibitem{Golden:2018gtk}
J.~Golden and A.~J. Mcleod, ``{Cluster Algebras and the Subalgebra
  Constructibility of the Seven-Particle Remainder Function},''
  \href{http://dx.doi.org/10.1007/JHEP01(2019)017}{{\em JHEP} {\bfseries 01}
  (2019) 017}, \href{http://arxiv.org/abs/1810.12181}{{\ttfamily
  arXiv:1810.12181 [hep-th]}}.

\bibitem{Baadsgaard:2015voa}
C.~Baadsgaard, N.~Bjerrum-Bohr, J.~L. Bourjaily, and P.~H. Damgaard,
  ``{Integration Rules for Scattering Equations},''
  \href{http://dx.doi.org/10.1007/JHEP09(2015)129}{{\em JHEP} {\bfseries 09}
  (2015) 129}, \href{http://arxiv.org/abs/1506.06137}{{\ttfamily
  arXiv:1506.06137 [hep-th]}}.

\bibitem{Baadsgaard:2015ifa}
C.~Baadsgaard, N.~Bjerrum-Bohr, J.~L. Bourjaily, and P.~H. Damgaard,
  ``{Scattering Equations and Feynman Diagrams},''
  \href{http://dx.doi.org/10.1007/JHEP09(2015)136}{{\em JHEP} {\bfseries 09}
  (2015) 136}, \href{http://arxiv.org/abs/1507.00997}{{\ttfamily
  arXiv:1507.00997 [hep-th]}}.

\end{thebibliography}\endgroup

\end{document}